\documentclass[prd,eqsecnum,aps,showpacs,twocolumn]{revtex4}

 1
 1
 1

\def\D0{D\O}

\usepackage{graphicx}
\usepackage{bm}
\begin{document}
\title{A Model of Electrons, Photons, and the Ether}
\author{Robert L. McCarthy}                                                                                                         
\affiliation{Stony Brook University, Stony Brook, New York 11794}                                                                                  
\email{robert.mccarthy@stonybrook.edu}

\begin{abstract}

     This is an attempt to construct a {\it classical} microscopic model of 
the electron which underlies quantum mechanics.  An electron is modeled, not
as a point particle, but as the end of an electromagnetic string, a line of 
flux.  These lines stretch across cosmic distances, but are almost unobservable
because they condense into pairs---which form the ether.  Photons are modeled to 
propagate on these line pairs, which act effectively as wave guides.  These line 
pairs are also responsible for the force of gravity---which is electromagnetic 
in character. 

    Using our classical interpretation of an electron as a quantized 
electromagnetic field, we find that we achieve a natural interpretation of not 
only quantum mechanics, but also mass, spin, Mach's principle, gravity, self 
interactions, Newton's second law of motion, the quantization of the photon, 
wave-particle duality, inflation, dark energy, AGN jets, the Pauli Exclusion 
Principle, superconductivity, and the quantum Hall effects.  Mass generation
and inflation are discussed without the use of scalar fields.  We attempt to 
calculate Newton's gravitational constant G, the mass of the 
electron, the spin of the electron, and the spin of the photon---all in terms 
of Planck's constant, the charge of the electron, the outer radius of a line, 
and the distribution of matter in the universe. 

     The predictions of this model and the Standard Model are both compared to experiment, 
emphasizing areas where the  predictions differ.

     This model's remaining  problems are discussed.

     This model is a classical field theory of interacting strings which explains a large 
number of diverse phenomena, even though it does not discuss  the strong 
or weak interactions.  Since the ether of this model is quite well determined, it might be 
useful in restricting the landscape of vacua in more complete string theories.

\end{abstract}


\maketitle
\tableofcontents

\section{Introduction}
\label{sec-introduction}

     In the year 2000 the Standard Model was believed to be consistent with all
known experimental facts (except the possible existence of a neutrino mass).
Yet there are many aspects of the Standard Model which are assumed but not 
explained: the wave nature of matter, the quantization of the photon ($E=h\nu$), 
and Newton's Second Law of Motion.  In addition gravity is not usually considered
to be part of the Standard Model, but a separate interaction.  With the clear 
measurement of dark energy by the WMAP Collaboration\cite{WMAP}, it became 
important to know what makes up the majority of our universe.  The explanation
apparently does not lie within the Standard Model.  This paper investigates the
possibility that dark energy is caused by the wave nature of matter.  The resulting
model apparently succeeds in explaining dark energy.  It also appears to naturally
explain the quantization of the photon and Newton's Second Law, as well as provide
an electromagnetic model of gravity.  

     According to the Standard Model~\cite{standard} electrons are believed to
be point objects when examined on scales as small as $10^{-18}$m.  Recent 
limits~\cite{d0comp} show that quarks are also not composites of other objects
when examined on similar scales.  However, we know that quarks have
{\it external} structure.  They are bound together by color flux 
tubes\cite{colorflux} to form 
hadrons, so that free quarks cannot exist.  We here discuss the postulate that
the electron is similarly bound to another object
by a line of magnetic flux---a quantized vortex in the electromagnetic field.  We 
suggest that this line may be key to understanding the full meaning of spin.
Since we model the structure of the electron, 
it is possible to attempt a calculation of the properties of the electron in 
terms of Planck's constant $h$, the electron charge $-e$, and the 
distribution of matter in the universe.

     This is a {\it classical} model in the sense that we attempt to follow the
detailed motion of vortex-lines as a function of position in space and time.
The detailed motions of lines are then averaged to find the behavior of 
waves and particles---normally discussed in quantum mechanics.  
We do not follow the 
Copenhagen interpretation\cite{Copenhagen}
of quantum mechanics, but rather take the view, in sympathy with 
Einstein\cite{Einstein}, that it should be  possible {\it in principle} 
(though it would take a rather large computer) to predict the future course 
of the universe given {\it perfect} knowledge of the positions and velocities of all lines (and 
other objects---hadrons...) at a given time.  
Thus, in this view, indeterminacy is not intrinsic but results from measurement uncertainties.
Some authors\cite{Mermin} would call the positions
and velocities of these lines {\it hidden variables}.  These quantities 
are difficult, but not impossible, to measure.  Of course, all measurement uncertainties 
are subject to limitations from the Heisenberg Uncertainty Principle,
but discussion of the motions of the lines is useful because 
it can provide underlying structure to relate and interpret more measureable 
quantities.

     This model is a classical model of a quantized superconducting object.  Hence it is intrinsically
a quantized model in which e.g. Newtonian gravity is obtained as an average over
the line-pair distribution.  The individual quantum fluctuations are automatically
built in.  The classical solution for the motion of
electrons and lines appears much like in standard quantum theory.
                                                                                     
     This model is based on the quantization of magnetic flux in free space (discussed
in Section~\ref{ssec-FQ}, and which we claim was first derived by Dirac\cite{Dirac2}), which follows
from the requirement that wave functions be single valued.  We take this result from 
quantum mechanics and use it as a general principle which reaches its most general form
in equation~\ref{eq-quantization}, which implies the relativistic London equations~\ref{eq-London} for our 
superconducting fields.  We make a major assumption in Section~\ref{ssec-electroncharge} about the origin of electric
charge.  Aside from this quantization condition and our assumption
about electric charge, this classical model of
a quantized object is based on the Principle of
Least Action, or Hamilton's principle.  The Principle of Least Action leads to the microscopic 
Maxwell equations which we take to be valid at arbitrarily small length scales:

\begin{equation}
   {\bf \nabla \cdot E} = 4 \pi \rho
\end{equation}
\begin{equation}
   {\bf \nabla \cdot B} = 0
\end{equation}
\begin{equation}
   {\bf \nabla \times E} = - {1\over c} {\partial {\bf B} \over {\partial t}} 
\end{equation}
\begin{equation}
   {\bf \nabla \times B} = {1\over c} {\partial {\bf E} \over {\partial t}} + {4 \pi \over c} {\bf J}
\label{eq-ampcirc}
\end{equation}

\noindent
We generally follow the conventions of Jackson\cite{Jackson}.  The electric and magnetic induction
fields can be defined in terms of the vector and scalar potentials
\begin{equation}
   {\bf E} = - {\bf \nabla} V - {1 \over c} {\partial {\bf A} \over {\partial t} }
\end{equation}
\begin{equation}
   {\bf B} = {\bf \nabla} \times {\bf A}
\end{equation}
\noindent
which we presume to satisfy the Lorentz gauge condition
\begin{equation}
{1 \over c} {\partial V \over {\partial t} } + {\bf \nabla \cdot A} = 0 = 
{{\partial A^\mu} \over {\partial x^\mu}}
\label{eq-Lorentzgauge}
\end{equation}
\noindent
so that the four-vector potential and current density take their standard form
\begin{equation}
   A^\mu = (V,\bf A)^\mu
\end{equation}
\begin{equation}
   J^\mu = (c \rho , \bf J)^\mu
\end{equation}
\noindent
for $ \mu = (0,3)$ .  However, we deviate from the standard assumptions
by asserting that gauge invariance is broken and the electromagnetic four-vector 
potential $A^\mu$, at least sometimes, is directly measureable.  Gauge invariance is already 
asserted \cite{gaugeTonomura} to be broken in superconductors, and this fact
is quite relevant to our present dicussion.  We prove below (Section~\ref{ssec-static}) that in
a very important special case, our line of flux, the vector potential is unique.

     Our default reference frame uses flat Minkowski space referred to the local rest
frame of the universe.  This is defined to be the frame in the region of our
solar systsem in which the 2.73K black-body radiation is isotropic\cite{bbframe}.

     We claim that this is a classical model even though it starts with the Dirac quantization
condition---a quantum mechanical result which implies the structure of our line of flux.  Since
electron wave functions in standard quantum mechanics have infinite extent, the only way in which
they can be single-valued in the presence of magnetic flux
is if the lines of flux are quantized and filamentary.  This implies
that the magnetic flux in the lines must be contained (equation~\ref{eq-ampcirc}).  Since normal charged particles
are not sufficient to contain this  flux in general, the most natural assumption we can make is that
the flux is contained by a superconducting line, with charge carriers of a new type.

     If we take this model seriously, it pays great dividends.  We presume that one end of a line
is an electron, the other a positron (or ends in a proton).  Thus space-time is crisscrossed by many such lines
(in pairs - Section~\ref{ssec-overview}).  Then, in turn, all fundamental quantization conditions---including the
Dirac quantization condition, the quantization of the photon, and the quantization of matter waves---can 
 be understood from the fact that space-time is a multiply connected superconductor (equation (\ref{eq-multisup})).
This, in turn, implies quantum mechanics itself, when the interaction of a bare electron  with the
flux lines is taken into account (Section~\ref{sec-quantmech}), and appropriate averages are taken.  Thus, the Dirac
quantization condition, our starting point, is understood at a deeper level and, in fact, implied
by our classical model in a self-consistent way.  The classical model implies quantum mechanics, but the
reverse is not true since the {\it classical} model is at the deeper level.  (We, of course, take
the value of $h$ from experiment and we define {\it classical} wave functions which are 
{\it not} continuous, before smoothing to find the quantum mechanical wave functions.)  
We suggest that at the birth of quantum mechanics the problem
which the founding fathers solved was not due to the inadequacy of classical mechanics.  They had the 
wrong model of the electron.  It's not a point particle, but rather the end of a string. 

     Since our method is totally new, it seems appropriate to summarize the highlights of our model
to orient the reader---to let the reader know what we will claim:

\begin{enumerate}
\item Our model, many superconducting line pairs crisscrossing in space-time, naturally explains
      six dominant aspects of everyday physics which are either unexplained or questionable
      in the Standard Model:
\begin{itemize}
\item quantization of the photon (Section~\ref{ssec-quantphoton}) and matter waves (Section~\ref{sec-quantmech}): $E=h\nu$
\item wave-particle duality (Section~\ref{sec-wpduality}) and the wave nature of matter (Section~\ref{sec-quantmech})
\item gravity (Section~\ref{sec-gravity})
\item Newton's $2^{nd}$ Law of Motion (Section~\ref{sec-Newton}):\ ${\bf F}=m{\bf a}$
\item mass of the electron (Section~\ref{ssec-masselectron})
\item spin of the electron (Section~\ref{ssec-spinelectron}) and spin of the photon (Section~\ref{ssec-spinphoton})
\end{itemize}
\item Our model gives a natural explanation of dark energy (Section~\ref{ssec-darkenergy}).  The line pairs are
      under tension or negative pressure (Section~\ref{ssec-tensiondarkenergy}).
\item We have modeled the propagation of the photon as a helical wave on the two lines of a pair
      (Section~\ref{ssssec-helicalphoton}).  After applying the photon quantization condition and making a binding
      energy correction (Section~\ref{ssec-energyphoton}), we calculate the spin of the photon (Section~\ref{ssec-spinphoton}) to 
      be $\hbar$ with no free parameters.
\item Our natural model of the bare electron (end of a line of flux) has the pole field of a Dirac
      monopole plus a line field (inside a superconducting sheath).  After
      considering interaction with the surrounding line pairs, the motion of this bare electron is
      {\it zitterbewegung} (Section~\ref{ssec-motionbpp}), very rapid motion  near the speed of light in rapidly 
      changing directions.  The line pairs completely screen the pole field due to the
      quantization of magnetic flux so that the magnetic induction field reaching macroscopic distances 
      is zero.
\item In order to agree with experiment we must add the electric charge of an electron by hand 
      (Section~\ref{ssec-electroncharge}) to our model of the bare electron.  This is consistent with the Principle of
      Least Action and does not change the character of the zitterbewegung motion (Appendix~\ref{ssec-motionelectron}).
      The unnatural part of this addition is that the electric charge and effective magnetic charge
      of the electron must have their relative separation {\it fixed} at a very small distance parallel to
      the line.  This direction is the direction of the spin of the electron, and the magnitude of the 
      spin is calculated to be $\hbar/2$.
\item Using a toy version of our model (which we hope to be accurate within a factor of 10), we have
      calculated formulas for Newton's constant $G$ (equation (\ref{eq-G})) and the mass of the electron 
      (equation (\ref{eq-me})), containing some parameters whose values we do not currently know.  We have also 
      calculated the dark energy density of the universe.  We then combine measurements
      of the dark energy density, $G$ and $m_e$ to calculate the line pair density near the Earth
      ($4\times 10^{39}\ line\ pairs\  /\  meter^2$) (equation (\ref{eq-fulldensity})), which is consistent with known
      constraints, and a factor of about $10^{14}$ larger than would be expected without inflation.
\item Our model predicts that the gravitational force will reverse under conditions of very high density.
      At least heuristically, this could explain inflation and AGN jets.
\item We offer a new explanation of the Pauli Exclusion Principle.  We need to do this since our
      model ties a string on every electron in the universe, and therefore electrons are {\it distinguishable}.
      The explanation occurs naturally in the model.
\item We offer a new explanation of the quantum Hall effects simply by suggesting that the flux 
      lines seen in these effects are real.

\end{enumerate}

    This model amounts to a classical field theory of interacting strings (or lines).  The assumption
that lines will condense into pairs avoids having to solve the general case.  
We presume that intercommutation is not an 
issue since all excitations are quantized and excitations with a small wavelengh must have a high energy.

    Since this model implies quantum mechaincs and ${\bf F}=m{\bf a}$ for the electron, differences from the Standard Model
should be hard to find.  (But see Section~\ref{sec-relevantexperiments}.)  In order to restrict the  scope of our discussion,
weak and strong interactions are essentially not discussed.

    We use the word $``ether"$ in the title of this paper with some trepidation, because it is an anathema
to most physicists today.  If the reader prefers he/she can substitute a euphemism such as $``vacuum"$.
But we have apparently succeeded in describing the photon propagating as a quantized electromagnetic wave 
on one of our line pairs---which was the original function of the ether.  We have also apparently succeeded
in modeling  dark energy, a uniform energy density in space which is widely believed to be inconsistent
with the Standard Model.  This real, measureable energy density is not consistent with the idea of $``vacuum"$.  
In Section~\ref{ssec-overview} we address Einstein's reasons for rejecting the ether concept.

     But (using the language of string theory), we may have found the 
``Vacuum Selection Principle"\cite{Douglas} relevant to our universe.  It makes sense that only
electromagnetic strings exist outside nuclei, because the nuclear force has a short range.  Our selection
principle is that every neutron, or electron-proton combination, is the source of a pair of flux lines, probably
stretching out to great distances in all directions.  Our model gives a reasonable description of dark energy,
gravity and the generation of mass without scalar fields.

\section{Mach's Principle}
\label{sec-Mach}

     Newton's second law of motion, ${\bf F} = m {\bf a}$ , is valid only in {\it inertial}
frames.  The fact that inertial frames at a given point
all move with constant velocity relative to each 
other implies that acceleration has absolute significance in our universe, unlike
velocity for which only relative motion has significance.  With respect to what must
acceleration be defined?  Mach pointed out\cite{Mach} that the mass of the Earth and the 
other celestial bodies must determine the inertial frames.  Inertial frames 
near the Earth do not seem to 
accelerate with respect to the average mass of the universe.  This paper was 
stimulated by the fact that somehow information on the the average local rest frame 
of the universe must be present {\it at} each object, to act as its absolute 
reference for acceleration.  This information could not be carried by a smooth
gravitational field.  If we were far from any stars, the Newtonian gravitational field 
would add to zero in a symmetric universe.  
Somehow the gravitational field of the universe must have {\it structure}.
It must be quantized.

\section{One Line}
\label{sec-one}

\subsection{Flux Quantization}
\label{ssec-FQ}

     Following Dirac\cite{Dirac1} we consider a region of space containing a wave 
function of a particle of charge $q$.  If a nodal line appears in that wave function, 
this nodal line can
contain magnetic flux as long as the wave function remains single valued\cite{single}
in the presence of the flux.  This requires that 

\begin{equation}
exp{\{ {i q \over \hbar c} \oint {\bf A \cdot dx}\} } = 1
\label{eq-singv}
\end{equation}

\noindent
where the path of the integral includes the line.  This implies the 
quantization of magnetic flux in the line

\begin{equation}
\oint {\bf A \cdot dx} = \int {\bf B \cdot dS} = 2 \pi {\hbar c \over q } n = n \Phi_0
\label{eq-FQ}
\end{equation}

\noindent
where $\Phi_0 = {h c / q}$ is the quantum of flux\cite{Londonflux} and $n$ is an 
integer.  This condition is usually\cite{Dirac2} taken to imply that charge is 
quantized if magnetic monopoles exist.  We reason that electron wave functions
fill the universe and therefore we take this condition to imply the
quantization of magnetic flux in free space\cite{Saffouri} (with {\it q = e}), 
which, as we will see, leads in a fairly natural way to the
quantization of not only electric charge, but also spin angular momentum---even
though true monopoles do not exist, at least in this model.  (The experimental situation
on this point is discussed in Section~\ref{ssec-existing}.)

\subsection{Filamentary Model}
\label{ssec-filament}

\begin{figure}
    \vbox to 2.6in{
\includegraphics[scale=0.65]{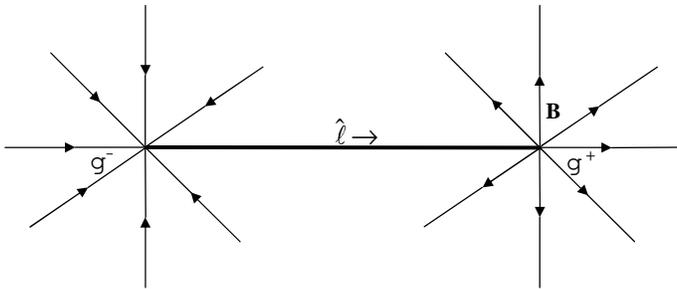} }
    \caption{\small A line of flux.}
    \label{fig-line}
\end{figure}

     Figure~\ref{fig-line} shows the building block of our model---a line of 
flux\cite{Faraday}---static, for the moment.  
We only consider the case $n=1$ in equation (\ref{eq-FQ}).  Here we consider the case 
$q = e$ appropriate if the
wave function is that of an electron (charge $-e$).  Magnetic induction 
field enters through one pole (labelled $g^-$), continues through the line
and exits through the other pole
($g^+$).  The total flux exiting $g^+$ is $\Phi_0 = 4 \pi g$, where 
\begin{equation}
g = {\hbar c \over 2 e}
\end{equation}
\noindent
is the strength of a Dirac monopole\cite{Dirac2}.  Hence the line of flux is a 
Dirac string.  Dirac introduced a singularity to exclude the line
from physical reality so that the  pole is a monpole.  Here we {\it include}
the line in our picture of reality and attempt to interpret one end of the 
line as an {\it electron} and the other as a {\it positron}. 
Since we assert that gauge invariance is broken, the position
of the line is observable\cite{Jackson2}.  Note that the 
total magnetic flux through the positive pole is zero, since we must include
the line field as well as the pole field.  We will speak of the total
configuration as a {\it vortex}.  It consists of one line field and two pole 
fields, one of each sign.  The {\it line} is a {\it string} and we will use 
both terms, depending on the context.

\begin{figure}
    \vbox to 2.0in{
\includegraphics[scale=0.75]{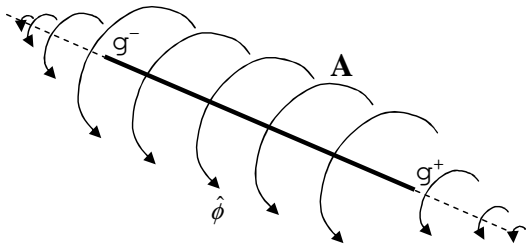}  }
    \caption{\small Vector potential of a vortex/line of flux.}
    \label{fig-vortex}
\end{figure}

Figure~\ref{fig-vortex} shows the vector
potential associated with a static vortex/line of flux and indicates why we call it a vortex.  
The differential dipole magnetic moment
of an element of the line is
given\cite{Jackson2} by

\begin{equation}
d{\bf m} = g d{\bf y}
\end{equation}

\noindent
where ${\bf y}$ denotes a point on the line and
$d{\bf y}$ points in the direction 
of the magnetic induction field along the line (which we call $\hat{\ell}$)
so that the total field of the vortex at the field point ${\bf x}$ is given by

\begin{equation}
{\bf A(x)} = -g \int_\ell  d{\bf y} {\bf \times \nabla}_x \Bigl( { 1 \over {\bf \mid x - y \mid} }\Bigl) \ \ \ .
\label{eq-agenline}
\end{equation}

\noindent
Taking the curl of ${\bf A(x)}$ we find the magnetic induction field of the vortex

\begin{equation}
{\bf B(x)} = 
    -\  g\ \Bigl({{{\bf x}-{\bf y}_{N}} \over{{\mid {\bf x} - {\bf y}_{N} \mid}^3 }}\Bigl)\ 
    +\ \Phi_0 \  \delta^{(2)}_T({\bf x} - {\bf y}) {\hat{\ell}}\  
    +\  g\ \Bigl({{{\bf x}-{\bf y}_{P}} \over{{\mid {\bf x} - {\bf y}_{P} \mid}^3 }}\Bigl)
\label{eq-singular}
\end{equation}

\begin{figure}
    \vbox to 1.5in{
\includegraphics[scale=0.75]{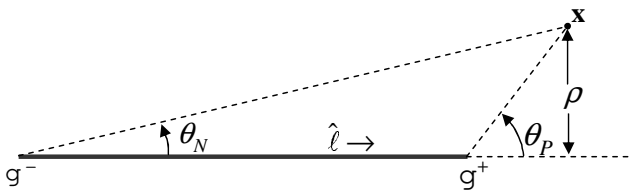}  }
    \caption{\small Geometrical quantities associated with a vortex/line of flux.}
    \label{fig-geom}
\end{figure}

\noindent
where ${\bf y}_{N}$ and ${\bf y}_{P}$ are the 
positions of the negative and positive poles at the ends of the
line.  The two-dimensional Dirac delta function operates in the plane perpendicular
to ${\hat{\ell}}$.  The line is presumed to have no sharp kinks, but it can
bend smoothly in an arbitrary manner.  In the particular case that the static line is straight,
the vector potential from the vortex at the field point ${\bf x}$ is given by

\begin{equation}
{\bf A(x)} = {2 g \over {\rho} }\ {[sin^2 (\theta_P/2) - sin^2 (\theta_N/2)]}\ {\hat{\phi}}
\label{eq-astline}
\end{equation}

\noindent
where $\theta_P$ and $\theta_N$ are the polar angles from the positive and negative poles
to the point ${\bf x}$ respectively, as shown in Figure~\ref{fig-geom}, where $\rho$ is the cylindrical radius 
to point ${\bf x}$ from the line (or extension of the line), and $\hat{\phi}$ is the azimuthal unit vector. 

\subsection{Classic Vortices}
\label{ssec-classic}

      Lamb\cite{Lamb} has summarized the classic work on vortices in the 1800's, principally
by Helmholtz\cite{Helmholtz} and Lord Kelvin\cite{Kelvin1}.  Most of this early work was stimulated 
by the circulation theorem of Helmholtz, which showed that in a perfect (nonviscous
and incompressible) fluid, the circulation
of a vortex $\oint {\bf v \cdot dx}$ is an absolute constant, where {\bf v} is the fluid 
velocity and the path of integration includes the vortex and moves with it.  Hence vortices retain
individual identity so that Kelvin proposed\cite{Kelvin2} that ``vortex atoms" might 
be the building blocks of matter (30 years before the discovery of the electron). 

     The quantized vortices discussed in this paper bear strong resemblance to the 
classic vortices with the identification that the vector potential ${\bf A}$
corresponds to the classical fluid velocity ${\bf v}$ and that the magnetic induction  
field ${\bf B}$ corresponds to the classic vorticity (${\bf \nabla \times v}$).
The circulation around one of our vortices $\oint {\bf A \cdot dx} = \Phi_0$ (far
from the ends to avoid flux returning from the poles) is quantized,
implying that they also retain a separate identity and are good candidates for
building blocks.  The classic vortices could not end, so that they either had to 
form rings or end on the walls of the container.  Our vortices, in the terminology
of the classic works, actually do not end either.  Since the fields of our vortices
satisfy the Maxwell equation ${\bf \nabla \cdot B} = 0$ everywhere, technically our vortices 
do not end at the poles.  Their cross sections just become very large at the poles, and
the magnetic field extends to infinity in all directions.

     There are two major differences, however, between our quantized vortices and the 
classic vortices.  Since our vortices are formed in the electromagnetic field, they
have no rest mass and do not sense pressure in the normal way, except at the poles.  (See Section~\ref{sec-Pauli}.)  
Consequently we must develop new techniques to follow the motions of our vortices.  

\subsection{Static Model of the Core}
\label{ssec-static}

     Since we wish to consider the energies of the lines of flux, the singular behavior of the
magnetic induction field (equation (\ref{eq-singular})) in the filamentary model is unacceptable.  We need to 
integrate ${\bf B(x)}^2$.  Consequently we develop a model of a line of flux with a 
finite radius $a$, called the {\it line radius}, which we take to be constant.  For the moment we consider only a straight 
line far from its ends.  We simply use a model of the Abrikosov vortex~\cite{Abrikosov} which has been 
successful\cite{TypeII,Devices} in modeling the vortices in type II superconductors
\begin{equation}
{\bf B(x)} = {2 g \over {a^2} } \  K_0(\rho/a) \ \hat{\ell}
\label{eq-bline}
\end{equation}
where $K_0$ is the hyperbolic Bessel function of the second kind\cite{Arfken} of order $0$.
The line radius plays the role of the penetration depth in a type II superconductor.
Since the magnetic flux is confined, there must be a current distribution associated 
with the line.  In the static case, from equation (\ref{eq-ampcirc}) we find
\begin{equation}
{\bf J(x)} = {c \over {4 \pi}} \ {\bf \nabla \times {\bf B}} = {{gc} \over {2 \pi a^3} }\  K_1(\rho/a)\ \hat{\phi} 
\end{equation}
where $K_1$ is the hyperbolic Bessel function of the second kind of order 1.  We assume 
that this current density is {\it superconducting}.  It is associated with the magnetic field
at zero voltage, as in a type II superconductor.  We assume the superconducting region 
around the line extends to a distance $R_a=k_aa$ from the centerline, where $k_a\gg1$,
so that the flux is contained to any required accuracy.  We will try to measure $R_a$.
We assume that the current density satisfies the 
London equations\cite{London2} since, if a coherence length is relevant to this 
situation, we take it to be negligible.  We treat this current density as a {\it field},
so we can take its charge carriers to be massless (Section~\ref{ssec-superconfield}).
Hence the current density is a new concept since it cannot be due to the motion of known
charged parcicles!  These assumptions seem required by our main
point, that, following (\ref{eq-FQ}), magnetic flux is quantized in free space.  
We will see that every wave function contains such lines, and therefore all 
wave functions superconduct.  This last statement is not surprising;
it is just diamagnetism\cite{diamag}.  Hence the surprising assumptions which we have been led to make
in this section have one reasonable consequence.  We need to see what else is implied.

     With this magnetic induction field, or magnetic flux density, 
the total flux through the line at a given point on the centerline is
\begin{equation}     
\int {\bf B \cdot dS} = {\Phi_0 \over {a^2}} \int_0^\infty d\rho\ \rho\ K_0(\rho/a) = \Phi_0 
\end{equation}
as required.  The self-energy per unit length of the line is given by
\begin{equation}
{dU \over {d\ell}} = {1 \over {8\pi}} \int d\rho\ \rho\ d\phi\ B^2 = {g^2 \over {2 a^2}} = T
\label{eq-T}
\end{equation}
where $T$ is the tension in the line.  The differential magnetic moment of an element of the 
line is given by
\begin{equation}
d{\bf m} = {1 \over {2c}} \int d^3x \ \bm{\rho} \times {\bf J} = g\ d{\bf y} 
\end{equation}
also as required.  Following the expected behavior\cite{Kittel} in superconductor we assume a trial solution
for the vector potential given by
\begin{equation}
{\bf A}_{trial} = - {{4\pi a^2} \over c} \  {\bf J} = - {{2g} \over a}\  K_1(\rho/a)\  {\hat \phi} \ \ \ .
\end{equation}
Taking the curl of ${\bf A}_{trial}$, we indeed find equation (\ref{eq-bline}).
But our trial solution has the wrong behavior for large $\rho$ (see equation (\ref{eq-astline}), for points far from the ends) 
 since $K_1(x) \rightarrow (\pi/{2x})^{1/2}\ e^{-x}$ for $x \gg 1$.
In order to obtain the proper solution we must add a field with zero curl (except at $\rho = 0$)
\begin{equation}
{\bf A}_{long-distance} = {{2g} \over \rho} \  {\hat \phi}
\end{equation}
which gives the correct long-distance field of the line, so that the total field is given by
\begin{equation}
{\bf A(x)} = {{2g} \over \rho} \ (1 - {\rho \over a} \ K_1(\rho/a) ) \ {\hat \phi} \ .
\label{eq-aunique}
\end{equation}
The curl of (\ref{eq-aunique}) gives the correct magnetic induction 
field of a static line (\ref{eq-bline}) at all points\cite{curlrho0} (not near the ends of the line).   Hence (\ref{eq-aunique}) gives the 
{\it unique} solution
for the vector potential of the line (for n=1, not near the ends).  It is determined by the azimuthal symmetry of 
the line and the quantization condition (\ref{eq-FQ}). 

\subsection{The Phase Field}
\label{ssec-phase}

     Actually equation (\ref{eq-singv}) is a special case.  The requirement that a particle's wave function
be single valued at a fixed time is stated more generally by
\begin{equation}
exp{\{ {i \over \hbar} \oint {\bf p}_s\ {\bf \cdot\ dx}\} } = 1
\label{eq-singvps}
\end{equation}
where
\begin{equation}
{\bf p}_s = m {\bf v}_s\ +\ {q \over c} {\bf A}
\label{eq-ps}
\end{equation}
is the total momentum of the particle, and where we use a notation common in the
case that the wave function represents superconducting pairs of electrons.  If we extrapolate
from the case of superconducting pairs of electrons
\begin{equation}
{\bf v}_s = { { \bf J} \over {n_s q} }
\label{eq-vs}
\end{equation}
(where $n_s$ is the number of superconducting pairs per unit volume), to our
line current density in terms of the penetration depth\cite{Kittel}
\begin{equation}
  \lambda_L =  {\Bigl( { {mc^2} \over {4\pi n_s q^2} } \Bigl) }^{1/2} = a
\label{eq-lambdaL}
\end{equation}
(which has been identified with the line radius) we find
\begin{equation}
{\bf v}_s = {\hbar \over {ma}}\ K_1(\rho/a)\ {\hat \phi}\
\end{equation}
and hence
\begin{equation}
{m\bf v}_s = {\hbar \over {a}}\ K_1(\rho/a)\ {\hat \phi}\ .
\end{equation}
From (\ref{eq-aunique})
\begin{equation}
{q \over c} {\bf A(x)} = {\hbar \over \rho}{\hat \phi} \ - \ {\hbar \over a} \ K_1(\rho/a) \ {\hat \phi} \ 
\end{equation}
we can find the generalization of ${\bf p}_s$ to our case
\begin{equation}
{\bf p}_s = {\hbar \over \rho}{\hat \phi}
\end{equation}
which we call the {\it phase field}\cite{Yang} of the vortex, because,
in the case of our vortex, it does not behave like a momentum.  It is independent of both $m$ and $n_s$.
We associate this phase field with the vortex supercurrent wave function.  
We say that this new type of 
current is due  to {\it aon} superconductivity, in order to distinguish it from superconduction by Cooper 
pairs.  We will see that this new phenomenon is completely characterized by the line radius, which we 
call {\it a}.

     Evidently, in our model of the core with a finite size, the phase field 
still has a filamentary curl, and in the general static case it is given by (see equation (\ref{eq-agenline}))
\begin{equation}
{\bf p}_s ({\bf x}) = - {\hbar \over 2}
 \int_\ell  d{\bf y} {\bf \times \nabla}_x \Bigl( { 1 \over {\bf \mid x - y \mid} }\Bigl)\ 
\label{eq-psx}
\end{equation}
which is just the law of Biot and Savart.   

     A supercurrent (i.e. aon) wave function will be single valued as long as
\begin{equation}
   {\bf \nabla\ \times\ } {\bf p}_s = 0
\end{equation}
throughout the region of interest.  In general, using the line of reasoning for
equation (\ref{eq-singular}), the curl of the phase field of a single static vortex 
not including the ends (where ${\bf p}_s$ is not smooth) is given by
\begin{equation}
{\bf \nabla\ \times\ } {\bf p}_s = h \  \delta^{(2)}_T({\bf x} - {\bf y}) {\hat{\ell}}\ \ \ .
\end{equation}
The line field itself is quantized precisely so that its supercurrent wave function encircling the
centerline is single-valued.   Since the curl of the phase field is filamentary, there is no
region of space in which the wave function of one line is not single-valued
and therefore two lines (otherwise isolated) can move through each other 
driven by their magnetic interaction---with 
no extra currents (quantization currents) needed to make their wavefunctions single-valued\cite{phaseindep}.  
We will see (Section~\ref{ssec-effectivemagcharge}) that this is {\it not} true for the ends of a line.

     The London equations\cite{London} may be expressed very simply in terms of the phase field
\begin{equation}
   {\bf \nabla\ \times}\ {\bf p}_s = 0
\end{equation}
\begin{equation}
   {\bf \nabla} {p^0_s}\ +\ {1\over c} {\partial {\bf p}_s \over {\partial t}} = 0
\end{equation}
where ${p^0_s}$ is the 0-component of the 4-vector
\begin{equation}
   {p^\mu_s}\ = \ m{u^\mu_s} \ + \ {q \over c} A^\mu
\label{eq-4ps}
\end{equation}
and ${u^\mu_s}$ is the 4-velocity (extrapolating from the case of
 superconducting pairs of electrons).  
The relativistic form of the London equations
\begin{equation}
    {  {\partial{p^\nu_s}} \over {\partial x_{\mu} } } \ 
  - \ {  {\partial{p^\mu_s}} \over {\partial x_{\nu} } } \ = 0
\label{eq-London}
\end{equation}
just implies that
\begin{equation}
     {p^\mu_s} = { {\partial \chi} \over {\partial x_{\mu}} }
\end{equation}
the phase field can be expressed as the gradient of a scalar ($\hbar$ times the phase).
Evidently the center of the vortex is outside the superconducting region,
as in a type II superconductor.  This implies a phase transition in the center
of the vortex.

     If we allow time variation, the requirement that a wave function be single-valued
can be written in all generality
\begin{equation}
exp{\{- {i \over \hbar} \oint {p^\mu_s} dx_{\mu} \} } = 1
\end{equation}
so that
\begin{equation}
     \oint {p^\mu_s} dx_{\mu}\ = \ 2 \pi n \hbar\ = \ nh\  .
\label{eq-quantization}
\end{equation}
The London equations imply that this requirement is automatically satisfied
\begin{equation}
     \oint {p^\mu_s} dx_{\mu}\  = \ {1 \over 2}\ \oint dS_{\mu \nu}\ \Bigl( { {\partial{p_s}^{\nu}} \over {\partial x_{\mu} } }\
  - \ {  {\partial{p_s}^{\mu}} \over {\partial x_{\nu} } } \ \Bigl) \ = \ 0
\label{eq-quantization0}
\end{equation}
{\it if} the path and surface are totally in superconductor.
(We use Stokes theorem in four dimensions and $dS_{\mu \nu}$
is an element of the area enclosed by the path.) 

\section{Two Lines}
\label{sec-two}

     We consider the interaction of two static lines (far from their ends), first in the
case in which the magnetic induction fields of both lines are aligned with the $z$-axis.  The
magnetic energy per unit length is 

\begin{equation}
   {dU \over dz}\ =\ { 1 \over {8\pi} } \int \ \rho\ d\rho\ d\phi\ ({\bf B}_1\ +\ {\bf B}_2)^2
\end{equation}
\begin{equation}
    =\ { 1 \over {8\pi} } \int \ \rho\ d\rho\ d\phi\ ({B_1}^2\ +\ {B_2}^2\ +\ 2{\bf B}_1 \cdot {\bf B}_2)
\end{equation}
\begin{equation}
    =\ {{g^2} \over {a^2}}\ +\ {{dU_{12}} \over {dz}}
\end{equation}
where $U_{12}$ is the interaction energy.  This interaction energy is easily 
calculated
\begin{equation}
{{dU_{12}} \over {dz}} = {{g^2} \over {\pi a^4}} \int \ \rho\ d\rho\ d\phi\ K_0(\rho/a)\ 
           K_0({\mid{\bm\rho}-{\bm\rho}^\prime \mid}/a)
\end{equation}
(where ${\bm\rho}^\prime$ locates the center of line $2$ relative to line $1$)
using the expression obtainable from Jackson\cite{Jackson3}
\begin{widetext}
\begin{equation}                      
K_0({\mid{\bm\rho}-{\bm\rho}^\prime \mid}/a)=
  I_0(\rho_</a)K_0(\rho_>/a)+2\sum_{m=1}^\infty cos[m(\phi-\phi^\prime)]I_m(\rho_</a)K_m(\rho_>/a) 
\label{eq-K0fact}
\end{equation}
\end{widetext}
for writing $K_0({\mid{\bm\rho}-{\bm\rho}^\prime \mid}/a)$ in
completely factorized form.  Here $\rho_>$($\rho_<$) is the greater (lesser) of $\rho$,${\rho}^\prime$.
If $u \equiv \rho/a$ and $u^\prime \equiv \rho^\prime/a$ we obtain
\begin{equation}
{{dU_{12}} \over {dz}}\ =\ {{g^2}\over{a^2}}{u^\prime}^2 K_1(u^\prime)[I_1(u^\prime)K_0(u^\prime)+I_0(u^\prime)K_1(u^\prime)]\ \ .
\end{equation}
Recognizing a Wronskian determinant\cite{Jackson3}
\begin{equation}
   I_1K_0\ +\ I_0K_1\ =\ W[K_0,I_0]\ =\ +{1 \over {u^\prime}}
\label{eq-Wronskian}
\end{equation}
we find that the interaction energy is given by
\begin{equation}
  {{ dU_{12}} \over dz}\ =\ {{g^2} \over {a^2}} u^\prime K_1(u^\prime)
\end{equation}
and the energy per unit length of two aligned lines is
\begin{equation}
  {dU \over dz}{\Biggl\vert}_{aligned}\ =\ {{g^2}\over{a^2}}\ \Bigl[ 1+{u^\prime} K_1(u^\prime) \Bigr]
                                        \ =\ {{g^2}\over{a^2}}\ \Bigl[ 1+{{\rho^\prime}\over a}K_1({\rho^\prime}/a) \Bigr] \ \ .  
\end{equation}
Since $u^\prime K_1(u^\prime) \rightarrow 1$ as $u^\prime \rightarrow 0$, the total
energy of two aligned lines is four times as great as the energy of one line (see equation (\ref{eq-T})), as expected.
Hence two lines of like orientation {\it repel} each other.

     From the above discussion, two lines with opposite orientation will attract
each other.  The interaction energy simply reverses sign.
\begin{equation}
  {dU \over dz}{\Biggl\vert}_{oppposite}\ =\ {{g^2}\over{a^2}}\ \Bigl[ 1-{u^\prime} K_1(u^\prime) \Bigr]
                                        \ =\ {{g^2}\over{a^2}}\ \Bigl[ 1-{{\rho^\prime}\over a}K_1({\rho^\prime}/a) \Bigr] \ \ .
\label{eq-twolinepot}
\end{equation}
The minimum of the twoline potential energy occurs at $u^\prime=0$, and has the
value $0$.  The two line self energies are cancelled by the interaction energy.
Since the lines are fields, the cancellation is complete---except, of course,
for excitations of the twoline system.  We note that the annihilation of two 
oppositely directed vortices has been observed\cite{Qworld} in type II superconductors.

\section{The Ether}
\label{sec-ether}

\subsection{Overview}
\label{ssec-overview}

     At the birth of the universe we envision a very large concentration of energy in
the universe in accord with standard ideas of the hot big bang model\cite{Weinberg}.
We envision that the energy is originally in the form of continuous quantized vortices (lines)
which have no ends.  When sufficient energy is available locally, some lines break---creating 
electron-positron pairs.  As the universe continues to expand, oppositely 
directed lines are attracted to each other and eventually bind.  Initially excitations
of the twoline systems are very large and lines frequently switch partners.
But as the universe cools excitations become smaller and smaller, and in our current 
universe the line pairs are assumed to be condensed with an energy very close to zero (equation (\ref{eq-twolinepot})).
By this assumption we avoid having to solve the general case.

     In order to limit the scope of our discussion, weak and strong interactions 
will not be discussed more than is absolutely necessary.  Presumably cooling processes occurred
with these interactions, which could be similar to the processes which we will discuss.  
But, first we need to mention how protons and neutrons fit into our
picture.  Noting the decays
\begin{equation}
     n\ \rightarrow p\ e^-\ {\bar \nu}_e
\label{eq-ntop}
\end{equation}
\begin{equation}
     p\ \rightarrow\ n\ e^+\ \nu_e
\label{eq-pton}
\end{equation}
(the latter occurring only in nuclei) we presume that if a negative electric charge 
(electron or antiproton) is at one end of a line, any positive charge (positron or proton) can be
at the other end.  From (\ref{eq-pton}) we infer that the proton must have the same type
of line as the positron (defined by positive charge), and from (\ref{eq-ntop}) we infer
that the neutron has {\it both} types of line, i.e. contains the end of a line pair.
Thus in today's universe, essentially all lines are  bound in pairs,
with the upper limit on the number of unbound lines set by the neutrality of matter.  
The stability of these line pairs is discussed in Section~\ref{ssec-tensionpair}.
From (\ref{eq-ntop}) we presume that one line pair can end at a hydrogen atom or at a neutron. 
The ends of the lines near the Earth are bound together in atoms.  In stars significant
ionization occurs, but from the net neutrality of plasma matter in stars and energy
considerations, it is reasonable to believe that the two ends of a line pair do not become greatly 
separated.  The threshold for stringbreaking ($\sim 10^{10} K$) is barely approached in the cores
of the most massive stars\cite{stringbreak}.

      Our picture offers hope of explaining the evolution of the universe from
the big bang to the present day\cite{Kolb}, without adding any scalar fields.  
The earliest epoch is thought to be dominated by strings---which may correspond to our lines.  
The evolution of this model seems natural because the strings are still here!
We suggest that inflation was due to the reversal of the force of gravity under the very high
densities of the early universe (which we will discuss), when our line pairs were 
radiation dominated and pressures were naturally positive.  In recent times, when 
the line pairs are cold, their tension (negative pressure) dominates and they become 
the source of dark energy which may explain the recent acceleration of the universe.
Photons are the differential excitations of a twoline system, and gravitational waves are common mode
excitations of a twoline system (differs from the Standard Model).  
The Higgs field of the Standard Model is replaced by the field of line pairs.

     The concept of an ether was rejected in the early part of the $20^{th}$ century for many 
reasons.  We quote Einstein's\cite{Einstein2} reasons for rejecting the ether concept, along with our reply to
each point for why we now think the concept is useful:
\begin{quotation}
``If light was to be interpreted as undulatory motion in an elastic body (ether),
this had to be a medium which permeates everything; because of the transversality
of the lightwaves in the main similar to a solid body, yet incompressible so that 
longitudinal waves did not exist."

\underline{reply}: In this model the ether does permeate essentially everything
but it is not a solid body, but rather a mesh of lines.  Light waves are only transverse
because they are waves on a string; they are not bound longitudinally.

``This ether had to lead a ghostly existence alongside the rest of matter, inasmuch
as it seemed to offer no resistance whatever to the motion of ``ponderable" bodies."

\underline{reply}:  The ether is ``ghostly" because it is made of magnetic fields.  It does offer
resistance to the motion of massive bodies, because we postulate that it is the origin of mass.  It offers
resistance to acceleration, but does not offer resistance to velocity in accord with
Newtons $2^{nd}$ law, and special relativity.

``In order to explain the refraction-indices of transparent bodies as well as the processes
of emission and absorption of radiation, one would have had to assume complicated
reciprocal actions between the two types of matter, something which was not even seriously
tried, let alone achieved."

\underline{reply}:  There is only one type of matter.  Electrons are literally connected
to the ether in this model.  If an electron accelerates it causes a wave on a string.
\end{quotation}

     It should be noted that there is a great similarity between Maxwell's {\it molecular vortices}\cite{Maxwell}
with which he constructed his model of the ether and the quantized vortices
considered here.  There is a big difference, however.  Following the prejudices
of his day, Maxwell attempted to explain electromagnetism in terms of mechanics.
We, however, attempt to explain mechanics in terms of his electromagnetism. 

\subsection{Average Line-Pair Length}
\label{ssec-avelpl}

     It is necessary to estimate the average length of a line pair, ${\ell}_{ave}$, in the  
universe today.  In  the early universe there were two competing processes that determined
the length of the average line.  During inflation the velocity of separation
of the two ends of a line exceeded the velocity of light, perhaps greatly\cite{Linde}.
This expansion competed with the breaking of lines---forming an electron-positron pair 
at each break.  Because
of these two processes there will be a statistical distribution of line lengths about some
mean, ${\ell}_{ave}$.  In order to avoid dependence on a specific model of inflation
we simply take, as a first guess,
\begin{equation}
{\ell}_{ave}\ \sim\ c\ (t-t_0) 
\end{equation}
${\ell}_{ave}$ to be the distance to the horizon (where $t_0$ is the time of the big 
bang).  Within a minute\cite{HindKib} after the big bang both rapid inflation
and string-breaking are expected to have stopped, so that the lines (now in pairs) 
have simply expanded
with the  universe since that time.  Thus we might expect our estimate for ${\ell}_{ave}$
to be close in the absence of inflation.  In this sense we view it as a {\it minimum} value.
The value of ${\ell}_{ave}$ is a major uncertainly
in these considerations and we will attempt to measure it.

\subsection{The Density of Line Pairs}
\label{densitylp}

     According to the picture we have developed, essentially all lines are paired and,
at least near the Earth, they
start and end at atoms.  There is on average throughout the universe about one hydrogen
atom (or neutron) per cubic meter\cite{Peebles},  so this minimum average length of line pairs
per unit volume is
\begin{equation}
   {{dL} \over {dV}}\ =\ {{{\ell}_{ave}} \over 2}\ \Big({m \over {H-atom}}\Big) \times 1\Big({{H-atom} \over {m^3}}\Big)\ \ \ .
\label{eq-dLdV}
\end{equation}
We associate each atom with one-half of ${\ell}_{ave}$ since a line pair has two ends.
We assume all points in the universe are equivalent and we eliminate edge effects by
taking the universe to be very large.  Taking the time since the big bang to be $10^{10}$ years we have
\begin{equation}
   {{\ell}_{ave}}\ =\ 10^{10}\ light\ years\ = 9.5 \times 10^{25}\ m
\label{eq-lave}
\end{equation}
and
\begin{equation}
{{dL} \over {dV}}\ =\ 4.7 \times 10^{25}\ \Big({{m\ of\  line\ pairs} \over {m^3}}\Big)\ \ \ .
\label{eq-dLdVnum} 
\end{equation}
We refer to this quantity as the line-pair density of the ether.
If the number of line pairs per unit area per unit solid angle is denoted by ${{dn} \over {dA d\Omega}}$\ ,
it can be shown that
\begin{equation}
\ {{dL} \over {dV}}\ =\ \int{{dn} \over {dA d\Omega}} d\Omega\ \equiv\ {{dn} \over {dA}}\ 
=\ 4.7 \times 10^{25}\  { {line\ pairs} \over {m^2} }
\label{eq-dndAnum}
\end{equation}
if the distribution of line pairs is isotropic, and that
\begin{equation}
 {{dn} \over {dA d\Omega}}\ = {1 \over {2\pi}} {{dL} \over {dV}}\ 
\label{eq-dndAdOmega}
\end{equation}
where the solid angle of line pairs only changes from $0$ to $2\pi$ to avoid double counting.
(We also refer to ${ {dn} \over {dA} }$ and  ${ {dn} \over {dAd\Omega} }$ as line pair densities.)
Thus within the first Bohr orbit of a hydrogen atom, with the above minimum value for ${\ell}_{ave}$
\begin{equation}
  n\ =\ {{dn} \over {dA}}\ \pi\ {{a_0}^2}\ =\ 4.2 \times 10^5\ 
\end{equation}
there are about $ 4 \times 10^5$ line pairs.  However, within the radius of the proton
($r_p\ \simeq\ 1.4\times10^{-15}m$ as determined by electron scattering\cite{Hofstadter}) there are only
\begin{equation}
  n\ =\ {{dn} \over {dA}}\ \pi\ {{r_p}^2}\ =\ 2.9 \times 10^{-4}\ 
\end{equation}
about $3 \times 10^{-4}$ line pairs.

    (In Section~\ref{ssec-estimatelpd} we will attempt to measure $\ell_{ave}$ and tentatively show that it should be increased 
over our assumed value (\ref{eq-lave}) by an estimated factor of $10^{14}$, presumably as a result of inflation.  But since this 
inflation factor is quite uncertain, we prefer to show these minimum estimates here.)

     We note that the net flux of line pairs is always zero because no net direction is 
defined.  If, however, the ether could be completely polarized (elimintate lines with {\bf B}
pointing from right to left), the magnetic induction field would be tremendous
(with the above value of $\ell_{ave}$).
\begin{equation}
 B_z\ =\ {\Phi_0}\int{{dn} \over {dA d\Omega}}cos\theta d\Omega\ =\ {{\Phi_0} \over 2} {{dL} \over {dV}}\ 
\end{equation}
\begin{equation}
 B_z\ =\ {{4.14\times10^{-7}(Gauss-cm^2)} \over 2} \times 4.7 \times 10^{25}\ \Big({{lines} \over {m^2}}\Big)
\end{equation}
\begin{equation}
 B_z\ = 9.8 \times 10^{14} Gauss\ \simeq\  10^{15} Gauss\ \ . 
\label{eq-Bunbalanced}
\end{equation}
This field indicates how cold (condensed) our universe has become.
Magnetic induction fields of this strength are thought to exist\cite{highfield} in neutron stars.

     A condensed line pair (far from its ends) has little energy density and no net magnetic field.
How can it be detected?  It can respond to a test
particle (electron) which does have an electromagnetic field, and cause it to have mass.

    The quantum nature of this field of line pairs should be emphasized.  Essentially every one
of these line pairs ends at a particular atom, or at least near a particular location, 
someplace in the universe.  Since the universe is
homogeneous and isotropic on the large scale, the field of line pairs, reflecting the universe as
a whole, should also be homogeneous and isotropic as well.  But, for instance, if a star approaches
the field point under our consideration, the line-pair field will vary both in time (as the star
approaches), and direction (strongest toward the star).  The star may be assumed to have one 
line pair per nucleon stretching off to great distances.  
As a first approximation, we assume the line pairs stretch out isotropically from a star.  
(See Section~\ref{sec-rotations} for a discussion of rotation.)  The density of line pairs is
obviously quantized, but with the above value for ${ {dn} \over {dA} }$, the density of line pairs is sufficiently
large for most purposes that we can average over the quantum fluctuations of this distribution,
and treat it as a smooth field.

     Our superconducting fields are {\it aon} superconductors.  They become polarized 
in an electric or magnetic field but do not conduct
electrons or Cooper pairs---since the electric and magnetic fields of the electron cannot exist inside the superconductor.  
(We will show (\ref{eq-poleforce}) that, at least near the Earth, the pole field of the electron dominates at close range and repels 
superconductor.)  Hence the line pairs do not short out all voltage differences in the universe.

\subsection{Intercommutation}
\label{ssec-intercommutation}

    The density of line pairs in the ether is very large and we must remember that these pairs are
{\it superconducting}.  Hence the line pairs certainly have effects on each other.  The line pairs cross and 
hence form closed superconducting circuits which cause the quantization of magnetic flux.  We will 
show that these circuits also cause the quantization of the photon.

    {\it Because} of these quantization effects it is reasonable to presume that cold, condensed 
line pairs can move freely through each other, i.e. intercommute freely, once this quantization
is taken into account.  We have seen that quantized lines can move through each other (Section~\ref{ssec-phase})
affected only by their magnetic interaction.
Excitation energies are quantized and low energy excitations (such as photons)
have very long wavelengths, which do not couple well to objects as small as another line pair
 (smaller than an atom).  With this presumption we avoid a problem which would otherwise cause 
great difficulty\cite{Vilenkin}, especially with the above density of line pairs (equations (\ref{eq-dLdVnum}) and (\ref{eq-dndAnum})).  
We presume that intercommutation
was important only in the early universe, and perhaps in the interior of stars. 

    We will show that {\it all} excitations of a line pair obey the photon quantization conditions.
When these conditions are satisfied, as indicated above, the line pair can {\it not} be 
considered isolated.  Supercurrent flows around closed circuits as necessary to maintain
the quantization conditions.  We call these supercurrents {\it quantization currents}.
We will also show that all line pairs in the universe (at the same time)  have
the same tension if they do not have an excitation.

\subsection{Dielectric Constant and Magnetic Permeability of the Ether}
\label{ssec-dielectric}

    In Section~\ref{sec-introduction} we defined the {\it microscopic} Maxwell equations, and until now we have had 
no need to deal with macroscopic electromagnetic quantities, which Jackson\cite{Jacksonmac} informs us 
involve a spatial average over a typical length of order $10^{-8}m$, appropriate for standard atomic matter.  
In our case by {\it microscopic} we mean on a length scale $\ll \ell_{typ}$ , where $\ell_{typ}$ is 
the typical separation of line pairs.  We take {\it macroscopic} to mean on a length scale $\gg \ell_{typ}$.
After inflation is taken into account we will find that $\ell_{typ}$ is very small.  (See Table~\ref{table-lpdatEarth}.)

    Now we will need to consider
macroscopic electromagnetic fields which we define as ${\bf E}^{mac}$ and ${\bf B}^{mac}$ in order to
distinguish them from the microscopic fields ${\bf E}$ and ${\bf B}$.   The standard macroscopic fields
are: ${\bf B}^{mac}$ - magnetic induction, ${\bf H}$ - magnetic field,
${\bf M}$ - magnetization, ${\bf E}^{mac}$ - electric field, ${\bf D}$ - displacement vector,
and ${\bf P}$ - polarization.  The fields ${\bf H}$, ${\bf M}$, ${\bf D}$, and ${\bf P}$ are only defined
as macroscopic fields.

    An electric field will certainly polarize the ether since dipole fields will be set up 
on the exterior of
every line pair, ensuring that the net electric field falls to zero several penetration depths
inside a line pair.  Hence, including this polarization ${\bf P}$, the ether has a displacement 
vector given by

\begin{equation}
    {\bf D}\ =\ {\bf E}^{mac}\ +\ 4\pi{\bf P}\ \equiv\ \varepsilon{\bf E}^{mac}
\label{eq-D}
\end{equation}

\noindent
where $\varepsilon > 1$ is the dielectric constant of the ether.  

    A similar expression holds for the magnetic permeability $\mu$

\begin{equation}
    {\bf B}^{mac}\ =\ {\bf H}\ +\ 4\pi{\bf M}\ \equiv\ \mu{\bf H} \ \ \ .
\label{eq-Bmac}
\end{equation}

\noindent
We can achieve an exact analogy to the electric polarization by 
rearranging (\ref{eq-Bmac}) to find

\begin{equation}
    {\bf H}\ =\ {\bf B}^{mac}\ -\ 4\pi{\bf M}\ \equiv\ \varepsilon{\bf B}^{mac} \ \ \ .
\label{eq-H}
\end{equation}

\noindent
Since a superconductor is a perfect diamagnet, the magnitization ${\bf M}$ is opposite to 
the applied field ${\bf B}$
in order to cancel ${\bf B}$ well within the superconductor.  This follows because a magnetic dipole
 ${\bf m}$ produces a ${\bf B}$-field parallel to ${\bf M}$, but an electric dipole ${\bf p}$ produces
an ${\bf E}$-field {\it opposite} to ${\bf P}$.  Hence with the negative sign in (\ref{eq-H}) we have an 
exact analogy with equation (\ref{eq-D}).  Electric field {\bf E} is excluded from each superconducting line 
pair in (\ref{eq-D}), and ${\bf B}$ excluded from the same line pairs in (\ref{eq-H}).  Hence

\begin{equation}
    {\bf H}\ =\ \varepsilon{\bf B}^{mac}\ =\ \varepsilon\mu{\bf H} 
\end{equation}

\noindent
so 

\begin{equation}
    \varepsilon\mu\ =\ 1 
\label{eq-epsmue1}
\end{equation}

\noindent
for the superconducting ether.

    We can achieve a fairly realistic calculation of the dielectric constant of the ether by considering
the case of normal matter where atoms are modeled by superconducting spheres of radius $R$, as originally 
proposed by Mosotti\cite{Beckersphere}.  (He called them perfect conductors before superconductivity was
known to exist.)  In this case the local field at the atom in question is given by

\begin{equation}
    {\bf E}_{local}\ =\ {\bf E}\ +\ { {4\pi} \over 3 } {\bf P}
\end{equation}

\noindent
and the atomic electric dipole moment is given by

\begin{equation}
    {\bf p} \ =\ \alpha\ {\bf E}_{local}
\end{equation} 

\noindent
where

\begin{equation}
    \alpha\ =\ R^3 \ \ \ .
\end{equation}

\noindent
    If $n$ is the number of such atoms per unit volume then

\begin{equation}
    \varepsilon_{spheres} - 1 \ =\ { {4\pi n \alpha} \over {1\ -\ {{4\pi} \over 3} } n \alpha} \ =\ { {3f} \over {1\ -\ f}}
\end{equation}

\noindent
where

\begin{equation}
    f\ \equiv\ n { {4\pi} \over 3 } R^3
\end{equation}

\noindent
is the fractional concentration of superconductor.

    We can approximate a line pair by a row of superconducting spheres, so we will simply take

\begin{equation}
    \varepsilon - 1 \ =\ { {3f} \over {1\ -\ f}}
\label{eq-epsminus1}
\end{equation}

\noindent
to be an expression for the dielectric constant of the ether, where $f$ is the fractional concentration of superconducting
fields.  This expression approaches $\infty$ as $f\ \rightarrow\ 1$ as it must
for a uniform superconductor.  There is a major difference, however, between
atomic spheres, which are subject to the Pauli Exclusion Principle (Section~\ref{sec-Pauli}) and line pairs, which can move through each 
other.  Because line pairs can move through each other,  they can have significant overlap.  As a result $\varepsilon$
is lower than it would be if the line pairs were impenetrable.  But if the density of line pairs becomes {\it very} large,
e.g. near a black hole, we expect $\varepsilon$ to approach infinity as the ether becomes a continuous superconductor.  
We presume that equation (\ref{eq-epsminus1}) is completely general and will apply it even if the ether is not homogeneous and isotropic.
This equation is particularly useful because of its simplicity.

\subsection{Charge of the Bare Electron}
\label{ssec-chargeofbareelectron}

    From the macroscopic Maxwell equation\cite{Jacksondeldotd}

\begin{equation}
  \nabla \cdot {\bf D}\ =\ 4 \pi \rho_{free}\ =\ 4 \pi e_0 \delta^3 ({\bf x}-{\bf y})\ =\ \varepsilon \nabla \cdot {\bf E}^{mac}\ =
  \ \varepsilon\  4 \pi e \delta^3 ({\bf x}-{\bf y})    
\end{equation}

\noindent
we see that the charge on the bare positron is {\it not} $e$ but

\begin{equation}
    e_0\ \equiv\ \varepsilon e
\end{equation}

\noindent
after correction for the dielectric constant of the ether.  This implies that the flux quantum is actually

\begin{equation}
    \Phi_0\ =\ 4 \pi g_0
\end{equation}

\noindent
where

\begin{equation}
    g_0\ \equiv \ { {\hbar c} \over {2 e_0} }\ =\ { g \over \varepsilon}\ =\ \mu g
\end{equation}

\noindent
if we use the correct value of the bare electron's charge.  Standard methods of measuring the flux quantum\cite{Tonomuraphi0}
use the measured charge of the electron, so will obtain $4 \pi g$.  Since the density of the ether (\ref{eq-dndAnum}) is large, 
$\varepsilon$ 
could be large.  We assume that $e_0$ and $g_0$ are fundamental constants but that $e$ is {\it not}, because the density of the
ether changes as the universe expands.

    Hence in all formulas to this point in this paper we should make the replacement 

\begin{equation}
        g\ \rightarrow \ g_0 \ \ \ .
\end{equation}

\noindent
In particular the tension of a line becomes

\begin{equation}
        T\ =\ { {g_0}^2 \over {2 a^2} } \ \ \ .
\label{eq-T0}
\end{equation}

\noindent
The phase field (\ref{eq-ps}) involves the product $q {\bf A}$, so does not change, and the fundamental quantization condition 
(\ref{eq-quantization}) is not affected.

    Hence we recognize that the large magnetic field in (\ref{eq-Bunbalanced}) could be too large if $\varepsilon>1$.  
But it is the field which would be measured using current techniques. 

\section{One Moving Line}
\label{sec-moving}

\subsection{Motion of a Filamentery Line---Neglecting Interactions} 
\label{ssec-neglecting}
     The classical solution for the motion of a filimentary light string has been illustrated by
Goddard, Goldstone, Rebbi, and Thorn\cite{Goddard}.   Following this reference
we take the action for a light string proportional to the 
area of the surface swept out by the string between fixed times.
If the string is parameterized as a function of $\sigma$, determining the position
along the string, and $\tau$ determining the position in time, then a general 
position along the string is given by
\begin{equation}
   y^{\mu} = y^{\mu}(\tau,\sigma)\ .
\end{equation}
The action is given by
\begin{equation}
   S = \int_{\tau_i}^{\tau_f}d\tau\ \int_{\sigma_i}^{\sigma_f}d\sigma\ {\cal L}
\end{equation}
where
\begin{equation}
   {\cal L} = {T\over c}\Bigl\{ (y^\prime \cdot {\dot y})^2 \ -\ (y^\prime)^2\ (\dot y)^2 \Bigl\}^{1/2}
\label{eq-Lagrangiandensity}
\end{equation}
\begin{equation}
   {y^\prime}^\mu \equiv {{ \partial y^\mu} \over {\partial \sigma }}
\end{equation}
\begin{equation}
   {\dot y}^\mu \equiv {{ \partial y^\mu} \over {\partial \tau }}
\end{equation}
and $T$ is the tension (\ref{eq-T0}), presumed constant.  

     For some purposes it is convenient to consider this parameterization arbitrary\cite{Barutarb}.  It is easy to see that the
action is independent of the particular parameterization.  In order to consider a detailed solution, it is convenient to take
$\tau\ =\ ct$.  In this case 

\begin{equation}
 {{ \partial y^\mu} \over {\partial \tau }}\ \equiv\ (\ 1,\ {\bm \beta})^{\mu} \ \ \ . 
\label{eq-taudef}
\end{equation}

\noindent
(Since we choose $\tau = ct$, ${\dot y}^\mu$ is {\it not} a 4-vector.)  
If $\ell$ (a function of $\sigma$) represents length 
along the string
\begin{equation}
{\hat \ell} = {{\partial{\bf y}} \over {\partial \ell}} = {{{\partial{\bf y}}/{\partial \sigma}}\over{d\ell /d\sigma}} 
\end{equation}
is a unit vector.  If
\begin{equation}
   {\bm \beta} = {{\partial {\bf y}} \over {\partial \tau}} = \beta_\ell\ {\hat \ell}\ +\ {\bm\beta}_\perp 
\label{eq-beta}
\end{equation}
\begin{equation}
   \gamma_\perp\ \equiv\ {1 \over {\sqrt{1-{\beta_\perp}^2}}}
\end{equation}
and
\begin{equation}
    {u^\mu_\perp}\ \equiv\ \gamma_\perp (1,{\bm \beta}_\perp)^{\mu}
\end{equation}
the generalized momenta are
\begin{equation}
{{\partial {\cal L}} \over {\partial y^\prime_\mu}}\ =\ -{T\over c}\Bigl\{\beta_\ell\ {u^\mu_\perp}\ +\ 
    \sqrt{1-{\beta_\perp}^2}\ (0,{\hat \ell})^{\mu} \Bigr\}
\label{eq-pLdymup}
\end{equation}
and
\begin{equation}
{{\partial {\cal L}} \over {\partial {\dot y}_\mu}}\ =\ {T\over c}\ {d\ell \over d\sigma}\ {u^\mu_\perp}\ \ \ .
\label{eq-pLdymud}
\end{equation}
Hence, from the variation of the action, the general equation of motion is
\begin{equation}
{{\partial \ } \over {\partial \sigma}}\ \Bigl\{-{T\over c}\{\beta_\ell\ {u^\mu_\perp}\ +\ 
     \sqrt{1-{\beta_\perp}^2}\ (0,{\hat \ell})^{\mu} \}\Bigl\}\ +\
{{\partial \ } \over {\partial \tau}}\Bigl\{{T\over c}\ {d\ell \over d\sigma}\ {u^\mu_\perp}\Bigr\}\ =\ 0
\label{eq-eqofmotion}
\end{equation}
with the boundary conditions
\begin{equation}
\Biggl\lbrack{{\partial {\cal L}} \over {\partial {y^\prime}^\mu}}\ \delta y^\mu {{\Biggr \vert}_{\sigma_i}^{\sigma_f}}\ =\ 0
\label{eq-bcendofline}
\end{equation}
where $\delta y^\mu$ is the variation in $y^\mu$, which is presumed to vanish at $\tau_i$ and $\tau_f$.

     The boundary conditions are satisfied if
\begin{equation}
     \delta y^\mu(\sigma_i)\ =\ 0\ = \delta y^\mu(\sigma_f)\ 
\end{equation}
the ends of the line are held fixed.  For some purposes this might be a good approximation since we will
see that the ends of many lines are bound in matter.  But in general we cannot require
the ends of the lines to be fixed.
In the more useful case that the ends of a line are free, the boundary conditions are
satisfied if
\begin{equation}
{{\partial {\cal L}} \over {\partial {y^\prime}^\mu}}(\sigma_i)\ =\ 0\ = 
{{\partial {\cal L}} \over {\partial {y^\prime}^\mu}}(\sigma_f)\ \ .
\end{equation}
All four of these boundary conditions at each end point can be satisfied if $\beta_\perp = 1$
and $\beta_\ell = 0$.  This implies that the effective tension at the end of
the line is zero and the end moves with the  speed of light transverse to the line.
We will interpret this motion as the {\it zitterbewegung}\cite{BJD} of a bare electron
familiar from the theory of the Dirac equation.  For the moment we simply assume that
the boundary conditions are satisfied so that we can proceed to discuss the motion
of a line.  We give a fuller discussion of the boundary conditions in Section~\ref{ssec-motionbpp}.

     It probably comes as no surprise that any wave traveling at the speed of light
is an exact solution to the equation of motion of the string (\ref{eq-eqofmotion}).  We presume that
the string stretches over space in the $z$-direction so that it is convenient to
take $\sigma = z$ as the parameterization along the string.  Then a solution of 
general shape is given by
\begin{equation}
   {\bf y}\ =\ {\hat y}\ F(\tau-z)\ +\ {\hat z}\ z
\label{eq-yzline}
\end{equation}
where $F$ is a smooth function of arbitrary shape.
We have
\begin{equation}
{{\partial \bf{y}} \over {\partial \sigma}}\ \equiv\ {{\partial \bf{y}} \over {\partial z}}\ =\ {\hat z}\ -\ {\hat y}\ F^\prime
\end{equation}  
\begin{equation}
{\bm \beta}\ =\ {{\partial \bf{y}} \over {\partial \tau}}\ =\  \ {\hat y}\ F^\prime
\label{eq-betayz}
\end{equation}
(where $F^\prime$ is the derivative of $F$ with respect to its argument)
\begin{equation}
{d\ell \over d\sigma}\ =\ {d\ell \over dz}\ =\sqrt{1\ +\ (F^\prime)^2}\ =\ \gamma_\perp
\end{equation} 
\begin{equation}
{\hat \ell}\ =\ {{\partial \bf{y}} \over {\partial \ell}}\ =\ {{{\hat z}\ -\ {{\hat y}\ F^\prime}} \over  
                                     \sqrt{1\ +\ (F^\prime)^2} }
\label{eq-lhat}
\end{equation}
\begin{equation}
{\beta_\ell}\ =\ {{-\ (F^\prime)^2 } \over \sqrt{1\ +\ (F^\prime)^2} }
\end{equation}
\begin{equation}
{\bm \beta}_\perp\ =\ {\ {\hat y}\ F^\prime\ +\ {{\hat z}\ (F^\prime)^2} \over {1\ +\ (F^\prime)^2}}
\end{equation}
so that
\begin{equation} 
{{\partial {\cal L}} \over {\partial y^\prime_\mu}}\ =\ - {T \over c} \biggl( -(F^\prime)^2\ ,\ -{\hat y}\ F^\prime\ +\ 
              {\hat z}\ [1-(F^\prime)^2] \biggr)
\end{equation}
\begin{equation}
{{\partial {\cal L}} \over {\partial {\dot y}_\mu}}\ =\ + {T \over c} \biggl(1 +(F^\prime)^2\ ,\ {\hat y}\ F^\prime\ +\
              {\hat z}\ (F^\prime)^2 \biggr) \ \ .
\end{equation}
Note that the wave is only a function of its argument
\begin{equation}
       \Phi\ =\ \tau-z
\end{equation}
so that, when acting on $F^\prime$
\begin{equation}
   {{\partial \ } \over {\partial \sigma}}\ \equiv\ {{\partial \ } \over {\partial z}}\ =\ -{{d \ } \over {d \Phi}}\
\end{equation}
and
\begin{equation}
   {{\partial \ } \over {\partial \tau}}\ =\ {{d \ } \over {d \Phi}}\ \  . 
\end{equation}
The equation of motion (\ref{eq-eqofmotion}) becomes
\begin{equation}
 {{d \ } \over {d \Phi}}\ \{ {T \over c} (1,{\hat z})^\mu \}\ =\ 0
\end{equation}
and we have a solution.

     Note that in our parameterization the wave is transverse to the ${\hat z}$-direction, the
average direction of the string, but it is {\it not} transverse to the instantaneous 
direction of the string as assumed in reference\cite{Goddard}, 
because the string bends.  Indeed $\beta_\ell$ is always negative
or zero.  The only solution with $\beta_{\ell} = 0$ everywhere requires $F^\prime = 0$,
i.e. no wave.

     The authors of reference\cite{Goddard} go on to discuss the quantum dynamics of the light
string and claim that in order for a reasonable solution to exist, the dimension of
spacetime must be 26.  This claim does not apply to our model because it is a classical
model---though we will construct from it the elements of quantum mechanics.  We work
in four spacetime dimensions, but do not exclude the existence of hidden, compactified
dimensions. 

     Using the above methods, it can be shown that a slightly more general wave traveling at 
the speed of light
\begin{equation}
   {\bf y}\ =\ {\hat x}\ E(\tau-z)\ +\ {\hat y}\ F(\tau-z)\ +\ {\hat z}\ z
\label{eq-circularline}
\end{equation}
is also a solution of (\ref{eq-eqofmotion}), where both $E$ and $F$ are smooth functions of $\Phi$ with
arbitrary shape.

\subsection{Motion of a Filimentary Line with Interactions}
\label{ssec-winteractions}

     Since the electromagnetic string can interact with itself, we must consider
interaction even when considering motion of a single line.  It has been shown by
Barut and Bornzin\cite{BarutBornzin} that the equation of motion for the filamentary
Dirac string with interactions is given by

\begin{equation}
{{\partial \ } \over {\partial \sigma}} \Bigl( { {\partial {\cal L}_m} \over {\partial {y^\prime}^\mu } }  \Bigr) \ +\
{{\partial \ } \over {\partial \tau}} \Bigl( { {\partial {\cal L}_m} \over {\partial {\dot y}^\mu } } \Bigr) \ =\ 
{{4\pi g_0} \over c^2}\  {\dot y}^\alpha {y^\prime}^\beta \epsilon_{\mu \nu \alpha \beta} J^\nu
\label{eq-filmotionwint}
\end{equation}

\noindent
where now ${\cal L}_m$ is the mechanical string Lagrangian density given by equation (\ref{eq-Lagrangiandensity}) and the total
Lagrangian density including interactions is given by

\begin{equation}
{\cal L}_{tot}\ =\ {\cal L}_m\ +\ {{g_0} \over c}\ {\dot y}^\alpha {y^\prime}^\beta {\tilde F}_{\alpha \beta}(y) 
\label{eq-Lagrangiandwint}
\end{equation}

\noindent
where ${\tilde F}_{\alpha \beta}(x)$ is the dual of the electromagnetic field tensor at point $x^\mu$.

\begin{equation}
{\tilde F}_{\alpha \beta}(x)\ =\ {1 \over 2}\  \epsilon_{\alpha \beta \mu \nu} F^{\mu \nu}(x)
\end{equation}

     Dirac\cite{Dirac2} showed that in the filamentary model, but otherwise in the general case,
that the dual field of the line is given by

\begin{equation}
{\tilde F}^{\alpha \beta}(x)\ =\ -4\pi g_0 \ \int d\tau d\sigma \Bigl(
{ {\partial y^\alpha} \over {\partial \tau} } { {\partial y^\beta} \over {\partial \sigma} }\ -\ 
{ {\partial y^\beta} \over {\partial \tau} } { {\partial y^\alpha} \over {\partial \sigma} } \Bigr)\ {\delta}^4(x-y)
\label{eq-filEMdual}
\end{equation}

\noindent
and therefore the electromagnetic field tensor of the line is given by

\begin{equation}
F_{\mu \nu}(x)\ =\ 4\pi g_0 \ \epsilon_{\mu \nu \alpha \beta} \int d\tau d\sigma 
{ {\partial y^\alpha} \over {\partial \tau} } { {\partial y^\beta} \over {\partial \sigma} }\ {\delta}^4(x-y)\ \   .
\label{eq-filEM}
\end{equation}

\noindent
Dirac's pole current formalism can be used (Section~\ref{ssec-Diracpolecurrent}) to find the field at the ends of the line.  
Using Maxwell's equations the line current density is given by

\begin{equation}
J_\nu(x)\ =\ g_0 c\ \epsilon_{\nu \mu \alpha \beta} \int d\tau d\sigma
{ {\partial y^\alpha} \over {\partial \tau} } { {\partial y^\beta} \over {\partial \sigma} }\ 
{ {\partial \ } \over {\partial y_\mu}  } {\delta}^4(x-y)\ \  .
\label{eq-fil4current}
\end{equation}

\subsection{Definition of a Superconducting Field}
\label{ssec-superconfield}

     We define a superconducting field as a region of spacetime within which the relevant 
phase field obeys the equation 

\begin{equation}
\oint {p^\mu_s}\ dx_\mu\ =\ 0\ \ \ \ \ (simply\ connected)
\label{eq-simply}
\end{equation}

\noindent
for all closed paths enclosing areas that are totally within that region, so that the 
corresponding wave function
is single-valued.  In this case we see from (\ref{eq-quantization0}) that the London equations (\ref{eq-London})
are exact.  If the superconducting field is multiply connected, we must allow flux quanta
to be outside the superconducting region but enclosed by the path

\begin{equation}
\oint {p^\mu_s}\ dx_\mu\ =\ nh\ \ \ \ \ (multiply\ connected)\ \ \ .
\label{eq-multisup}
\end{equation}

\noindent
Here the phase field corresponds to the momentum field of Cooper pairs (\ref{eq-4ps}) in 
a type II superconductor.  In the case of a superconducting line we extrapolate all results
to the limit $m\rightarrow 0$ and $n_s q^2\rightarrow 0$ such that the line radius $a$ (\ref{eq-lambdaL})
remains finite, at its characteristic constant value.

\subsection{General Motion of a Superconducting Line}
\label{ssec-generalmotion}

     The phase field must also satisfy the general relationship between ${u^\mu_s}$ and $A^\mu$
implied by Maxwell's equations with sources

\begin{equation}
{ {\partial F^{\mu \nu}} \over {\partial x^\mu} }\ =\ { {4\pi} \over c }\ J^\nu\ \ .
\label{eq-Maxwellwsources}
\end{equation}

\noindent
In terms of the four-vector potential we obtain

\begin{equation}
{ {\partial^2 A^\nu} \over { \partial x^\mu \partial x_\mu } }\ =\ { {4\pi} \over c }\ J^\nu\ \ .
\label{eq-4potto4current}
\end{equation}

\noindent
This is the usual equation relating the electromagnetic potential to its source current.

     In a superconducting field the London equations (\ref{eq-London}) give

\begin{equation}
m\ \Bigl( { {\partial {u^\nu_s} } \over { \partial x_\mu } }\ -\ 
          { {\partial {u^\mu_s} } \over { \partial x_\nu } }\ \Bigr)\ +\ {q \over c}\ F^{\mu \nu}\ =\ 0
\end{equation}

\noindent
or, with the use of

\begin{equation}
J^\mu\ =\ n_s q {u^\mu_s}\ \ 
\label{eq-4supercurrent}
\end{equation}

\noindent
we obtain

\begin{equation}
{ {4\pi a^2} \over c }\ \Bigl( { {\partial {J}^\nu } \over { \partial x_\mu } }\ -\
          { {\partial {J}^\mu } \over { \partial x_\nu } }\ \Bigr)\ +\ F^{\mu \nu}\ =\ 0
\label{eq-EMgensupercurrent}
\end{equation}

\noindent
implying that the supercurrents respond to the electromagnetic field at the same point in
spacetime.  Using the equation of continuity (which follows from Maxwell's equations)

\begin{equation}
{ {\partial J^\mu} \over {\partial x^\mu} }\ =\ 0 
\label{eq-continuity}
\end{equation}

\noindent
we obtain

\begin{equation}
{ {4\pi a^2} \over c }\\ \Bigl[ { {\partial^2 {J}^\nu} \over {\partial x^\mu \partial x_\mu}}\ -
            { {\partial \ } \over {\partial x_\nu} } \Bigl( {{\partial {J}^\mu} \over {\partial x^\mu} } \Bigr) \Bigr] \ +\ 
{{4\pi} \over c}\ J^\nu\ = 0
\end{equation}

\noindent
so that

\begin{equation}
{ {\partial^2 {J}^\nu} \over {\partial x^\mu \partial x_\mu}}\ =\ -{ {J}^\nu \over a^2 }\ \ .
\label{eq-dalJeJ}
\end{equation}

\noindent
This equation depends only on the London equations and Maxwell equations, 
so it is {\it always} true for a supercurrent obeying (\ref{eq-4supercurrent}).

    The general solution of (\ref{eq-4potto4current}) in a superconducting field is simply

\begin{equation}
A^\nu\ =\ -{ {4\pi a^2} \over c }\ J^\nu\ +\ {c \over q} {p^\nu_s}
\label{eq-generalsol}
\end{equation}

\noindent
where we have used (\ref{eq-4ps}) and (\ref{eq-4supercurrent}).  
Using (\ref{eq-Lorentzgauge}), (\ref{eq-London}), (\ref{eq-continuity}) and (\ref{eq-dalJeJ}) we find

\begin{equation}
{ {\partial^2 {A}^\nu} \over {\partial x^\mu \partial x_\mu}}\ =\ 
-\ {{4\pi a^2} \over c }\ {{\partial^2 {J}^\nu} \over {\partial x^\mu \partial x_\mu}}\ =\ 
{{4\pi} \over c}\ J^\nu\   
\end{equation}

\noindent
so we indeed have the solution.

    This general solution (\ref{eq-generalsol}) is useful because ${p^\mu_s}$ can often be determined.  In a simply connected
superconducting field far from other objects

\begin{equation}
{p^\nu_s}\ =\ 0
\end{equation}

\noindent
as is well known for the 3-vector ${\bf p}_s$ from static situations\cite{Kittel}, so that

\begin{equation}
A^\nu\ =\ -{ {4\pi a^2} \over c }\ J^\nu\ \ \ .
\end{equation}

    Barut and Bornzin\cite{BarutBornzin}
claim that the relation between the four-vector potential and the source current density uses the standard
retarded Green function.  This is {\it not} the case in a  superconducting field
because the electromagnetic field generates supercurrents (\ref{eq-EMgensupercurrent}), and this is not taken into account
in the standard formalism.  Thus in a superconducting field
the electromagnetic field does {\it not} propagate outward from the source supercurrents like a wave.
The 4-vector potential is proportional to the supercurrent density at the {\it same} position in 
spacetime, with the possible addition of a field with zero curl---which cannot affect the 
electromagnetic field.

\subsection{Violation of the Principle of Causality on the Microscopic Scale}
\label{ssec-causality}

     The principle of causality holds that no information
can travel faster than the speed of light.  This principle is {\it not} required
by special relativity since it is violated by an object that moves with a finite
size---that does not propagate as a wave.  We have seen that in a superconducting line 
the electromagnetic field does not propagate like a wave.  Motion as an object of finite size
satisfies (\ref{eq-4potto4current}) and (\ref{eq-generalsol}) because  $A^\nu$, $J^\nu$ and ${p_s}^\nu$ move together.
So we assert that the entire cross section of a line moves together as a unit, continuing to satisfy 
the quantization condition (\ref{eq-multisup}) for a multiply connected region.  This is a violation of
the principle of causality on the microscopic scale.

     Using the phase field we can illustrate this violation directly.  For a path 
which includes the centerline of a superconducting line, we obtain from (\ref{eq-multisup}) in the transverse
rest frame of an element of the line centered at the point ${\bf y}^{\ast}$

\begin{equation}
\oint { {\bf p}^\ast_s} \ {\bf \cdot\ } {\bf dx}^{\ast} \ =\ h
\label{eq-intpseh}
\end{equation}

\noindent
if the line contains one quantum of flux (taking $n=-1$ for convenience).  
Here the $\ast$ refers to the transverse rest frame at ${\bf y}^{\ast}$, moving at ${\bm\beta}$
(see (\ref{eq-beta})) relative to the lab (default rest frame, see Section~\ref{sec-introduction}).
Because of equation (\ref{eq-intpseh}) we can no longer have ${p^\mu_s}=0$ in the 
presence of the centerline.  But the London equations (\ref{eq-London}) must hold outside of the 
centerline, so within the line

\begin{equation}
{\bf \nabla\ \times\ } {{\bf p}^\ast_s} = h \  \delta^{(2)}_T({\bf x}^{\ast} - {\bf y}^{\ast}) {\hat{\ell}}^{\ast}\ \ .
\end{equation}

\noindent
We presume that ${\bf \nabla\ \cdot\ } {{\bf p}^\ast_s} = 0$, since ${{\bf p}^\ast_s}$ would be equal to $0$
if not for the flux quantum at the core of the line.

     We now consider motion in the lab but restrict consideration to nonrelativistic motion transverse to 
the average direction of the line
($\beta^2 \ll 1$)  see (\ref{eq-beta}).  We then use the fact that an appropriate vector field
can be specified by its divergence and curl\cite{Jackson4} to find\cite{polefield} 
that the phase field is completely specified by its behavior on the centerline 

\begin{equation}
{\bf p}_s({t,\bf x})\ =\ {1 \over {4\pi}}\ {\bf \nabla\ \times\ } \int d^3{y^\ast}
  { { {\bf \nabla^\ast_{y^\ast} \times\ } {\bf p}_s(t,{\bf y^\ast}) } \over {\bf \mid x - y^\ast \mid} }\ \ \ .
\label{eq-pstx} 
\end{equation}

\noindent
Evaluating (\ref{eq-pstx}) we simply obtain
(\ref{eq-psx}), the static result.  But this is valid for the moving line as long as relativistic
effects of the transverse motion can be neglected.  This indicates that the whole 
transverse cross section of the line moves together---as an object with finite size.
This type of motion is required by the quantization condition (\ref{eq-intpseh}) and
by the nature of the superconducting line.  

    This type of motion is theoretically convenient because the Lagrangian density for 
the filamentary line with interactions (\ref{eq-Lagrangiandwint}) can be used to derive the equations of
motion for the superconducting line with only minor modifications.  There is no need
to consider added degrees of freedom for the transverse size of the line.  The state of
the entire line can still be completely specified as a function of only two parameters,
which can still be taken as $\tau$ and $\sigma$.  Now $y^\mu(\tau,\sigma)$ refers to the
position in time and space of a point on the centerline of the line.  
     Because this model is {\it acausal}, it is also
{\it nonlocal}; the response at a point is affected by a field over an extended
region of spacetime.  (See section~\ref{ssec-selfintline}.)
The nonlocality of this model makes even self interactions
finite, and indeed self interactions are an important part of the model.
 
\subsection{Electromagnetic Field of the Moving Line}
\label{ssec-EMfield}

    In order to consider the superconducting line, we generalize the filamentary model
of (\ref{eq-filEMdual}-\ref{eq-fil4current}).  For instance the electromagnetic field (\ref{eq-filEM}) is a rank 2 tensor in 4 dimensions,
because $y^\alpha$ (and $y^\beta$) are position 4-vectors, the $\delta$-function is 
Lorentz invariant, and the integral formalism shown is independent of the type of parameterization
as long as $\sigma$ covers the full length of the line, and $\tau$ covers the full
time under consideration.  We must modify the $\delta$-function to give the line a finite
transverse size.  Prescriptions for doing this are frame dependent, so we must start in the
transverse rest frame of an element of the line.  Then we can obtain more general results
via Lorentz transformation to the lab.

    We evaluate (\ref{eq-filEM}) in the transverse rest frame of an element of the 
superconducting line by making the replacement

\begin{equation}
\delta^4 (x-y)\ \rightarrow\ \delta(\tau^\ast_x - \tau^\ast_y) \delta(\ell^\ast_x - \ell^\ast_y)
                             \Delta^2_T ({\bf x}^\ast - {\bf y}^\ast)
\label{eq-deltatrans}
\end{equation}

\noindent
where

\begin{equation}
\Delta^2_T ({\bf x}^\ast - {\bf y}^\ast)\ \equiv\ {1 \over {2\pi a^2} } 
           K_0({\rho^\ast}/ a )
\end{equation}

\noindent
approaches a Dirac $\delta$-function in the dimensions transverse to the line as $a \rightarrow 0$.
Here ${\rho^\ast} {\hat {\rho^\ast} }$ is the perpendicular vector from the centerline to point 
${\bf x}^\ast$ and this perpendicular intersects the centerline at ${y^\ast}^\mu(\tau^\ast,\sigma^\ast)$.
(If there is more than one such point, we must add the contributions in order to find the total field.)
In the transverse rest frame

\begin{equation}
    { {\partial {y^\ast}^\mu} \over {\partial {\tau}^\ast} }\ =\ (1,0)^\mu
\end{equation}

\begin{equation}
    { {\partial {y^\ast}^\mu} \over {\partial \sigma^\ast} }\ =\ 
                                  (0,{\hat \ell}^\ast { {d \ell^\ast} \over {d \sigma^\ast} } )^\mu
\end{equation}

\noindent
where both of these quantities are 4-vectors because ${d {\tau}^\ast}$ and ${d {\sigma}^\ast}$ represent
the proper time and a length element transverse to ${\bm \beta}$ in the
rest frame of an element of the line. 
Hence the electromagnetic field in the transverse rest frame of this element of the string is given by

\begin{widetext}
\begin{equation}
F^\ast_{\mu \nu}(x^\ast)\ =\ 4\pi g_0 \ \epsilon_{\mu \nu \alpha \beta} \int d{\tau^\ast} d{\sigma^\ast}
{ {\partial {y^\ast}^\alpha} \over {\partial {\tau}^\ast} }\  
{ {\partial {y^\ast}^\beta} \over {\partial {\sigma^\ast} } }\ 
\delta(\tau^\ast_x - \tau^\ast_y) \delta(\ell^\ast_x - \ell^\ast_y)
\Delta^2_T ({\bf x}^\ast - {\bf y}^\ast)
\end{equation}

\begin{equation}
F^\ast_{\mu \nu}(x^*)=4\pi g_0 \epsilon_{\mu \nu \alpha \beta} \int d{\tau^\ast} 
{{d{\sigma^\ast}} \over {d {\ell}^\ast}} d {\ell}^\ast
{(1,0)}^\alpha {(0,{\hat \ell}^\ast { {d \ell^\ast} \over {d \sigma^\ast} } ) }^\beta
\delta(\tau^\ast_x - \tau^\ast_y) \delta(\ell^\ast_x - \ell^\ast_y)
\Delta^2_T ({\bf x}^\ast - {\bf y}^\ast)
\end{equation}
\end{widetext}

\begin{equation}
F^\ast_{\mu \nu}(x^\ast)=4\pi g_0 \epsilon_{\mu \nu \alpha \beta}
{(1,0)}^\alpha {(0,{\hat \ell}^\ast) }^\beta
{1 \over {2\pi a^2} } 
K_0({\rho^\ast}/ a )
\end{equation}

\begin{equation}
F^\ast_{\mu \nu}(x^\ast)= { {2 g_0 } \over {a^2} }
K_0({\rho^\ast} / a )
\left( \matrix{0&0&0&0\cr
               0&0&-{{{\hat \ell}^{\ast 3}}}&{{{\hat \ell}^{\ast 2}}}\cr
               0&{{{\hat \ell}^{\ast 3}}}&0&-{{{\hat \ell}^{\ast 1}}}\cr
               0&-{{{\hat \ell}^{\ast 2}}}&{{{\hat \ell}^{\ast 1}}}&0\cr} \right)_{\mu \nu}
\end{equation}

\noindent
where the matrix contains components of the vector ${\hat \ell}^\ast$ giving the direction of
the centerline in the transverse rest frame of the line at the point 
${\bf y^\ast}(\tau^\ast,\sigma^\ast)$.  
So the electromagnetic field in this frame is given by

\begin{equation}
{\bf E^\ast(x^\ast)} = 0
\label{eq-eline*}
\end{equation}

\begin{equation}
{\bf B^\ast(x^\ast)} = { {2 g_0} \over {a^2} } \  K_0({\rho^\ast} /a) \ {\hat{\ell}}^\ast\ \ \ .
\label{eq-bline*}
\end{equation}

\noindent
This looks like the point at which we started (\ref{eq-bline}), but now the unit vector ${\hat{\ell}}^\ast$ 
does not necessarily point along the ${\hat z}$ axis and its direction can change as a function
of $\tau$ and $\sigma$ .

     By performing a tranverse Lorentz transformation from the rest frame of the above element 
of the line to the lab we find the electromagnetic field of the line (for the line (\ref{eq-yzline}) moving in the $y-z$
plane) in the lab

\begin{equation}
{\bf E}(\tau,{\bf x})\ =\ -\gamma \beta B^\ast {\ell_z} {\hat x}
\label{eq-Efieldline}
\end{equation}

\begin{equation}
{\bf B}(\tau,{\bf x})\ =\ \gamma B^\ast {\hat \ell}
\label{eq-Bfieldline}
\end{equation}

\noindent
where ${\hat \ell}$ is the result of Lorentz transformation of ${\hat \ell^\ast}$, selecting events
simultaneous in the lab to describe the moving field.  The extra $\gamma$ factors arising from the 
transformation of the field  enable ${\hat \ell}$ to take the direction of the field when expressed 
in lab coordinates.  In fact the net result of the Lorentz transformation, expressed in lab 
coordinates is the change ${\hat \ell^\ast} \rightarrow \gamma {\hat \ell}$.  Since ${\hat \ell^\ast}$
and ${\hat \ell}$ are unit vectors, they are, in fact, equal.  The directions of the line in its rest
frame and the lab are the same.  Here we use

\begin{equation}
B^\ast(\rho^\ast/a)\ =\ {{2 g_0 } \over a^2}\ K_0(\rho^\ast/a)
\end{equation} 

\begin{equation}
{\bm \rho}\ \equiv\ {\bf x}\ -\ {\bf y}\  \equiv\ \Delta x {\hat x}\ +\ \Delta y {\hat y}\ +\ \Delta z {\hat z}
\end{equation}

\noindent
and

\begin{equation}
\rho^\ast\ =\ \sqrt{(\Delta x)^2\ +\ \gamma^2 (\Delta y)^2\ +\ (\Delta z)^2}
\end{equation}

\noindent
is the pancaking cylindrical radius of the line expressed in lab coordinates.

     By our use of a Lorentz transformation we are ignoring the fact that the rest frame of the element
of the line is an accelerating reference frame.  According to Meissner, Thorne, and Wheeler\cite{MTW} this 
approximation, using a comoving inertial frame, is permissible as long as inertial forces are small and 
typical accelerations are much less than $c^2/d$ where $d$ is the amplitude of a typical oscillation.  We will 
assume that these conditions hold true except near the ends of a line.

\subsection{Supercurrent of the Moving Line}
\label{ssec-supercurrent}

     We wish to use Maxwell's equations to find the supercurrent in the lab.  
Maxwell's equations act at the {\it field} point ${\bf x}$ and therefore we do not consider the explicit
space and time dependence of the string coordinate ${\bf y}$, which acts effectively as a
source point.  Neither do we consider the explicit space and time dependence of $\gamma$ and $\beta$
which is determined by ${\bf y}$.  
However, the field at {\bf x} can change with time due to motion of the string.
Since the electromagnetic field does not change (is {\it quantized}) 
in the rest frame of the string, all time dependence is due to motion.  So at a fixed field point ${\bf x}$, 
the change of the magnetic induction field with time is given by

\begin{equation}
d{\bf B}\ =\ (-{\bm \beta} d\tau ) \cdot {\bf \nabla B}
\end{equation}

\begin{equation}
{ {\partial {\bf B}} \over {\partial \tau} }\ =\ -({\bm\beta} \cdot {\bf \nabla}){\bf B}\ 
\end{equation}

\noindent
so for the line (\ref{eq-yzline}) moving in the $y-z$ plane

\begin{equation}
{ {\partial {\bf B}} \over {\partial \tau} }\ =\ -\beta {{\partial {\bf B}} \over {\partial y}}\ \ \ .
\end{equation}

\noindent
Here ${\bm \beta}$ is the velocity of the line at the field point ${\bf x}$ which is the 
same as the velocity of the point on the centerline ${\bf y}$ associated with ${\bf x}$, since the cross
section of the line moves together.
For similar reasons the change of the electric field with time is given by

\begin{equation}
{ {\partial {\bf E}} \over {\partial \tau} }\ =\ -({\bm\beta} \cdot {\bf \nabla}){\bf E}\
\end{equation}

\noindent
and if the line satisfies (\ref{eq-yzline})

\begin{equation}
{ {\partial {\bf E}} \over {\partial \tau} }\ =\ -\beta {{\partial {\bf E}} \over {\partial y}}\ \ \ .
\end{equation}

\begin{widetext}
     Using Maxwell's equations we find

\begin{equation}
{J}^\nu(x)\ = { {g_0 c}  \over {2\pi a^3} }{ {\gamma K_1(\rho^*/a)} \over \rho^*}\ 
     (l_z \beta \Delta x \ ,\ {\hat \ell} \times {\bm \rho})^\nu \ 
   + { {g_0 c}  \over {2\pi a^2} }{ {\gamma K_0(\rho^*/a)} \over R}\  (1 - \ell_y^2\beta^2)\  (0,\hat{x})^\nu 
\label{eq-supercurrentyz}
\end{equation}

\noindent
where $R$ is the instantaneous radius of curvature at $x$.
This current satisfies the equation of continuity (\ref{eq-continuity}) exactly.  Details of these calculations are 
included in Appendix~\ref{ssec-instradius}, where the derivatives of $\hat{\ell}$ 
are explicitly evaluated.  It is also shown in Appendix~\ref{ssec-calcEMfield} that if we apply equation (\ref{eq-generalsol}) 
and attempt to derive the electromagnetic field (\ref{eq-Efieldline}) and (\ref{eq-Bfieldline}) 
from (\ref{eq-supercurrentyz}) we do not get precise agreement.  We get the expected result plus
remainder terms proportional to $a/R$ and $(a/R)^2$ which are small for large radii of curvature.  
We take this to imply that our simple 
procedure, which involves giving the delta function a finite size and using a Lorentz transfomation from a comoving inertial frame, 
does not provide an exact solution for the electromagnetic field.  It will be sufficient, however, for our present purposes. 
 
Performing an inverse Lorentz transformation we find the supercurrent in the rest frame of an element of the line:

\begin{equation}
{J^\ast}^\nu(x^\ast)\ = { {g_0 c}  \over {2\pi a^3} } \Bigl\{ {K_1(\rho^*/a)}\ 
    (0\ ,\ {\hat \ell}^\ast \times {\hat \rho}^\ast)^\nu \ 
   + { a \over {R^\ast}} (1-{\ell^\ast_y}^2 \beta^2 / \gamma^2) {K_0(\rho^*/a)}\  (0,\hat{x})^\nu \Bigr\} 
\label{eq-supercurrentyz*}
\end{equation}
\end{widetext}

\noindent
The parameters $\gamma$ and $\beta$ enter this expression because we have defined $R$ in the lab.  
(See Appendix~\ref{ssec-instradius})  The charge density is zero in the rest frame. 

     It is shown in Appendix~\ref{ssec-energyofbending} that the energy of the finite size line is independent of the instantaneous 
radius of curvature $R$ at a given point.  This fact is important because we do not include any term in the Lagrangian density for the 
energy of bending.    

     So in the absence of interactions, general wave solutions of the type (\ref{eq-yzline}) and (\ref{eq-circularline}) traveling in 
the positive $z$-direction could occur with the superconducting line.  Replacing the argument $\tau - z$ with $\tau + z$ produces 
solutions traveling in the negative $z$-direction.

\subsection{Self Interaction of the Moving Line}
\label{ssec-selfintline}

     Following the sign conventions of Bjorken and Drell\cite{BjDrellaction} (which are consistent with Jackson)
the action (except for weak and hadronic modes) is

\begin{widetext}
\begin{equation}
   S\ =\ - { 1 \over {16\pi} }\ \int { {d^4x} \over c } F_{\mu \nu} F^{\mu \nu}\ 
      -\ { 1 \over {c^2} } \int {d^4x}\ (J^e_\mu + J^s_\mu) \ A^\mu\ 
      +\ \sum_j \ {1 \over c}\ \int d\tau_j d\sigma_j {\cal L}_{mj}
\label{eq-actionnoh}
\end{equation}

\noindent
where no mass term appears in the Lagrangian because we assert that mass is not fundamental.  Here ${\cal L}_{mj}$ is the 
mechanical string Lagrangian density (\ref{eq-Lagrangiandensity}) 
for line $j$, $J^e_\mu$ is the electromagnetic current due to electric charges, 
and $J^s_\mu$ is the electromagnetic current due to superconducting lines.
Applying the principle of least action to the superconducting line with finite size, using (\ref{eq-deltatrans}) in (\ref{eq-fil4current}), 
it can be shown that the equation of motion for a given line, instead of (\ref{eq-filmotionwint}), is given in the case of interactions by

\begin{equation}
{{\partial \ } \over {\partial \sigma}} \Bigl( { {\partial {\cal L}_m} \over {\partial {y^\prime}^\mu } }  \Bigr) \ +\
{{\partial \ } \over {\partial \tau}} \Bigl( { {\partial {\cal L}_m} \over {\partial {\dot y}^\mu } } \Bigr) \ =\
{{4\pi g_0} \over c^2}\  {\dot y}^\alpha {y^\prime}^\beta \epsilon_{\mu \nu \alpha \beta} \bigl\langle J^\nu (y) \bigr\rangle\ 
\equiv \vartheta_\mu (y)
\label{eq-eqofmotionwint}
\end{equation}

\noindent
where

\begin{equation}
\bigl\langle J^\nu (y) \bigr\rangle\ \equiv\ {1 \over {2\pi a^2}} 
\int d^2{\rho}^\ast\ K_0(\rho^\ast/a)\ 
J^\nu(\tau_y^\ast,\ell_y^\ast,{\bf x}^\ast)
\label{eq-Jnuave}
\end{equation}

\noindent
is the average current over the cross section of the line weighted by the shape of the line field.  This averaging must be
performed in the transverse rest frame of the line.  But the resulting current is a 4-vector, and can be transformed to any frame
(within the approximations which have been discussed\cite{MTW}).  Indeed, since the transverse rest frame varies with the line
element in question for solutions of the type (\ref{eq-yzline}),  
it is most convenient for these solutions to think of the quantities ${\dot y}^\alpha$, ${y^\prime}^\beta$, and 
$\bigl\langle J^\nu \bigr\rangle$ transformed from the rest frame to the lab.  Then if we wish to make a change of parameterization
to the particular parameterization of Section~\ref{ssec-neglecting}, we can do so.  (It is only important that the parameterization be the same on
both sides all equations.)  

    For a given line we work in the large radius of curvature approximation, $R_0 \gg a$, where $R_0$ is
the radius of curvature of the centerline---so that all points which make a significant contribution to the average are
perpendicular to the line at a single unique point (always true for points closer to the line than $R_0$).

     To evaluate the interaction of the line with itself, we use equation (\ref{eq-supercurrentyz*}) in the interaction term 
and obtain (with $u\ \equiv\ \rho^\ast/a$)

\begin{displaymath}
\vartheta_\mu\ =\ -\ { {g_0^2} \over {\pi c a^3}} { {d\ell^\ast} \over {d\sigma^\ast}}
\Biggl\lbrack\ \int du\ u\ K_0(u)K_1(u)\ \int d\phi^\ast\ (0, {\hat \rho}^\ast)_\mu
\end{displaymath}

\begin{equation}
-\ (1-{\ell^\ast_y}^2 \beta^2 / \gamma^2)\ 
(0,{\hat \ell}^\ast\times{\hat x})_\mu\ \int du\ u\ {K_0^2(u)} \int d\phi^\ast 
{ 1 \over { ({R^\ast_0}/a) + u cos(\phi^\ast)} } \Biggr\rbrack
\label{eq-selfintline}
\end{equation}
\end{widetext}

\noindent
where we have used the fact that the radius of curvature of the centerline can be related to the shape 
of the trajectory (see Appendix~\ref{ssec-radiusandshape}) according to the equation

\begin{equation}
{1 \over {R_0}}\ =\ -\ \ell^3_z\ {{d^2y} \over {dz^2}}
\end{equation}

\noindent
in order to learn how to transform the radius of curvature from the lab to the line's rest frame

\begin{equation}
R\ =\ \gamma R^\ast\ \ \ .
\end{equation}

     Because of azimuthal symmetry the first term in (\ref{eq-selfintline}) gives a result of zero.  Unless the radius of curvature is infinite,
 the second term in (\ref{eq-selfintline}) is easily seen to be finite but not equal to zero.  In fact if $ {R^\ast_0}\  \gg\  a$, we obtain 

\begin{equation}
\vartheta_\mu\ =\  +\  { {2T} \over { c R^\ast_0 } }\  
 { {(0,{\hat \ell}^\ast \times {\hat x})_\mu }  \over { \vert {\ell^\ast_z} \vert} }\ 
 (1-{\ell^\ast_y}^2 \beta^2 / \gamma^2)\ \ \ .  
\end{equation}

    Hence we have evaluated the self interaction of the superconducting line.  It is finite and calculable in spite of the fact that
the field diverges along the centerline.  The averaging required by the Principle of Least Action leads to a finite result.

    Because the self-interaction term in equation (\ref{eq-eqofmotionwint}) is not zero, and traveling waves are an exact solution of this equation
without interactions, there are {\it no} traveling wave solutions propagating at  the velocity of light for the motion of a single 
isolated superconducting line once self-interactions are included.  Wave motion of a single superconducting line may still be 
possible when the photon quantization condition is satisfied---but then the line is not isolated, and we are currently
unable to describe the motion.

\section{The Photon}
\label{sec-photon}

\subsection{Wave Function of the Electron}
\label{ssec-wfelectron}

    Now we are ready to define what we mean by the wave function of the electron in this model 
which plays such a prominent role in the Dirac
quantization condition (\ref{eq-singv}).  We define the electron wave function to be the change in the ether caused by the 
bare electron, i.e. by
the end of the appropriate line.  This change will be the response of the ether to the 
bare electron's electromagnetic field.
(See Section~\ref{ssec-masselectron}.)  Since the 
bare electron's electromagnetic field has infinite extent, the wave function of the electron also has infinite 
extent---as in standard quantum theory, and consistent with our reasoning in Section~\ref{ssec-FQ}.

\begin{figure}
    \vbox to 2.6in{
\includegraphics[scale=0.75]{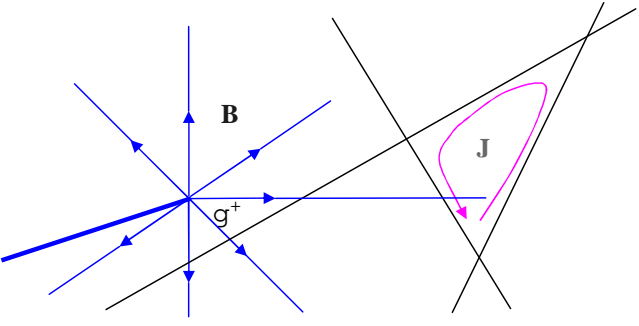}  }
    \caption{\small  (Color online.)  If the net external flux through a complete circuit of line pairs is less than a half a flux
                     quantum, a supercurrent will flow bringing the net flux to zero.}
\label{fig-fluxlhalf}
\end{figure}

\begin{figure}
    \vbox to 2.6in{
\includegraphics[scale=0.75]{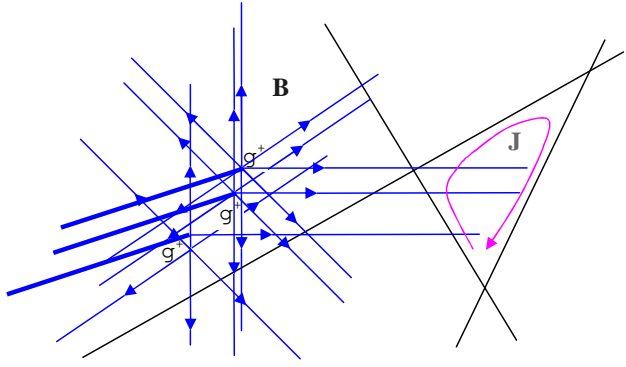}  }
    \caption{\small (Color online.)  If the net external flux through a complete circuit of line pairs is greater than
                    half a flux quantum, a supercurrent will flow bringing the net flux up to a flux quantum.}
\label{fig-fluxghalf}
\end{figure}

    On the microscopic scale, this wave function is not continuous (differing from the Standard Model).  
It is a mesh of line pairs.  
Each line pair is continuous.  We should realize that the cross sections of our line pairs are small on
the atomic scale and mostly fit into the holes in matter.  The outer radius of a line pair (Section~\ref{ssec-static})
must be much smaller than the Bohr radius of the hydrogen atom: $R_a=k_a a \ll 5.29 \times 10^{-11}m$. 
Hence there
are many holes in the wave function which constitute a larger version of 
Dirac's ``nodal lines".  If three of the line pairs form a complete circuit
around such a hole, and if magnetic field links the circuit, the circuit will superconduct in response to the vector potential 
(Figure~\ref{fig-fluxlhalf}),
causing quantization of magnetic flux (\ref{eq-FQ}) through the circuit.  If the applied flux through the complete circuit is less than
half a flux quantum, the net flux will be brought to zero.  If the applied flux through the circuit is greater than half a 
flux quantum,
experimental evidence\cite{Qworld2} indicates that the supercurrent in the complete circuit will {\it increase} the flux as
necessary to match the nearest flux quantum (Figure~\ref{fig-fluxghalf}).

    We will see that the photon can be considered as an excitation of the wave function of the electron.

\subsection{Interactions of a Line Pair}
\label{ssec-intoflp}

    We seek the types of waves which can propagate at the speed of light 
on the two lines of an isolated pair.  We attempt to find them using the known
solutions (\ref{eq-yzline}) and (\ref{eq-circularline}) without the interaction terms.  
From the Principle of Least Action for the twoline system we find

\begin{widetext}
\begin{displaymath}
    \delta\ S\ =\ 0
\end{displaymath}

\begin{displaymath}
=\int d\tau d\sigma \Biggl \{ \biggl [ 
 {{4\pi g_0 } \over c^2}\ \epsilon_{\mu \nu \alpha \beta}
 {{\dot y}_1}^\alpha {y^\prime_1}^\beta  \bigl\langle {J_1}^\nu + {J_2}^\nu \bigr\rangle_1
 - \biggl \{
{{\partial \ } \over {\partial \sigma}} \Bigl( { {\partial {\cal L}_{m1}} \over {\partial {y^\prime_1}^\mu } }  \Bigr) \ +\
{{\partial \ } \over {\partial \tau}} \Bigl( { {\partial {\cal L}_{m1}} \over {\partial {{\dot y}_1}^\mu } } \Bigr) \biggr \} 
\biggr ] \delta {y^\mu_1}
\end{displaymath}

\begin{equation}
\ \ \ \ \ \ \ \ \ \   +\ \biggl [
 {{4\pi g_0 } \over c^2}\ \epsilon_{\mu \nu \alpha \beta}
 {{\dot y}_2}^\alpha {y^\prime_2}^\beta  \bigl\langle {J_1}^\nu + {J_2}^\nu \bigr\rangle_2
 - \biggl \{
{{\partial \ } \over {\partial \sigma}} \Bigl( { {\partial {\cal L}_{m2}} \over {\partial {y^\prime_2}^\mu } }  \Bigr) \ +\
{{\partial \ } \over {\partial \tau}} \Bigl( { {\partial {\cal L}_{m2}} \over {\partial {{\dot y}_2}^\mu } } \Bigr) \biggr \} 
\biggr ] \delta {y^\mu_2} \Biggr \}  \ \ \ .
\end{equation}

\noindent
Due to the interaction terms the variations in $y_1$ and $y_2$ are coupled so we cannot set the coefficients of
$\delta y_1$ and $\delta y_2$ separately to zero, without first knowing the solutions to the equations of motion.

To handle this situation, we define the 4-vectors

\begin{equation}
     y\ \equiv\ {1 \over 2}\ (y_1\ +\ y_2)\ \ \ \ \ \ \ \ \ \ d\ \equiv\ {1 \over 2}\ (y_1\ -\ y_2)
\label{eq-yd}
\end{equation}

\noindent
as the average and half the difference between the wave solutions of line 1 and line 2.  Then we find

\begin{equation}
     y_1\ =\ y\ +\ d\ \ \ \ \ \ \ \ \ \ \ \ y_2\ =\ y\ - \ d\ \ \ .
\end{equation}

\noindent
We express the mechanical string Lagrangian densities (\ref{eq-Lagrangiandensity}) for line 1 and line 2 in terms of $y$ and $d$:

\begin{equation}
 {\cal L}_{m1} = {T\over c}\Bigl\{(y^\prime\cdot{\dot y}+y^\prime\cdot{\dot d}+d^\prime\cdot{\dot y}+d^\prime\cdot{\dot d})^2 
- ({y^\prime}^2+2y^\prime\cdot d^\prime+{d^\prime}^2)({\dot y}^2+2{\dot y}\cdot{\dot d}+{\dot d}^2) \Bigl\}^{1/2}
\label{eq-Lm1}
\end{equation}

\begin{equation}
 {\cal L}_{m2} = {T\over c}\Bigl\{(y^\prime\cdot{\dot y}-y^\prime\cdot{\dot d}-d^\prime\cdot{\dot y}+d^\prime\cdot{\dot d})^2
- ({y^\prime}^2-2y^\prime\cdot d^\prime+{d^\prime}^2)({\dot y}^2-2{\dot y}\cdot{\dot d}+{\dot d}^2) \Bigl\}^{1/2}
\label{eq-Lm2}
\end{equation}
\end{widetext}

\subsubsection{Common Mode Solutions---Gravitational Waves}
\label{sssec-commonmodesol}

    To find the common mode solutions we set

\begin{equation}
   d\ =\ d_c
\end{equation}

\noindent
where $d_c$ is a constant 4-vector.  Note that $d$ has no time component and no
$z$-component by construction.  Thus  

\begin{equation}
    y_1\ =\ y_2\ +\ 2 d_c
\label{eq-ydif}
\end{equation}

\noindent
and the twoline system exhibits no excitation.  It moves together as one system.  Since the common mode solutions
exhibit no excitations, and hence no net electromagnetic field at a given $(\tau,{\hat z}z)$, 
we interpret them as gravitational waves.  (See Section~\ref{sec-generalrel}.)

    Since $d$ is constant, ${\dot d}=0$ and $d^\prime=0$, hence the mechanical Lagrangian density of the system becomes

\begin{equation}
 {\cal L}_m ={\cal L}_{m1}+{\cal L}_{m2}
= {2T\over c}\Bigl\{ (y^\prime \cdot {\dot y})^2 -(y^\prime)^2(\dot y)^2 \Bigl\}^{1/2}
=2{\cal L}_{m1}=2{\cal L}_{m2}
\label{eq-Lm}
\end{equation} 

\noindent
just twice the Lagrangian density (\ref{eq-Lagrangiandensity}).  
Since for common mode solutions $\delta {y^\mu_1}=\delta {y^\mu_2}=\delta {y}^{\mu} $
the variation of the action becomes

\begin{widetext}
\begin{displaymath}
    \delta\ S\ =\ 0
\end{displaymath}

\begin{displaymath}
=\int d\tau d\sigma \Biggl \{ 
 {{4\pi g_0 } \over c^2}\ \epsilon_{\mu \nu \alpha \beta}
\biggl [ {{\dot y}_1}^\alpha {y^\prime_1}^\beta  \bigl\langle {J_1}^\nu + {J_2}^\nu \bigr\rangle_1
\ +\     {{\dot y}_2}^\alpha {y^\prime_2}^\beta  \bigl\langle {J_1}^\nu + {J_2}^\nu \bigr\rangle_2 \biggr ]
\end{displaymath}

\begin{equation}
 - \biggl \{
{{\partial \ } \over {\partial \sigma}} \Bigl( { {\partial {\cal L}_{m}} \over {\partial {y^\prime}^\mu } }  \Bigr) \ +\
{{\partial \ } \over {\partial \tau}} \Bigl( { {\partial {\cal L}_{m}} \over {\partial {\dot y}^\mu } } \Bigr) \biggr \}
 \Biggr \} \delta {y}^{\mu}
\end{equation}

\noindent
and the equation of motion becomes

\begin{displaymath}
{{\partial}\over{\partial \sigma}}\Bigl\{-{2T\over c}\{\beta_\ell\ {u_\perp}_\mu +
     \sqrt{1-{\beta_\perp}^2} \ (0,{\hat \ell})_\mu\}\Bigl\}+
{{\partial }\over{\partial \tau}}\Bigl\{{2T\over c}{d\ell \over d\sigma}{u_\perp}_\mu\Bigr\}\ 
\end{displaymath}

\begin{equation}
 =\ {{4\pi g_0 } \over c^2}\ \epsilon_{\mu \nu \alpha \beta} 
\Bigl\{ {{\dot y}_1}^\alpha {y^\prime_1}^\beta  \bigl\langle {J_1}^\nu + {J_2}^\nu \bigr\rangle_1
    +   {{\dot y}_2}^\alpha {y^\prime_2}^\beta  \bigl\langle {J_1}^\nu + {J_2}^\nu \bigr\rangle_2 \Bigl\}
\label{eq-intcommode}
\end{equation}
\end{widetext}

\noindent
where $y_1, y_2$ and $y$ must satisfy equations (\ref{eq-yd}) and (\ref{eq-ydif}), and the left hand side of the equation refers to $y$.

    We label the 8 interaction terms ${\vartheta_\mu}^{ijk}$ where $i=1$ or 2 corresponding to the line where the average
is performed, $j=1$ or 2 corresponding to the line field which is averaged, and $k=1$ or 0 corresponding to the term in the
current (containing either a $K_1$ or a $K_0$, see (\ref{eq-supercurrentyz*}) ).  For the common mode solutions, line 1 and line 2
have the {\it same} instantaneous rest frame, which simplifies the averaging calculation.

    As in Section~\ref{ssec-selfintline}, we find that ${\vartheta^\ast_\mu}^{111} = 0 = {\vartheta^\ast_\mu}^{221}$.  In order to obtain
wave solutions with the remaining interaction terms we specify the constant 4-vector to be in the ${\hat x}$-direction

\begin{equation}
    d_c\ =\ ( 0,{\hat x} d_c)
\end{equation}

\begin{figure}
    \vbox to 1.8in{
\includegraphics[scale=0.65]{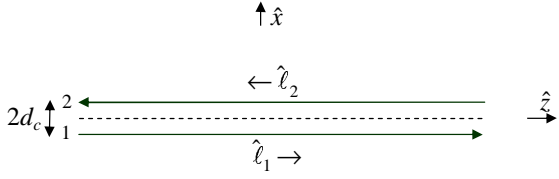}  }
    \caption{\small Line pair common mode geometry.  The lines bend in the $y-z$ plane.}
\label{fig-commonmodegeom}
\end{figure}

\noindent
(labeling the constant length also $d_c$) if the motion is in the $y-z$ plane.  
Note that ${{\hat \ell}_2}^\ast=-{{\hat \ell}_1}^\ast$
since the fields are in opposite directions.  (See Figure~\ref{fig-commonmodegeom}.)  Using the large radius of curvature 
approximation and equations (\ref{eq-K0fact}), (\ref{eq-Wronskian}) and the derivative of (\ref{eq-Wronskian}) we find

\begin{equation}
{\vartheta^\ast_\mu}^{121} \ = \ { {4T} \over {ca} } { { (0,-{\hat x})_{\mu} } \over {\vert \ell^\ast_z \vert} }
   \ \Bigl \{ {u_c \over 2} K_2(u_c)\ -\ K_1(u_c) \Bigr \}
\label{eq-I121}
\end{equation}

\noindent
where the normalized separation of the two lines in the ${\hat x}$ direction is given by

\begin{equation}
    u_c\ =\ { {2 d_c} \over a}\ \ \ .
\end{equation}

\noindent 
(We quote the covariant components $(0,-{\hat x})_{\mu}$, so the interaction is {\it attractive}.)
As expected, this interaction term is zero if $u_c = 0$ and attractive for all $u_c$ greater than 0.
From rotational symmetry we see from Figure~\ref{fig-commonmodegeom} that

\begin{equation}
{\vartheta^\ast_\mu}^{211}\ =\ -\ {\vartheta^\ast_\mu}^{121} \ \ \ .
\end{equation}

    After taking advantage of the fact that for this common mode geometry the instantaneous rest frames
and instantaneous radii of curvature of the two lines are the same 
if the separation is perpendicular to the motion, we find that

\begin{widetext}
\begin{equation}
{\vartheta^\ast_\mu}^{110}\ +\ {\vartheta^\ast_\mu}^{120}\ =
  \ +{{2T} \over {c {R^\ast_0}} } { {(0, {\hat \ell}^\ast_1 \times {\hat x} )_{\mu} } \over {\vert \ell^\ast_z \vert} }
  \ (1\ -\ { { {\ell^\ast_y}^2 \beta^2 } \over \gamma^2 } )\ (1\ -\ u_c K_1(u_c))
\label{eq-I110pI120}
\end{equation}
\end{widetext}

\noindent
if $ \vert R^\ast_0 \vert \gg a$ .
Since ${{\hat \ell}_1}^\ast = - {{\hat \ell}_2}^\ast$ for common mode solutions, we find that

\begin{equation}
{\vartheta^\ast_\mu}^{110}\ +\ {\vartheta^\ast_\mu}^{120}\ 
+\ {\vartheta^\ast_\mu}^{210}\ +\ {\vartheta^\ast_\mu}^{220}\ =\ 0
\end{equation}

\noindent
and the sum of all interaction terms for this common mode geometry is zero.

    Hence, before considering the quantization conditions, common mode solutions---gravitational waves---of types (\ref{eq-yzline}) 
can propagate freely at the speed of light if ${\hat d}$, the direction of the 3-vector 
associated with $d_c$, is
perpendicular to the motion.  It is shown in Appendix~\ref{sec-generalmotionincommonmode} that the orientation
of ${\hat d}$ does not affect common mode motion.  So common mode solutions of type (\ref{eq-circularline})
can also propagate at the speed of light.  We will see in Section~\ref{ssec-generalityquant} that the photon 
quantization conditions must still be satisfied, even in this case.  But the quantization currents have no role
beyond their role in keeping the separation $d_c$ constant.  (See Section~\ref{ssec-tensionpair}.)

    Since the orientation of ${\hat d}$ does not matter in the common mode, and the sum of the 
interaction terms is zero in common mode, electromagnetic waves with ${\hat d}$ spinning (helical polarization)
can propagate at the speed of light on top of gravitational waves---as long as the photon quantization conditions
are satisfied.

\subsubsection{Differential Mode Solutions---Electromagnetic Waves}
\label{sssec-differentialmodesol}
    To find differential mode solutions we set

\begin{equation}
    y\ =\ (\tau,{\hat z} z)
\end{equation}

\noindent
without variation, and therefore

\begin{equation}
    {\dot y}\ =\ (1,0) \ \ \ \ \ \ \ \ \ \ \ y^{\prime}\ =\ (0,{\hat z}) \ \ \ .
\end{equation}

\noindent
We consider the case where $d$ has only transverse components (unlike Appendix~\ref{sec-generalmotionincommonmode}) 
so $d$, ${\dot d}$ and $d^\prime$ are all
orthogonal to $y$, ${\dot y}$ and $y^\prime$.  The total mechanical Lagrangian density becomes

\begin{widetext}
\begin{equation}
 {\cal L}_m  ={\cal L}_{m1}+{\cal L}_{m2}= 
{2T\over c}\Bigl\{ (d^\prime \cdot {\dot d})^2 \ -\ (-1+{d^\prime}^2)\ (1+{\dot d}^2) \Bigl\}^{1/2}\ 
=\ 2{\cal L}_{m1}\ =\ 2{\cal L}_{m2}
\end{equation}

\noindent
If we simply define

\begin{equation}
    y_d\ \equiv\ y\ +\ d
\end{equation}

\noindent
the mechanical Lagrangian density becomes

\begin{equation}
 {\cal L}_m  = {2T\over c}\Bigl\{ ({y^\prime}_d \cdot {\dot y_d})^2 \ -\ ({y^\prime}_d^2)\ ({\dot y_d}^2)  \Bigr\}^{1/2}
\end{equation}

\noindent
with the standard solutions of the form (\ref{eq-yzline}) and (\ref{eq-circularline}) without interaction.  We refer to this differential
solution as $y_d$ (rather than $y_1$) to indicate generality.
Both $y_1$ and $y_2$ are solutions to this mechanical Lagrangian
without interaction.

    So for the differential mode

\begin{equation}
    y_{1d}\ =\ y\ +\ d\ \ \ \ \ \ \ \ \ \ \ \ \ \ y_{2d}\ =\ y\ -\ d\ \ \ .
\label{eq-y1dy2d}
\end{equation}

\noindent
We require

\begin{equation}
    \delta y^{\mu}\ =\ 0
\end{equation}

\noindent
and therefore

\begin{equation}
    \delta {y_{1d}}^\mu\ =\ +\ \delta d^\mu \ \ \ \ \ \ \ \ \ \ \ \ \ \ \  \delta {y_{2d}}^\mu\ =\ -\ \delta d^\mu\ \ \ .
\end{equation}

\noindent
The variation of the action becomes

\begin{displaymath}
    \delta\ S\ =\ 0
\end{displaymath}

\begin{displaymath}
=\int d\tau d\sigma \Biggl \{ \biggl [
 {{4\pi g_0 } \over c^2}\ \epsilon_{\mu \nu \alpha \beta}
 {{\dot y}_{1d}}^\alpha {y^\prime_{1d}}^\beta  \bigl\langle {J_1}^\nu + {J_2}^\nu \bigr\rangle_1
 - \Bigl \{
{{\partial \ } \over {\partial \sigma}} \Bigl( { {\partial {\cal L}_{m1}} \over {\partial {d^\prime}^\mu } }  \Bigr) \ +\
{{\partial \ } \over {\partial \tau}} \Bigl( { {\partial {\cal L}_{m1}} \over {\partial {\dot d}^\mu } } \Bigr) \Bigr \} 
\biggr ]  (+\delta d^\mu)
\end{displaymath}

\begin{equation}
\ \ \ \ \ \ \ \ \ \   +\ \biggl [
 {{4\pi g_0 } \over c^2}\ \epsilon_{\mu \nu \alpha \beta}
 {{\dot y}_{2d}}^\alpha {y^\prime_{2d}}^\beta  \bigl\langle {J_1}^\nu + {J_2}^\nu \bigr\rangle_2
 + \Bigl \{
{{\partial \ } \over {\partial \sigma}} \Bigl( { {\partial {\cal L}_{m2}} \over {\partial {d^\prime}^\mu } }  \Bigr) \ +\
{{\partial \ } \over {\partial \tau}} \Bigl( { {\partial {\cal L}_{m2}} \over {\partial {\dot d}^\mu } } \Bigr) \Bigr \}
\biggr ](- \delta d^\mu)   \Biggr \}
\end{equation}

\begin{displaymath}
=\int d\tau d\sigma \Biggl \{
 {{4\pi g_0 } \over c^2}\ \epsilon_{\mu \nu \alpha \beta}
\biggl [ {{\dot y}_{1d}}^\alpha {y^\prime_{1d}}^\beta  \bigl\langle {J_1}^\nu + {J_2}^\nu \bigr\rangle_1
\ -\     {{\dot y}_{2d}}^\alpha {y^\prime_{2d}}^\beta  \bigl\langle {J_1}^\nu + {J_2}^\nu \bigr\rangle_2 \biggr ]
\end{displaymath}

\begin{equation}
 - \biggl \{
{{\partial \ } \over {\partial \sigma}} \Bigl( { {\partial {\cal L}_{m}} \over {\partial {y^\prime_d}^\mu } }  \Bigr) \ +\
{{\partial \ } \over {\partial \tau}} \Bigl( { {\partial {\cal L}_{m}} \over {\partial {{\dot y}_d}^\mu } } \Bigr) \biggr \}
 \Biggr \} \delta {{y_d}}^{\mu}
\end{equation}

\noindent
and we see that the differential mode solutions involve the {\it difference} between the interaction terms of the two lines.
The equation of motion for differential mode solutions becomes:

\begin{displaymath}
{{\partial}\over{\partial \sigma}}\Bigl\{-{2T\over c}\{\beta_\ell\ {u_\perp}_\mu +
     \sqrt{1-{\beta_\perp}^2} \ (0,{\hat \ell})_\mu\}\Bigl\}+
{{\partial }\over{\partial \tau}}\Bigl\{{2T\over c}{d\ell \over d\sigma}{u_\perp}_\mu\Bigr\}\
\end{displaymath}

\begin{equation}
\ \ \ \  =\ {{4\pi g_0 } \over c^2}\ \epsilon_{\mu \nu \alpha \beta}
\Bigl\{ {{\dot y}_{1d}}^\alpha {y^\prime_{1d}}^\beta  \bigl\langle {J_1}^\nu + {J_2}^\nu \bigr\rangle_1
    -   {{\dot y}_{2d}}^\alpha {y^\prime_{2d}}^\beta  \bigl\langle {J_1}^\nu + {J_2}^\nu \bigr\rangle_2 \Bigl\}
\label{eq-eqofmdiff}
\end{equation}
\end{widetext}

\noindent
where $y_{1d}$ and  $y_{2d}$ must satisfy (\ref{eq-y1dy2d}) and the left hand side of the equation refers to $y_d$ (hence either
$y_{1d}$ or $y_{2d}$).      

\ \ \ \ \ {\it a.  Radial Excitation---Linear Polarization}
\label{ssssec-radialexcitation}

\begin{figure}
    \vbox to 2.6in{
\includegraphics[scale=0.75]{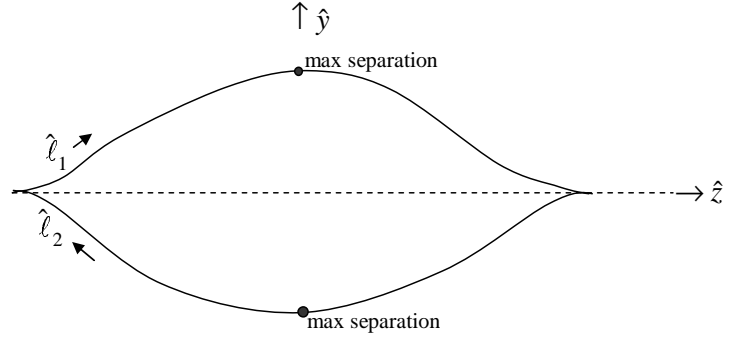}  }
    \caption{\small Geometry of a differential mode wave with radial excitation.}
\label{fig-radialexcitation}
\end{figure}

    Now we let $d^\mu = (0,{\hat y} F(\tau-z))^\mu$ in order to look for differential solutions of the type (\ref{eq-yzline}).
We consider the moment in space and time shown in Figure~\ref{fig-radialexcitation} when the two lines reach 
maximum separation.  At this
moment the two lines will be at rest in the lab, hence they have common mode geometry in their joint rest frame.
Accordingly we can use (\ref{eq-I121}) and (\ref{eq-I110pI120}) to calculate the interaction terms for differential mode waves 
with radial excitation

\begin{equation}
{\vartheta^\ast_\mu}^{111}\  =\ 0\ =\ {\vartheta^\ast_\mu}^{221}
\end{equation}

\begin{equation}
{\vartheta^\ast_\mu}^{121} \ -\ {\vartheta^\ast_\mu}^{211} \ 
= \ { {8T} \over {ca} }\  { { (0,+{\hat y}^\ast)_{\mu} } \over {\vert \ell^\ast_z \vert} }
   \ \Bigl \{ {u_s \over 2} K_2(u_s)\ -\ K_1(u_s) \Bigr \}
\label{eq-I121m211}
\end{equation}

\begin{displaymath}
{\vartheta^\ast_\mu}^{110}\ +\ {\vartheta^\ast_\mu}^{120}\
-\ {\vartheta^\ast_\mu}^{210}\ -\ {\vartheta^\ast_\mu}^{220}\ =\ 
\end{displaymath}

\begin{equation}
  \ +{{4T} \over {c {R^\ast_0}} } { {(0, {\hat \ell}^\ast_1 \times {\hat x} )_{\mu} } \over {\vert \ell^\ast_z \vert} }
  \ (1\ -\ { { {\ell^\ast_y}^2 \beta^2 } \over \gamma^2 } )\ (1\ -\ u_s K_1(u_s))
\label{eq-I110p120m210m220}
\end{equation}

\noindent
at the distance $d_s$ of maximum separation (with $u_s=2d_s/a$).

    We keep the sign conventions of the common mode interaction terms for clarity.  Since we must take the {\it difference} of
these terms, it is very easy to see that the differential mode interaction does not vanish for the case of radial excitation.

    Although we have only considered the point of maximum separation, clearly the nonvanishing of the
interaction terms is general throughout the motion with radial excitation.  
The interaction terms (\ref{eq-I121m211}) and (\ref{eq-I110p120m210m220})
point in the $(0,-{\hat y}^\ast)^\mu$ direction, but this is no longer true away from maximum separation since
${\hat \ell}^\ast_1 \times {\hat x} $ and ${\hat \ell}^\ast_2 \times {\hat x} $ are {\it not} in opposite directions.

    Since motion with radial excitation is in the same direction as nonzero interaction terms, 
we cannot model differential mode waves 
with radial excitation propagating at the speed of light.  If the photon quantization conditions are  
satisfied, such motion could still occur, but a model would require knowledge of 
the quantization currents which we currently do not understand.

\ \ \ \ \ {\it b.  Model of the Photon---Helical Polarization}
\label{ssssec-helicalphoton}

    To model the photon we use solutions of the type (\ref{eq-circularline}) with sinusoidal functions $E$ and $F$, i.e. one Fourier
component, where $d_0$ is assumed constant.

\begin{equation}
{\bf y}_1\ =\ {\hat z}z\ -\ {\hat x} d_0\ sink(\tau-z)\ +\ {\hat y}d_0\ cosk(\tau-z)\ =\ {\hat z}z\ +d_0 {\hat y}^\ast
\end{equation}

\begin{equation}
{\bf y}_2\ =\ {\hat z}z\ +\ {\hat x} d_0\ sink(\tau-z)\ -\ {\hat y}d_0\ cosk(\tau-z)\ =\ {\hat z}z\ -d_0 {\hat y}^\ast
\end{equation}

\noindent
Here ${\hat y}^\ast$ is the direction in the rotating frame (which is the rest frame of {\it both} lines along their entire
length),
which is initially vertical (${\hat y}$ direction) in the lab, $k = {{2\pi} / \lambda} = {\omega / c}$ and the wave
has a frequency $\nu = {\omega / {2\pi}}$.  With this choice we can achieve lack of radial excitation since

\begin{equation}
{\bf y}_1\ -\ {\bf y}_2\ =\ 2d_0\ {\hat y}^\ast
\end{equation}

\begin{figure}
    \vbox to 2.6in{
\includegraphics{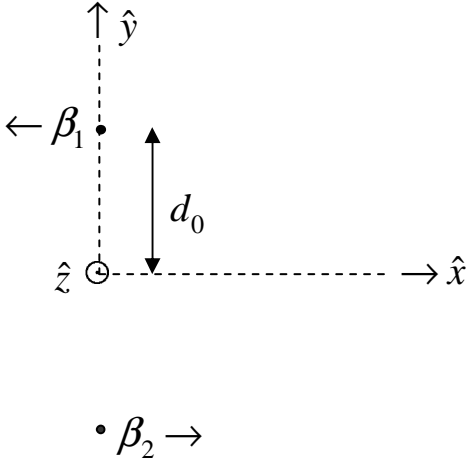}  }
    \caption{\small Geometry of a differential helical wave, model of the photon, at $z=0$, $\tau=0$.}
\label{fig-helicalphoton}
\end{figure}

\noindent
(see Figure~\ref{fig-helicalphoton}).  The whole system rotates with positive helicity.  The velocities are given by

\begin{widetext}
\begin{equation}
{\bm \beta}_1 \ =\ { {\partial {\bf y}_1} \over {\partial \tau} }\ 
=\ -\ {\hat x} kd_0\ cosk(\tau-z)\ -\ {\hat y}kd_0\ sink(\tau-z)\ = -\ kd_0\ {\hat x}^\ast 
\label{eq-beta1}
\end{equation}

\begin{equation}
{\bm\beta}_2 \ =\ { {\partial {\bf y}_2} \over {\partial \tau} }\
=\ +\ {\hat x} kd_0\ cosk(\tau-z)\ +\ {\hat y}kd_0\ sink(\tau-z)\ = +\ kd_0\ {\hat x}^\ast
\label{eq-beta2}
\end{equation}

\noindent
and the magnetic field directions are given by the following expressions:

\begin{equation}
{\hat \ell}_1\ =\ +\ { { {\partial {\bf y}_1} \over {\partial z} } \over { \vert {d\ell \over dz} \vert } }\ =\ 
{ { {+\hat z} + {\hat x} kd_0 cosk(\tau-z) + {\hat y}kd_0 sink(\tau-z) } \over {\sqrt{1\ +\ (kd_0)^2} }} \ 
=\ { {+{\hat z}+kd_0 {\hat x}^\ast} \over {\sqrt{1\ +\ (kd_0)^2} } }
\label{eq-lhat1}
\end{equation}

\begin{equation}
{\hat \ell}_2\ =\ -\ { { {\partial {\bf y}_2} \over {\partial z} } \over { \vert {d\ell \over dz} \vert } }\ =\
{ { {-\hat z} + {\hat x} kd_0 cosk(\tau-z) + {\hat y}kd_0 sink(\tau-z) } \over {\sqrt{1\ +\ (kd_0)^2} }} \ 
=\ { {-{\hat z}+kd_0 {\hat x}^\ast} \over {\sqrt{1\ +\ (kd_0)^2} } }
\label{eq-lhat2} 
\end{equation}

    Since both lines are at rest in the rotating frame, we calculate the currents in this frame directly.
We use (\ref{eq-eline*}-\ref{eq-bline*}) for the electromagnetic fields of both lines.  
Then from Appendix~\ref{ssec-generalcurl} and \ref{ssec-helicalmodegeom} we find the following expressions:
    
\begin{equation}
    {\bf \nabla}\ \times\ {{\hat \ell}_1}^\ast\ =\ {1 \over {\sqrt{1+(kd_0)^2}}} {({\hat x}^\ast-{\hat z}kd_0) \over {R^\ast_1}}\ 
=\ { {(+{\hat y}^\ast) \times {\hat \ell}^\ast_1 } \over {R^\ast_1} }
\end{equation}

\begin{equation}
    {\bf \nabla}\ \times\ {{\hat \ell}_2}^\ast\ =\ {1 \over {\sqrt{1+(kd_0)^2}}} {({\hat x}^\ast+{\hat z}kd_0) \over {R^\ast_2}}\ 
=\ { {(-{\hat y}^\ast) \times {\hat \ell}^\ast_2 } \over {R^\ast_2} }
\end{equation}

\begin{equation}
    { 1 \over {R^\ast_0} }\ =\ k^2 d_0
\end{equation}

\noindent
where ${R^\ast_0}$ is the instantaneous radius of curvature of the centerline, which is constant for both lines.  Hence we find:

\begin{equation}
{J^\ast_1}^\mu\ =\ 
{{g_0 c} \over {2\pi a^3}} \bigg [ K_1({{\rho}_1}^\ast/a)\  \Bigl (0,{{\hat \ell}_1}^\ast \times {{\hat \rho}_1}^\ast \Bigr )^\mu
\ +\ {a \over {R^\ast_1}} K_0({{\rho}_1}^\ast/a)\  \Bigl (0,{(+{\hat y}^\ast) \times {\hat \ell}^\ast_1 } \Bigr )^\mu \bigg ]
\end{equation}

\begin{equation}
{J^\ast_2}^\mu\ =\
{{g_0 c} \over {2\pi a^3}} \bigg [ K_1({{\rho}_2}^\ast/a)\  \Bigl (0,{{\hat \ell}_2}^\ast \times {{\hat \rho}_2}^\ast \Bigr )^\mu
\ +\ {a \over {R^\ast_2}} K_0({{\rho}_2}^\ast/a)\  \Bigl (0,{(-{\hat y}^\ast) \times {\hat \ell}^\ast_2 } \Bigr )^\mu \bigg ]
\end{equation}

\noindent
These currents satisfy the equation of continuity (\ref{eq-continuity}) exactly, but we still find a small inconsistency if we try to use
(\ref{eq-generalsol}) to recalculate the electromagnetic field (see Appendix~\ref{ssec-calcEMfield}).

    We find the interaction terms in helical mode

\begin{equation}
{\vartheta^\ast_\mu}^{110}\ +\ {\vartheta^\ast_\mu}^{120}\ 
= \ { {g_0^2} \over {\pi c a^4} }\  { { (0,+{\hat y}^\ast)_{\mu} } \over {\vert \ell^\ast_z \vert} }
  \int d^2 \rho^\ast_1 \biggl [ { {K_0^2(\rho^\ast_1/a)} \over {R^\ast_1} }\ 
+\  { {K_0(\rho^\ast_1/a) K_0(\rho^\ast_2/a) } \over {R^\ast_2} } 
  \Bigl ( { {1-(kd_0)^2} \over  {1+(kd_0)^2} } \Bigr ) \biggr ]
\end{equation}

\noindent
where we take into account that the radii of curvature for line 1 and line 2 point in opposite directions.
(Hence $R^\ast_1$ and $R^\ast_2$  are different functions of position though the magnitude of $R^\ast_0$ 
is the same on both centerlines.)  We make a correction for the fact that $K_0(\rho^\ast_1/a)$ and $K_0(\rho^\ast_2/a)$ 
have different axes by changing $\int d^2 \rho^\ast_1$ into $\int d^2 \rho^\ast$ where $\rho^\ast$  is the cylindrical radius
perpendicular to the ${\hat z}$-axis, and symmetric for both Bessel functions.  We obtain

\begin{equation}
{\vartheta^\ast_\mu}^{110}\ +\ {\vartheta^\ast_\mu}^{120}\ 
= \ { {2T} \over {c R^\ast_0} }\  { { (0,+{\hat y}^\ast)_{\mu} } \over {\vert \ell^\ast_z \vert} }\ 
\biggl \{ 1\ +\ u_0 K_1(u_0) \Bigl ( { {1-(kd_0)^2} \over  {\sqrt{1+(kd_0)^2} } } \Bigr ) \biggr \}
\label{eq-tiltBessel}
\end{equation}

\noindent
where

\begin{equation}
    u_0\ =\ { {2 d_0} \over a }
\end{equation}

\noindent
and in a similar fashion

\begin{equation}
{\vartheta^\ast_\mu}^{121}\ 
= \ { {4T} \over {c a} }\  (0,+{\hat y}^\ast)_\mu \ \Bigl ( 1-(kd_0)^2 \Bigr )\ 
\biggl \{ { {u_0} \over 2} K_2(u_0)\ -\ K_1(u_0) \biggr \} \ \ \ .
\end{equation}

\noindent
So the sum of the interaction terms (including the negative signs for differential mode)

\begin{displaymath}
 \sum  {\vartheta^\ast_\mu}^{ijk}\ 
= \ { {8T} \over {c a} }\  (0,+{\hat y}^\ast)_\mu \ \Bigl ( 1-(kd_0)^2 \Bigr )\
\biggl \{ { {u_0} \over 2} K_2(u_0)\ -\ K_1(u_0) \biggr \} 
\end{displaymath}

\begin{equation}
+ \ { {4T} \over {c R^\ast_0} }\  { (0,+{\hat y}^\ast)_{\mu} } \ 
\biggl \{ {\sqrt{1+(kd_0)^2} } +\ u_0 K_1(u_0) \Bigl (  {1-(kd_0)^2} \Bigr ) \biggr \}
\label{eq-inttermsdiff}
\end{equation}
\end{widetext}

\noindent
is in the radial $(0,-{\hat y}^\ast)^\mu$ direction and attractive for the helical differential mode of the photon.

    Since the motion of the helical photon is orthogonal to the radial direction, it can propagate 
freely at the velocity of light.  The quantization currents must play a particularly simple role.
The left hand side of (\ref{eq-eqofmdiff}) is zero in the radial direction, 
but the right hand side of (\ref{eq-eqofmdiff}), given by (\ref{eq-inttermsdiff}),
 is not zero in the radial direction.
The quantization currents (which have not been considered so far) 
must freeze the radial separation of the two lines of the photon at
$2d_0$.  This allows us to isolate the effects of the quantization currents, which we do not fully 
understand, and model the helical photon.

\subsection{Quantization of the Photon}
\label{ssec-quantphoton}

\begin{figure}
    \vbox to 1.5in{
\includegraphics{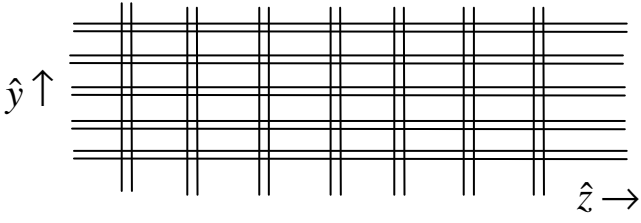}  }
    \caption{\small The wave function of an electron can be envisioned as a mesh of line pairs with each line pair oscillating
    coherently with the others.}
\label{fig-wfelectron}
\end{figure}

    In Figure~\ref{fig-wfelectron} we consider the wave function of an electron with 4-momentum $p^\mu_e=(E_e/c,{\bf p}_e)^\mu$.
The wave function is a mesh of line pairs, each of which (from quantum mechanics, as 
we will see in Section~\ref{sec-quantmech}) oscillates at frequency $\nu=E_e/h$ and
with a wavelength $\lambda=h/p_e$ if ${\bf p}_e$ points in the direction of the line pair and 
$\lambda=h/{\vert {\bf p}_e \cdot {\hat p} \vert }$ in the more general case, where ${\hat p}$ is the direction of the line pair.  
We use the absolute value sign to indicate that no positive sense for ${\hat p}$ is defined.  Hence $p^\mu_e=p^\mu_s$ is
the phase field of this wave function.

    We have pictured the mesh as real and drawn it with regular spacing for clarity.  
In general, of couse, a wave function is complex. In reality neither the spacing nor the directions will be
regular.  Closed superconducting circuits will be formed with many shapes (triangles, rectangles, pentagons, ...) and sizes, 
and in many different orientations.

    In each circuit

\begin{equation}
\oint {p^\mu_s} dx_{\mu}\ =\ \oint {p^\mu_e} dx_{\mu}\ =  \ 0
\end{equation}

\noindent
since $p^\mu_e$ is the phase field for this (single-valued) wave function.  
(We are using a plane-wave electron for simplicity.  The Fourier
composition of wave functions will be considered in Section~\ref{sec-quantmech}.)

    We see that the phase field of an electron wave is more than just the electron's 4-momentum.  It represents the entire 
wave function of a plane-wave electron.  (It is probably useful to think of it as a differential form\cite{MTWform} .)

\begin{figure}
    \vbox to 1.5in{
\includegraphics{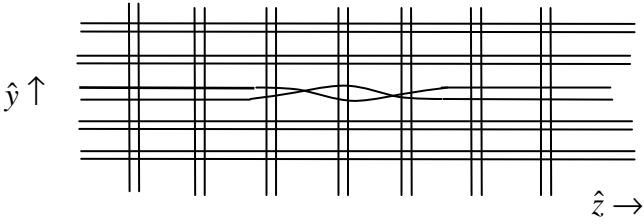}  }
    \caption{\small A photon is propagating on one of the line pairs.  View at fixed time.}
\label{fig-photonfixedt}
\end{figure}

    But now consider Figure~\ref{fig-photonfixedt}.  A photon is propagating on one of the line pairs, hence on only one leg of 
one or more closed circuits.

For {\it this} line pair, the number of oscillations per unit length has increased and so the phase field is

\begin{equation}
   p^\mu_s\ =p^\mu_e\ +\ p^\mu_\gamma
\end{equation}

\noindent
the sum of the 4-momenta of the electron and the photon.  Hence, since this photon is propagating along only 
one of the line pairs, we must have (from (\ref{eq-multisup}))

\begin{equation}
\oint ({p^\mu_e+p^\mu_\gamma}) dx_{\mu}\ =\ \oint {p^\mu_\gamma} dx_{\mu}\ =  \ nh
\label{eq-pepg}
\end{equation}

\noindent
so that the photon's momentum and energy are quantized.  For the electron's wave function to be single valued in the presence of
the photon, (\ref{eq-multisup}) must hold.  Thus (\ref{eq-multisup}) must hold for {\it all} quanta, not just flux quanta.

    Consider a circuit at {\it fixed time} (Figure~\ref{fig-photonfixedt}).  With ${\bf p}_\gamma$ (and hence the direction of the relevant line pair) 
defined to be in the ${\hat z}$-direction, equation (\ref{eq-pepg}) becomes

\begin{equation}
-\ \oint {\bf p}_\gamma \cdot d{\bf x}\ =\ -p_\gamma\ \lambda\ =\ (-1)h
\end{equation}

\noindent 
so that

\begin{equation}
 {p}_\gamma =\ { h \over \lambda }  \ \ \ .
\end{equation}

\noindent
We presume that the length of one photon is the wavelength $\lambda$, since it is the smallest length satisfying periodic
boundary conditions---bringing the two lines back to their quiescent condition.

\begin{figure}
    \vbox to 1.5in{
\includegraphics{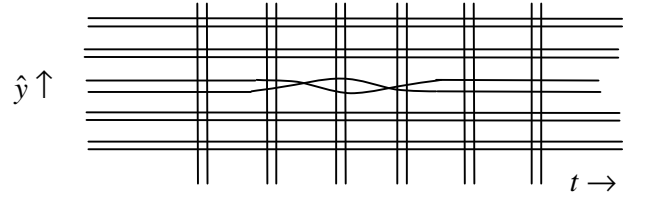}  }
    \caption{\small A photon is propagating on one of the line pairs.  View at fixed $z$.}
\label{fig-photonfixedz}
\end{figure}

    Consider a circuit at {\it fixed z} (Figure~\ref{fig-photonfixedz}, with ${\bf p}_\gamma$ still in the ${\hat z}$-direction)

\begin{equation}
    \oint {E_\gamma} dt\ = E_\gamma T\ =\ (1) h\
\end{equation}

\noindent
where we presume the photon passes in a time interval of one period $T=1/\nu$ (the smallest time duration satisfying
periodic boundary conditions).  So

\begin{equation}
    E_\gamma \ =\ h\nu\
\end{equation}

\noindent
and the photon must be {\it quantized} if it is to propagate on a line pair through the superconducting ether.   

    To consider whether or not a photon can exist we must integrate over the relevant line pair for at least a wavelength
$\lambda$ or a period of time $T$ and close the integral on nearby line pairs without excitation.  If the quantization 
condition (\ref{eq-pepg}) is satisfied, apparently the constant radial separation of the two lines of the helical photon 
is maintained at $2d_0$.  This explains how a sine wave can start and stop instantly.  The two lines of the photon
are in general {\it not isolated} if the quantization condition (\ref{eq-pepg}) is satisfied.  The situation is similar to 
Figure~\ref{fig-fluxghalf}, where currents in the ether maintain a flux quantum.

    We observe from equations (\ref{eq-pepg}) and (\ref{eq-multisup}) that the quantization of the photon 
really does not depend on the photon being an
excitation of any particular electron's wave function.  It is a property of the mesh of superconducting lines that comprises the
ether---and forms the wave functions of all electrons.  
 
\subsection{The Electromagnetic Field of the Helical Photon}
\label{ssec-EMfieldhelicalphoton}

    Even though we will show that the transverse motion of the helical photon is nonrelativistic, 
we attempt to make no approximations
(beyond the large radius of curvature approximation, which is well satisfied), because we will obtain precise results.
We perform separate local Lorentz transforms (using (\ref{eq-beta1}) or (\ref{eq-beta2}) as appropriate) 
on the electromagnetic field (\ref{eq-eline*}-\ref{eq-bline*}) in
the rotating frame to find the electromagnetic fields of line 1 and line 2 in the lab:

\begin{equation}
{\bf E}_1 = -\gamma {\bm \beta}_1 \times {\bf B}^\ast (1)
\ =\ \gamma \vert \ell_z \vert kd_0 B^\ast(1) (-{\hat y}^\ast)
\end{equation}

\begin{equation}
{\bf B}_1\ =\ \gamma \vert \ell_z \vert B^\ast (1) (+{\hat z}+kd_0 {\hat x}^\ast)
\end{equation}

\begin{equation}
{\bf E}_2  =  -\gamma {\bm \beta}_2 \times {\bf B}^\ast (2)
\ =\ \gamma \vert \ell_z \vert kd_0 B^\ast(2) (-{\hat y}^\ast)
\end{equation}

\begin{equation}
{\bf B}_2\ =\ \gamma \vert \ell_z \vert B^\ast (2) (-{\hat z}+kd_0 {\hat x}^\ast)
\end{equation}

\begin{equation}
\gamma\ =\ { 1 \over \sqrt{1-(kd_0)^2} }\ \ \ \ \ \ \ \ \ \ \vert \ell_z \vert\ =\ { 1 \over \sqrt{1+(kd_0)^2} }
\end{equation}

    The velocity of the entire cross section of each line is the velocity of its centerline in accord with Section~\ref{ssec-causality}.
We note that the transverse components of {\bf E} and {\bf B} are equal for both lines 
and the transverse fields of the two lines add---but
the net field in the ${\hat z}$-direction is zero.

\subsection{The Energy of the Helical Photon}
\label{ssec-energyphoton}
    The Poynting vector of the helical photon, expressing the flow of energy per unit area per unit time, is given by

\begin{equation}
{\bf S}\ =\ {c \over {4\pi}}  {\bf E \times B}
\end{equation}

\begin{widetext}
\begin{equation}
{\bf S}\ =\ {c \over {4\pi}} \gamma^2 {\vert \ell_z \vert }^2 kd_0 \biggl \{
 {\hat z} kd_0 \Bigl [B^{\ast}(1)^2 + 2 B^{\ast}(1) B^{\ast}(2) + B^{\ast}(2)^2 \Bigr ]
- {\hat x}^\ast \Bigl [B^{\ast}(1)^2 - B^{\ast}(2)^2 \Bigr ] \biggr \} 
\end{equation}
\end{widetext}

\noindent
which shows that some energy is directed along the ${\hat z}$-axis and some energy circulates about the ${\hat z}$-axis.
At fixed $z$, the energy flow per unit time is given by

\begin{equation}
{\hat z} { {dE} \over {dt} }\ =\ 
\int d^2 \rho \ {\bf S}\ =\ {\hat z}\  2 c \gamma \vert \ell_z \vert (kd_0)^2\  { {g_0^2} \over a^2 }\  (1 + u_0 K_1(u_0) )
\end{equation}

\noindent
with

\begin{equation}
u_0\ =\ { {2d_0} \over a }
\end{equation}

\noindent
and is totally in the ${\hat z}$-direction.  
(As for (\ref{eq-tiltBessel}) we take into account the tilt of both Bessel functions with respect to the ${\hat z}$-axis.)
We wish to find the energy in one photon, $E_\gamma$.
It should be given by the energy passing a fixed $z$ in one period.

\begin{equation}
E_\gamma\ =\ \int^T_0 dt { {dE} \over {dt} } 
= 2 c T \gamma \vert \ell_z \vert \ { {4\pi^2} \over \lambda^2 }\ \Bigr ( { {d_0} \over a} \Bigl)^2\  {g_0^2}\  (1 + u_0 K_1(u_0) )
\end{equation}

\begin{equation}
E_\gamma\ =\  
{ {\gamma \vert \ell_z \vert} \over \lambda } \ \Bigr ( { {d_0} \over a} \Bigl)^2\  
{ {h^2 c^2} \over {2 e_0^2}}\ (1 + u_0 K_1(u_0) )
\ \ \ .
\label{eq-egwoverlap}
\end{equation}

    If we want to compare with measured photon energies we must make a {\it binding energy correction}, because when the line pair
transfers the photon energy to an atom, the photon must provide the energy necessary to separate the lines,
because one line will go to an electron and the other to the nucleus.  So we see that the
overlap term containing $u_0 K_1(u_0)$ clearly does not belong in the energy transferred to an atom.  
We assert that the corrected energy $E_{\gamma-corr}$ transferred to an atom is just (\ref{eq-egwoverlap}) 
without the overlap term, because 
it gives the sum of the electromagnetic energies transferred by each line separately.
We presume that in the process of
the lines separating, angular momentum about the common mode centerline is conserved, because the force between the lines is
central---between the centers.  So we can check our method by calculating the angular momentum about the common mode center---i.e.
the spin of the photon.   This binding energy correction is approximately a factor of two decrease in the energy.

    Using similar reasoning we can also neglect the effects of the photon quantization currents on the energy of the traveling
photon because this energy is not transferred to the absorbing atom.  Quantization current effects are also part of the photon
binding energy.

    Using the above correction we find

\begin{equation}
E_{\gamma-corr}\ =\
{ {\gamma \vert \ell_z \vert} \over \lambda } \ \Bigr ( { {d_0} \over a} \Bigl)^2\  { {h^2 c^2} \over {2 e_0^2}}\ =\ h\nu\ =\ 
{ {hc} \over \lambda }
\ \ \ .
\end{equation}

\noindent
so

\begin{equation}
{\gamma \vert \ell_z \vert}\ \Bigr ( { {d_0} \over a} \Bigl)^2\ 
=\ { {hc} \over {8 \pi^2 g_0^2} }\ =\ { 1 \over \pi } { {e_0^2} \over {\hbar c} } \ \ \ .
\end{equation}

\noindent
We will show that when the large radius of curvature approximation is satisfied $\gamma \vert \ell_z \vert$ is 
very close to $1$.
{\it {Hence the quantization of the helical photon implies that the amplitude of the photon wave is an absolute constant!} }

    If we assume that $\varepsilon=1$ near the Earth (as we will infer from Section~\ref{ssec-parameters}), we have

\begin{equation}
    \Bigr ( { {d_0} \over a} \Bigl)^2\ =\ { 1 \over \pi } { {e^2} \over {\hbar c} }\ =\ 2.32 \times 10^{-3}\ \ \ \ \ \ \ \ \ \ 
   { {d_0} \over a}\ =\ 0.0482 \ \ \ .
\end{equation}
Since the line radius $a$ would give a finite size to the electron, from Section~\ref{sec-introduction} we have $a < 10^{-18}m$.
So for a green photon ($\lambda\ =\ 500 nm$) we have

\begin{equation}
kd_0\ <\ 6.0 \times 10^{-13}
\end{equation}

\begin{equation}
(kd_0)^2 <\ 3.7 \times 10^{-25}\ \ \ .
\end{equation}

\noindent
In addition we find

\begin{equation}
\gamma \vert \ell_z \vert\ =\ { 1 \over \sqrt{1-(kd_0)^4} }\ \simeq\ 1\ +\ {1 \over 2} (kd_0)^4
\end{equation}

\noindent
so

\begin{equation}
\gamma \vert \ell_z \vert\\ -\ 1\ <\ 6.7 \times 10^{-50} \ \ \ .
\end{equation}

\noindent
For helical geometry the large radius of curvature approximation

\begin{equation}
{ a \over {R^\ast_0} }\ =\ k^2 d_0 a\ =\ 0.303 \Bigl ( {a \over \lambda} \Bigr )^2\ < \  1.2\times 0^{-24}\ \ll\ 1
\label{eq-largeroc}
\end{equation}

\noindent
is well justified, for a visible photon.

    In Figures~\ref{fig-currentsfortogether} and \ref{fig-currentsforapart}, 
current distributions are shown with which the quantization condition could be maintained.  In Figure~\ref{fig-currentsfortogether}
currents running on the outside surfaces of the line pairs tend to decrease the radial separation of the central pair,
and in Figure~\ref{fig-currentsforapart} currents are shown which tend to separate the central pair.  
The dominant contributions are likely to come from
the currents running on the outer surfaces of the central pair.  Currents in the same direction attract.  Currents in the opposite
direction repel.  Hence in their simplest configuration, the photon quantization currents could form a loop around the outer
surfaces of the central line pair for the length of the photon (one wavelength).

\subsection{The Spin of the Helical Photon}
\label{ssec-spinphoton}
    The angular momentum about a given point on the common mode centerline of the pair is given by

\begin{equation}
{\bf L}_{em} \ =\ \int d^3x\  {\bf x} \times \Bigl ( { 1 \over {4\pi c} } {\bf E} \times {\bf B} \Bigl )\ =\ 
\int d^3x\  {\bf x} \times { {\bf S} \over {c^2} }
\end{equation}

\noindent
so at fixed $z$ the angular momentum per unit length is

\begin{equation}
{ d {\bf L}_{em} \over {dz} }\ =\ \int d^2\rho \ {\bf x} \times { {\bf S} \over {c^2} } \ \ \ .
\label{eq-dLEMdz}
\end{equation}

\noindent
Perhaps the easiest was to evaluate (\ref{eq-dLEMdz}) is to realize that 
in taking moments of momentum, the entire momentum of each line
can be considered to act at its centerline (center of momentum).  So, for example, for a term involving $B^\ast(1)^2$ we have

\begin{displaymath}
\int d^2\rho\  {\bf x} \times { 1 \over {4\pi c} } \gamma^2 {\vert\ell_z\vert}^2 kd_0 (-{\hat x}^\ast B^\ast(1)^2 )
\end{displaymath}
\begin{equation}
=\ { 1 \over {4\pi c} }\ \gamma^2 {\vert\ell_z\vert}^2 kd_0 \biggl [ \int d^2 \rho\ {\bf x} B^\ast(1)^2 \biggr ] 
\times (-{\hat x}^\ast )
\end{equation}

\begin{equation}
=\ { 1 \over {4\pi c} }\ \gamma {\vert\ell_z\vert} kd_0 \biggl [ \int d^2 \rho_1 \ {\bf x} B^\ast(1)^2 \biggr ]
\times (-{\hat x}^\ast )
\end{equation}

\begin{equation}
=\ { 1 \over {4\pi c} }\ \gamma {\vert\ell_z\vert} kd_0 \biggl [d_0 {\hat y}^\ast  \int d^2 \rho_1 \ B^\ast(1)^2 \biggr ]
\times (-{\hat x}^\ast )
\end{equation}

\begin{equation}
=\ { g_0^2 \over {c} }\ \gamma {\vert\ell_z\vert} k \Bigl ( {{d_0} \over a} \Bigr )^2 {\hat z} \ \ \ .
\end{equation}

\noindent
Therefore the spin of the photon is given by the angular momentum in one wavelength:

\begin{equation}
{ d {\bf L}_{em} \over {dz} }\ =\ 2{ {g_0^2} \over {c} }\ \gamma {\vert\ell_z\vert} k \Bigl ( {{d_0} \over a} \Bigr )^2 {\hat z}
\end{equation}

\begin{equation}
{\bf S}pin_\gamma\ =\ { d {\bf L}_{em} \over {dz} }\lambda\ =\ { {4\pi g_0^2} \over c }\ { {hc} \over {8\pi^2 g_0^2} } {\hat z}\ 
=\ \hbar {\hat z} 
\end{equation}

\noindent
Assuming the binding energy correction is valid, this result is exact
except for the the large radius of curvature approximation (\ref{eq-largeroc}).

\subsection{Generality of the Quantization Conditions}
\label{ssec-generalityquant}

\begin{figure}
    \vbox to 1.5in{
\includegraphics{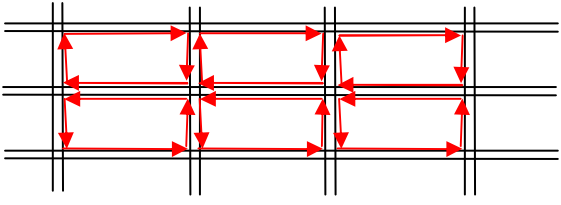}  }
    \caption{\small (Color online.)  A current distribution which pushes the central line pair together.  All currents flow on the 
    outer surfaces of the line pairs.}
\label{fig-currentsfortogether}
\end{figure}

\begin{figure}
    \vbox to 1.5in{
\includegraphics{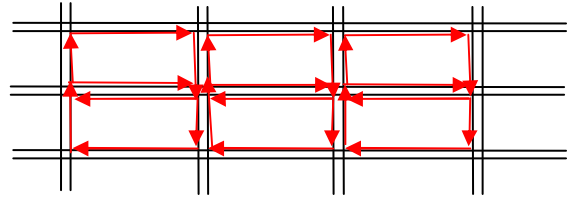}  }
    \caption{\small (Color online.)  A current distribution which pulls the central line pair apart.  All currents flow on the outer surfaces of
    the line pairs.}
\label{fig-currentsforapart}
\end{figure}

    Figures \ref{fig-wfelectron}, \ref{fig-photonfixedt}, and \ref{fig-photonfixedz} are simplifications.  
We will see (Section~\ref{ssec-motionbpp}) that the typical spacing of line pairs is less than
$10^{-13}m$, whereas the wavelength of a green photon is $5\times10^{-7}m$.  
Thus the quantization of the photon involves many 
closed circuits.  From Figures \ref{fig-currentsfortogether} and \ref{fig-currentsforapart}, 
the currents which maintain the quantization of the helical photon, 
i.e. the constancy of the photon
amplitude, can be complex and, as with Figure~\ref{fig-fluxghalf}, we do not have a detailed 
understanding, beyond the quantization condition itself, of
what makes the currents flow as they do.  But apparently such flow is required for single-valued wave functions, and so it happens.
We have made progress with our classical picture but are now meeting difficulty in understanding the quantization condition itself.

    An atom generates each photon on the relevant line pair, which acts as a
transmission line.  Once the photon exists and satisfies the quantization condition (\ref{eq-pepg}), apparently it is maintained by the
quantization currents as it transits the universe until it is absorbed by another atom, eventually.  Apparently these photon
quantization currents persist forever, like currents in a superconducting loop, which is apparently what they are.

    We see from the fact that (\ref{eq-pepg}) and (\ref{eq-multisup}) are identical,
that this quantization condition is a completely general result applying to the phase field of any particle traveling in the 
superconducting ether.  The ether is a multiply connected superconductor.
In particular this quantization condition must hold for any excitation of a line pair.  
It is {\it not} restricted to particles traveling at the speed of light, and must hold for matter waves and gravitational waves, 
as well as electromagnetic waves.  We can now identify a matter wave as a gravitational (i.e. common mode)
wave which is stimulated by a particle, 
so that it does not travel at the velocity of light, but rather travels with the particle.  It is the particle's wave function.

\subsection{The Tension of a Line Pair}
\label{ssec-tensionpair}

    In the case that there are no excitations of the line pairs under consideration we have

\begin{equation}
     \oint {p^\mu_s} dx_{\mu}\ =  \ 0\ =\ -\ \oint {\bf p}_s \cdot d{\bf x}
\end{equation}

\noindent
for any closed loop at a fixed time.  If there were a change in the radial separation of any line pair in the circuit,
this change would necessarily be quantized, contrary to our assumption.  Hence the radial separation in quiescent 
common mode and the tension of a line pair

\begin{equation}
   \tau_0\ =\ {{g_0^2}\over{a^2}}\ \Bigl[ 1-{u_q} K_1(u_q) \Bigr]\ +\ \tau_{qc}
\label{eq-tensiontau0}
\end{equation}

\noindent
must be constant {\it for the whole universe} at a given time.  
Here $\tau_{qc}$ is the tension (energy per longitudinal length) associated with the quantization currents for this line pair. 
It is convenient to think of the quantization currents 
in this case as flowing in opposite directions on the outer surfaces of each pair.  These currents are likely to be the dominant
quantization currents.  We assume that they also persist forever.  The radial separation, the line-pair tension and the
quantization currents, however, must change with time as the universe expands, 
reflecting the conservation of energy.  Apparently this line pair tension and radial
separation is a measure of world time\cite{Peeblesworldtime}.

\section{The Electron}

\subsection{Dirac's Pole Current}
\label{ssec-Diracpolecurrent}
    We use Dirac's pole current formalism\cite{Dirac2} to find the electromagnetic field of the poles 
at the ends of a line.  With our line of finite size the dual field of the line (\ref{eq-filEMdual}) becomes 

\begin{widetext}
\begin{equation}
{\tilde F}^{\alpha \beta}(x)\ =\ -4\pi g_0 \ \int d\tau d\sigma \Bigl(
{ {\partial y^\alpha} \over {\partial \tau} } { {\partial y^\beta} \over {\partial \sigma} }\ -\
{ {\partial y^\beta} \over {\partial \tau} } { {\partial y^\alpha} \over {\partial \sigma} } \Bigr)\ {\Delta}^4(x-y)
\label{eq-lineEMdual}
\end{equation}

\noindent
with from (\ref{eq-deltatrans})

\begin{equation}
{\Delta}^4(x-y)\ =\ \delta(\tau^\ast_x - \tau^\ast_y) \delta(\ell^\ast_x - \ell^\ast_y)\ 
{1 \over {2\pi a^2} } K_0({\rho^\ast}/ a ) \ \ \ .                                                         
\end{equation}

\noindent
The pole current is defined by the extension of Maxwell's equations if magnetic poles were to really exist:

\begin{equation}
{ {4\pi} \over c }K^\beta\ =\ { {\partial {\tilde F}^{\alpha \beta} } \over {\partial x^\alpha} }\ =\ 
-4\pi g_0 \ \int d\tau d\sigma \Bigl(
{ {\partial y^\alpha} \over {\partial \tau} } { {\partial y^\beta} \over {\partial \sigma} }\ -\
{ {\partial y^\beta} \over {\partial \tau} } { {\partial y^\alpha} \over {\partial \sigma} } \Bigr)\ 
{ {\partial {\Delta}^4(x-y)} \over {\partial x^\alpha} }
\end{equation}

\begin{equation}
=\ +4\pi g_0 \ \int d\tau d\sigma \Bigl(
{ {\partial y^\alpha} \over {\partial \tau} } { {\partial y^\beta} \over {\partial \sigma} }\ -\
{ {\partial y^\beta} \over {\partial \tau} } { {\partial y^\alpha} \over {\partial \sigma} } \Bigr)\
{ {\partial {\Delta}^4(x-y)} \over {\partial y^\alpha} }
\end{equation}

\begin{equation}
=\ +4\pi g_0 \ \int d\tau d\sigma \Bigl(
{ {\partial y^\beta} \over {\partial \sigma} }{ {\partial {\Delta}^4(x-y) } \over {\partial \tau} }\ -\
{ {\partial y^\beta} \over {\partial \tau} } { {\partial {\Delta}^4(x-y) } \over {\partial \sigma} } \Bigr)\
\end{equation}

    After an integration by parts we obtain the pole current

\begin{equation}
K^\beta\ =\ g_0 c\ \Biggl\lbrack \int d\sigma 
{ {\partial y^\beta} \over {\partial \sigma} } {\Delta}^4(x-y) {{\Biggr\vert}_{\tau_i}^{\tau_f}}
-\int d\tau 
{ {\partial y^\beta} \over {\partial \tau} } {\Delta}^4(x-y) {{\Biggr\vert}_{\sigma_i}^{\sigma_f}}\ \ \ \Biggr\rbrack \ \ \ .
\end{equation}

\noindent
If we presume that $\tau_i$ and $\tau_f$ are sufficiently distant that they are of no concern, the pole current becomes

\begin{equation}
K^\beta\ =\ -g_0 c\ \int d\tau { {\partial y^\beta} \over {\partial \tau} }(\sigma_f) {\Delta}^4(x-y)
\ +\ g_0 c\ \int d\tau { {\partial y^\beta} \over {\partial \tau} }(\sigma_i) {\Delta}^4(x-y)
\end{equation}

\begin{equation}
K^\beta\ =\ -g_0 c\ { {\partial y^\beta} \over {\partial \tau} }(\sigma_f) \delta(\ell^\ast_x - \ell^\ast_{yf})
{ 1 \over {2\pi a^2} } K_0(\rho^\ast_f/a)
+\ g_0 c\ { {\partial y^\beta} \over {\partial \tau} }(\sigma_i) \delta(\ell^\ast_x - \ell^\ast_{yi})
{ 1 \over {2\pi a^2} } K_0(\rho^\ast_i/a) \ \ \ .
\label{eq-dualcurrent}
\end{equation}
\end{widetext}

    We get our expectation for a pole current from each end of the line of finite size {\it except} for the signs of both poles.
The reason is that we actually cut off the integral over the line too soon.  The dual field actually extends to infinity beyond the
ends of our line.  So by cutting off the dual field at the mechanical end of the line, 
we are {\it absorbing} the dual field field at the end of the
line.  This is the reason for the incorrect signs.

    Contrary to our view Dirac, excluded the line from physical reality because he wanted the pole field to appear as a monopole,
in the space {\it external} to the line.  Therefore we take our dual field as the {\it negative} of Dirac's dual field.

    In our model Maxwell's equations are satisfied everywhere, and the pole currrent is actually $0$.

\begin{equation}
    { {\partial {\tilde F}^{\alpha \beta} } \over {\partial x^\alpha} }\ =\ 0
\end{equation}

\noindent
But we can use Dirac's pole current method to find the effective pole current outside the ends of our lines 
just by reversing the signs of the pole currents:

\begin{widetext}
\begin{equation}
K^\beta_{eff} =\ +g_0 c\ { {\partial y^\beta} \over {\partial \tau} }(\sigma_f) \delta(\ell^\ast_x - \ell^\ast_{yf})
{ 1 \over {2\pi a^2} } K_0(\rho^\ast_f/a)
-\ g_0 c\ { {\partial y^\beta} \over {\partial \tau} }(\sigma_i) \delta(\ell^\ast_x - \ell^\ast_{yi})
{ 1 \over {2\pi a^2} } K_0(\rho^\ast_i/a)
\end{equation}
\end{widetext}

\noindent
The field which appears to be absorbed in (\ref{eq-dualcurrent}), because we cut off the integral too soon, actually exits the line
and spreads out in all directions.  We assume, without justification, that there is simply a sharp, clean break 
at the end of the line---the superconducting field and the supercurrent simply stop.  The implications of this simple 
phenomenological assumption are rich and interesting.

    Using Maxwell's equations and (\ref{eq-generalsol}) we can show that in the superconducting field:

\begin{equation}
    \nabla^2 {\bf E}\ -\ { {\partial^2 {\bf E} } \over {\partial \tau^2} }\ =\ {1 \over {a^2}} {\bf E }
\end{equation}

\begin{equation}
    \nabla^2 {\bf B}\ -\ { {\partial^2 {\bf B} } \over {\partial \tau^2} }\ =\ {1 \over {a^2}} {\bf B }
\end{equation}

\noindent
Therefore the electromagnetic field at the break at the end of a line should act as a source for the electromagnetic field
beyond the line, which satisfies the homogeneous wave equation.

\subsection{The Electromagnetic Field of the Bare Electron-Pole}
\label{ssec-EMfieldbareelectronpole}
    We will see that the bare positron-pole (using positive charge for convenience) 
bounces around ({\it zitterbeweging}) so rapidly that a proper treatment of
the shape of the field of the bare positron-pole is unnecessary, as long as the total outward flux is correct.  In the limit 
$a \rightarrow 0$ we simply find

\begin{equation}
K^\beta_{eff} =\ +g_0 \ (c,{\bf v}_f)\ \delta^3({\bf x}-{\bf y}_f)
                     \ -g_0 \ (c,{\bf v}_i)\ \delta^3({\bf x}-{\bf y}_i)
\label{eq-poles} 
\end{equation}

\noindent
the current of point poles.  Therefore we can use the standard Lienard-Wiechert potentials\cite{JacksonLW} with the 
substitutions

\begin{equation}
    e \rightarrow g_0\ \ \ \ \ \ \ \ \ \ {\bf E} \rightarrow {\bf B}\ \ \ \ \ \ \ \ \ \ \ {\bf B} \rightarrow -{\bf E}
\end{equation}

\noindent
to find the electromagnetic field of the bare positron-pole

\begin{equation}
    {\bf B}\ =\ g_0 \ \biggl [{ {({\hat r}- {\bm \beta}) (1-\beta^2)} \over {\kappa^3 r^2} } \biggr ]_{ret}
+\ {g_0} \ \biggl [{ { {\hat r} \times \Bigl \{ ({\hat r}-{\bm \beta}) \times 
{\dot {\bm \beta}} \Bigr \} } \over {\kappa^3 r}}  \biggr ]_{ret}
\label{eq-bppbfield}
\end{equation}

\begin{equation}
{\bf E}\ =\ -\ \bigl [ {\hat r} \times {\bf B} \bigr ]_{ret}
\label{eq-bppefield}
\end{equation}

\noindent
with

\begin{equation}
    {\bf r}\ \equiv\ {\bf x}\ -\ {\bf y}
\end{equation}

\noindent
and where the expressions in brackets are evaluated at the retarded time

\begin{equation}
    t^\prime\ =\ t\ -\ {{r(t^\prime)} \over c}
\end{equation}

\noindent
and

\begin{equation}
   { 1 \over c} { {\partial {\bm \beta} } \over {\partial t^\prime} } \equiv {\dot {\bm \beta}}
\end{equation}

\begin{equation}
    \kappa\ \equiv \ 1\ -\ {\hat r} \cdot  {\bm \beta} (t^\prime) \ \ \ .
\end{equation}

\noindent
(The bare positron-pole will acquire it's electric charge in Section~\ref{ssec-electroncharge} and then be called the bare positron.)

    The field of the bare electron-pole is similar, coming from the negative pole in (\ref{eq-poles})
The line field is, of course, present, but 
it is quantized and shielded by a superconducting sheath, so we can often neglect its interaction (Section~\ref{ssec-intercommutation})
with the surrounding line pairs.

    In Section~\ref{ssec-neglecting}, we saw that without considering interactions, the end of a line moves 
at the speed of light transverse
to the line.  It is interesting to note that, if the bare positron-pole moves at the speed of light, it's electromagnetic
field (\ref{eq-bppbfield}-\ref{eq-bppefield}) is well behaved.  In fact, we can see that in the absence of acceleration 
(${\dot {\bm \beta}}=0$), the electromagnetic field of the bare positron-pole is actually $0$ at the speed of
light.

    In this universe, however, electromagnetic fields are always present.  We will see that in the presence of external fields the
speed of the bare positron-pole is less than the speed of light---but, unless the bare positron-pole is in a {\it very}
strong field, its speed will be close to the speed of light.

    So the natural state of the end of a line is to move near the speed of light transverse to the line.  This is in great contrast
to our concept of a point particle whose natural state is to remain at rest.  Evidently, this is because the end of a line 
carries no inertia.

\subsection{Self Interaction of the Bare Electron-Pole}
\label{ssec-selfintbareelectronpole}

    To solve the equations of motion for the bare positron-pole, we need to find the average of the line's dual field (\ref{eq-lineEMdual}) 
over the cross section of the line at the end points.  In the rest frame of the bare positron-pole we find

\begin{equation}
\bigl\langle {\tilde F}^{\ast\alpha \beta} (\sigma^\ast_f) \bigr\rangle\ =
\ -{ {g_0} \over {a^2}} { {d\sigma^\ast} \over {d\ell^\ast} } 
({\dot y}^{\ast \alpha} y^{\prime \ast \beta} - {\dot y}^{\ast \beta} y^{\prime \ast \alpha} ) \ \ \ .
\end{equation}

\noindent
Since $\tau^\ast \equiv s$ is the proper time, the vector

\begin{equation}
    {\dot y}^{\ast \alpha} \ =\ { {\partial y^{\ast \alpha} } \over {\partial s} }
\end{equation}

\noindent
is a 4-vector.  After transformation to the lab, the transformed vector is

\begin{equation}
    { {\partial y^{\alpha} } \over {\partial s} } \ 
=\  \gamma { {\partial y^{\alpha} } \over {\partial \tau} } \ 
=\ \gamma {\dot y}^\alpha
\end{equation}

\noindent
remembering our definition of $\tau$ (see (\ref{eq-taudef})).  Hence in the lab we find

\begin{equation}
\bigl\langle {\tilde F}^{\alpha \beta} (\sigma_f) \bigr\rangle\ =
\ -{ {g_0} \over {a^2}} { {d\sigma^\ast} \over {d\ell^\ast} } 
\gamma
({\dot y}^{\alpha} y^{\prime \beta} - {\dot y}^{\beta} y^{\prime \alpha} ) \ \ \ .
\end{equation}

    In order to deal with the end of a line we {\it pick} the directions defined by 
$\sigma^\ast$ and $\sigma$ to be transverse to the motion.  (So they are no longer necessarily
in the ${\hat z}$ direction.)  Then using the fact that 
${ {\partial y^{\ast \alpha} } / {\partial \sigma^\ast} }$ and
${ {\partial y^{\alpha} } / {\partial \sigma} }$  are both 4-vectors, we find

\begin{equation}
    { {d \ell^\ast} \over {d \sigma^\ast} } \ =\ 
    { {\sqrt{1-{\beta_\perp}^2} } \over {\sqrt{1-{\beta}^2} } }
\     { {d \ell} \over {d \sigma} } 
\end{equation}

\noindent
and so

\begin{equation}
\bigl\langle {\tilde F}^{\alpha \beta} (\sigma_f) \bigr\rangle\ =
\ -{ {g_0} \over {a^2}}  { {d\sigma} \over {d\ell} } \gamma_\perp
({\dot y}^{ \alpha} y^{\prime \beta} - {\dot y}^{\beta} y^{\prime \alpha} ) \ \ \ .
\end{equation}

\subsection{Motion of the Bare Electron-Pole}
\label{ssec-motionbpp}
    In the presence of interactions, the boundary conditions (\ref{eq-bcendofline}) which must be satisfied at the ends of a line become

\begin{equation}
\Biggl\lbrack{{\partial {\cal L}_{tot}} \over {\partial {y^\prime}^\mu}}\ \delta y^\mu {{\Biggr \vert}_{\sigma_i}^{\sigma_f}}\ =\ 0
\end{equation}

\noindent
and since $\delta y^\mu$ is in general not zero, we obtain

\begin{equation}
\Biggl\lbrack{{\partial {\cal L}_{m}} \over {\partial y^{\prime \mu}}}\ 
+\ { {g_0} \over c}\  {\dot y}^\alpha \bigl\langle {\tilde F}_{\alpha \mu} \bigr\rangle\ 
 {{\Biggr \vert}_{\sigma_i}^{\sigma_f}}\ =\ 0
\label{eq-motionep}
\end{equation}

\noindent
or at each end point

\begin{equation}
{{\partial {\cal L}_{m}} \over {\partial y^\prime_\mu} }\
+\ { {g_0} \over c}\  {\dot y}_\alpha \bigl\langle {\tilde F}^{\alpha \mu} \bigr\rangle_{self}\ 
+\ { {g_0} \over c}\  {\dot y}_\alpha \bigl\langle {\tilde F}^{\alpha \mu} \bigr\rangle_{external}\
=\ 0
\label{eq-motionepparts}
\end{equation}

\noindent
where ${{\partial {\cal L}_{m}} / {\partial y^\prime_\mu} }$ is given by (\ref{eq-pLdymup}).  The self interaction term is simply 
twice the mechanical term, so the equation of motion for {\it either} end point becomes

\begin{equation}
\beta_\ell\ {u^\mu_\perp}\ +\ \sqrt{1-{\beta_\perp}^2}\ (0,{\hat \ell})^\mu \ 
=\ -{ 1 \over {B_0}}\  ( {\bm \beta} \cdot {\bf B}, 
\ {\bf B} - {\bm \beta} \times {\bf E} ) 
\end{equation}

\noindent
where

\begin{equation}
    B_0\ \equiv\ { {3g_0} \over {2a^2} }
\end{equation}

\noindent
is a very strong field (3/2 times the average of the self field).  We define ${\hat e}_3$ as the direction
of the component of the electric field which is perpendicular to the line (${\hat \ell}$)

\begin{equation}
    {\bf E}\ =\ E_\ell{\hat \ell}\ +\ E_3 {\hat e}_3
\end{equation}

\noindent
and we also define

\begin{equation}
     {\hat e}_\perp\ =\ {\hat e}_3\ \times {\hat \ell} \ \ \ .
\end{equation} 

\noindent
We assume that the velocity of the bare positron-pole is perpendicular to ${\hat e}_3$ 

\begin{equation}
     {\bm \beta} \ =\ \beta_\ell {\hat \ell} \ +\ \beta_\perp {\hat e}_\perp
\end{equation}

\noindent
so the equations of motion become

\begin{equation}
time:\ \ \ \ \ \ \ \ \ \ { {\beta_\ell} \over \sqrt{1-{\beta_\perp}^2} }\ =\ -{ 1 \over {B_0}}\ 
     (\beta_\ell B_\ell\ +\ \beta_\perp B_\perp)
\label{eq-bpptmotion}
\end{equation}

\begin{equation}
{\hat \ell}: \ \ \ \ \ \ \ \ \ \ \sqrt{1-{\beta_\perp}^2}\ =\ -{ 1 \over {B_0}}\
     (B_\ell\ -\ \beta_\perp E_3)
\label{eq-bpplmotion}
\end{equation}

\begin{equation}
{\hat e}_\perp:\ \ \ \ \ \ \ \ \ \ { {\beta_\ell \beta_\perp} \over \sqrt{1-{\beta_\perp}^2} }\ =\ -{ 1 \over {B_0}}\ 
    (B_\perp\ +\ \beta_\ell E_3)
\label{eq-bpppmotion}
\end{equation}
                                                         
\begin{equation}
{\hat e}_3:\ \ \ \ \ \ \ \ \ \ \ \ \ \ 0\ =\ -{ 1 \over {B_0}}\ (B_3\ +\ \beta_\perp E_\ell)\ \ \ \ \ .
\label{eq-bpp3motion}
\end{equation}

\noindent
Usually we can neglect the averaging requirement for the external electromagnetic field, if it does not vary too rapidly
over the cross section of the line.  We use a simplified notation without averaging, but, in principle, we always have to 
take the average in the rest frame of the bare positron-pole and transform the resulting field to the lab (as we did in 
Section~\ref{ssec-selfintbareelectronpole}).

    We first consider the case where the electric field is zero.  In this case we find from (\ref{eq-bpplmotion})

\begin{equation}
\sqrt{1-{\beta_\perp}^2}\ =\ -{ B_\ell \over {B_0}} \ \ \ .
\end{equation}

\noindent
So it is obvious in this case that $B_\ell$ must be negative, i.e. the line orients both of its ends so that the external
magnetic field is opposite the magnetic field of the line, as we would expect from energy considerations.

    From (\ref{eq-bpptmotion}) and (\ref{eq-bpppmotion}) we find that $\beta_\ell$ is proportional to $B_\perp$.

\begin{equation}
    \beta_\ell\ =\ { {B_\perp ({1-{\beta_\perp}^2}) } \over { \beta_\perp B_\ell - E_3 } }
\end{equation}

\noindent
Hence we can solve both (\ref{eq-bpptmotion}) and (\ref{eq-bpppmotion}) by presuming that the ends of the line both 
align ${\hat \ell}$ opposite the magnetic field so that $B_\perp = 0$.

    Next we consider the case where the electric field and the magnetic field are perpendicular as is often the case with
electromagnetic waves and is true of the field of another moving pole (\ref{eq-bppbfield}-\ref{eq-bppefield}).  
In this case we can solve the equations of motion with the following presumptions:

\begin{quotation}
1)  At each end of the line, ${\hat \ell}$ aligns opposite the external magnetic field.

2)  The external electric field defines the direction of ${\hat e}_3$, perpendicular to ${\hat \ell}$ ($E_3 > 0$).

3)  Then $\beta_\ell = 0$ and the velocity of the bare positron-pole is in the direction ${\hat e}_\perp$ 
    (so ${\bf E} \times {\bf B}$) perpendicular to the line, and the value of $\beta_\perp$ can be found from (\ref{eq-bpplmotion}),

4)  Equation (\ref{eq-bpp3motion}) is satisfied because $B_3$ and $E_\ell$ are both zero.
\end{quotation}
So we have the solution for the motion of the bare positron-pole in the case that the external electric and magnetic 
induction fields
are perpendicular.  The bare positron-pole carries no inertia so it responds to the field conditions instantaneously, as it
moves and the external fields change.  In the general case where $E_\ell \not= 0$, the end of the line must orient itself such that
the ${\hat e}_3$ component of the Lorentz force on the moving pole is $0$.  This requires a nonzero value of $B_3$.
If ${\bf E}\cdot{\bf B}=EB sin\alpha$, it is easy to find the appropriate orientation of ${\hat \ell}$ if $\alpha$ is small.
The general case requires the solution of a transcendental equation.  (See Appendix~\ref{ssec-orientation}.)

    In all cases we use (\ref{eq-bpplmotion}) to find the velocity of the moving pole:

\begin{equation}
\beta_\perp\ =\ { {B_\ell E_3 \pm B_0 \sqrt{B^2_0 + E^2_3 - B^2_\ell} } \over {B^2_0 + E^2_3} }
\end{equation}

\noindent
and

\begin{equation}
\sqrt{1-\beta^2_\perp}\ =\ { {-B_\ell B_0 \pm E_3 \sqrt{B^2_0 + E^2_3 - B^2_\ell} } \over {B^2_0 + E^2_3} }
\label{eq-1ogamp}
\end{equation}

\noindent
where the $\pm$ sign takes the same value in both equations.  We see from (\ref{eq-1ogamp}) that 
we must have the $+$ sign because
$B_\ell$ and $E_3$ are independent, i.e. $B_\ell$ could be zero.  Hence the velocity of the bare positron-pole is given by

\begin{equation}
\beta_\perp\ =\ { {B_\ell E_3 + B_0 \sqrt{B^2_0 + E^2_3 - B^2_\ell} } \over {B^2_0 + E^2_3} }
\label{eeq-betaperp}
\end{equation}

\begin{equation}
\sqrt{1-\beta^2_\perp}\ =\ { {-B_\ell B_0 + E_3 \sqrt{B^2_0 + E^2_3 - B^2_\ell} } \over {B^2_0 + E^2_3} } \ \ \ .
\label{eq-1ogamp+}
\end{equation}

    Using the minimal value of the line-pair 
density of the ether given by (\ref{eq-dLdVnum}) we can estimate $\ell_{typ}$, the typical distance between line pairs

\begin{equation}
    \sqrt{{dn} \over {dA}}\ =\ {1 \over {\ell_{typ} } }\ \ \ .
\label{eq-ltyp}
\end{equation}

\noindent
We find
\begin{equation}
    \ell_{typ}\ =\ 1.46 \times 10^{-13} m
\end{equation}

\noindent
and hence
\begin{equation}
     \Delta t_{typ}\ =\ 4.87 \times 10^{-22} s
\end{equation}

\noindent
is the typical time between collisions of the bare positron-pole with line pairs 
at the minimal density of (\ref{eq-dLdVnum}).  During
each collision the external electromagnetic field changes radically in both magnitude and direction
and the bare positron-pole travels near the velocity of light in a new direction.  The main causes of 
the external field are

\begin{quotation}

\noindent
(1) reflection of the pole field (\ref{eq-bppbfield}-\ref{eq-bppefield}) by the superconducting line pairs 
(see (\ref{eq-pealEmac}) and (\ref{eq-memalBmac})), and

\noindent
(2) currents in closed superconducting loops, stimulated by the pole magnetic field, bringing the
    magnetic flux linking each loop to zero because of flux quantization
\end{quotation}

\noindent
Thus the motion of the bare positron-pole is {\it zitterbewegung}, jittery random-walk motion with a
very short time between collisions.  Even though the bare positron-pole is traveling near the velocity 
of light, for many purposes it can be considered at rest in a frame traveling at a much smaller velocity.

    Contrary to the general rule (Section~\ref{ssec-generalityquant}) this zitterbewegung motion is {\it not} 
quantized, because the end of a line (electron) is not part of a closed loop.

\subsection{Charge of the Electron}
\label{ssec-electroncharge}

    At this point we are forced to make an assumption which does not follow from the Principle of Least Action 
applied to the superconducting line.
In order to agree with experiment we need to essentially put the appropriate electric charges at the ends 
of each line.  We will try to do this so that the electron receives the correct charge and also the correct 
spin.  We can do this while avoiding questions about how we attach the charge by assuming that in exiting the 
end of a line the electromagnetic field undergoes a phase transition.  
We are implying that the core of a line is a third
phase, in addition to the superconducting field and normal space outside the line.

    In the rest frame of the bare positron-pole (still using posititve charge for convenience), 
this phase transition is given by

\begin{widetext}
\begin{equation}
    (magnetic\ field, electric\ field)\ =\ ({\bf B}_{pole}\ ,\ 0)_{core}\ \rightarrow\ 
( {\bf B}_{pole}\ ,\ {{e_0} \over {g_0} } {\bf B}_{pole})_{outside}
\label{eq-phasetrans}
\end{equation}

\noindent
where, outside the line,

\begin{equation}
    {\bf B}_{pole}\ =\ { {g_0} \over {r^2} }\ {\hat r}
\label{eq-bpole}
\end{equation}

\noindent
is the expected magnetic induction field of a point pole.
The magnetic induction ${\bf B}$ is constant across the
phase boundary, so that the Maxwell equations are satisfied in this phase transition.   
The conservation of charge is not violated since an electron and positron are assumed to be
pair produced when a line breaks. 

    Assumption (\ref{eq-phasetrans}) is equivalent to the transformation on the charges

\begin{equation}
    (magnetic\ charge,\ electric\ charge)\ =\ (g_0,\ 0)_{core}\ \rightarrow\ (g_0,\ e_0)_{outside}
\label{eq-origincharge}
\end{equation}

\noindent
and so is our model for the origin of electric charge.  

    We can take our new assumption about charges conveniently into account with the substitution

\begin{equation}
    g_0\ \rightarrow\ f_0\ \equiv\ g_0\ +\ ie_0\ =\ \mu g\ +\ i\varepsilon e
\end{equation}

\noindent
in equation (\ref{eq-bppbfield}).  Then if ${\bf B}_{g_0}$ is equation (\ref{eq-bppbfield}) 
and ${\bf E}_{e_0}$ is equation (\ref{eq-bppefield}) with the replacement 
$g_0 \rightarrow e_0$, we can express the total electromagnetic field ({\it except} the line field)
of the bare positron as a complex vector in the form

\begin{equation}
    {\bf F}_{Total}\ \equiv\ {\bf B}_{Total}\ +\ i\ {\bf E}_{Total}\ =\ 
    ({\bf B}_{g_0} + i {\bf E}_{e_0})\ -\ i\ {\hat r} \times ({\bf B}_{g_0} + i {\bf E}_{e_0} )
\end{equation}

\begin{equation}
    =\ {\bf F}_{f_0}\ -\ i\ {\hat r} \times {\bf F}_{f_0}
\end{equation}
\end{widetext}

\noindent
where ${\bf F}_{f_0}$ is just (\ref{eq-bppbfield}) with the substitution $g_0 \rightarrow f_0$.
It will be convenient 
to use this complex notation.  The real field comes from the magnetic charge, and the imaginary field (which, of course,
 is actually physical) comes from the electric charge.   Thus we are saying that the bare positron is almost a 
{\it dyon}\cite{Schwingerdyon}, but with a {\it spin-dependent} magnetic charge.

    In Section~\ref{ssec-motionbpp} the motion of the bare positron-pole was discussed entirely on the basis 
of the Principle of Least Action applied to the superconducting line, 
before making our assumption on the origin of electric charge (\ref{eq-origincharge}).  We note that the equations of motion
(\ref{eq-motionep}) or (\ref{eq-motionepparts}) do {\it not} involve the bare positron-pole's pole field
if we evaluate the self field just inside the end of the line.  They do involve the pole field
if we evaluate the self field just outside the line.   Our assumption (\ref{eq-origincharge}) 
makes a difference only outside the line.  With or 
without this assumption the dominant external electromagnetic field will be due to the reflection (including the role of current
loops) of the electromagnetic field of the bare positron.  We, therefore, presume that the character of the motion of the bare
positron ({\it zitterbewegung}) is unchanged by our assumption on the origin of electric charge (\ref{eq-origincharge}).  
We prove that this is true in Appendix~\ref{ssec-motionelectron}.  
However, by adding an electric charge, we have added angular momentum (spin) to the end of each line.  
This will affect the orientation of each end of the line.  So far we do {\it not} have a model of the magnetic moment of the 
electron or positron, so we expect our results for the {\it orientation} of the end of each line to be {\it incorrect}.
Nevertheless we presume that the microscopic motion of the bare positron is still zitterbewegung.

    From Section~\ref{ssec-motionbpp} we expect rapid changes in the external electromagnetic field felt by the bare positron
on a time scale of about $10^{-22} s$ or less.  (With a higher line-pair density of the ether, this time would be shorter.)
No measurement device in human possession can measure this time dependence, so we consider the electromagnetic 
field averaged over a time interval of

\begin{equation}
    \Delta t_{meas}\ =\ 10^{-15} s
\end{equation}

\noindent
during which there are at least $10^6$ collisions, enough to form a reliable measurement of the average field.

    Note that although the bare positron is rapidly accelerating, it {\it does not radiate} because
the magnetic induction field is blocked by the closed loops maintaining flux quantization.  Hence we can see immediately that

\begin{equation}
\bigl \langle  {\bf B}^{mac}_{bare\ positron} \bigr \rangle \ =\ 0    
\end{equation}

\noindent
the time-avearaged magnetic induction field is zero at macroscopic distances from the positron.  
The magnetic induction field is reflected back toward the bare positron
with no energy loss (by superconducting loops), where it is reflected again by other current loops, and again, and again until the
sense of direction of the field is lost, because the current loops are in random orientations.  
Hence, effectively, all the magnetic induction field from the bare positron is contained in a very small region
near the moving positron, and so in this region, it is {\it very} intense.  

    We work in the effective rest frame of the bare positron---the frame in which it is at rest over $\Delta t_{meas}$
and hence

\begin{equation}
\bigl \langle  {{\bm \beta}} \bigr \rangle _{\Delta t_{meas}} \ =\ 0\ \ \ .
\end{equation}

\noindent
If we average over the bare positron's spin and are far from any masses
the only unique vector is in the direction

\begin{equation}
\bigl \langle  {\hat r} \bigr \rangle _{\Delta t_{meas}} \ \simeq\ {\hat r}
\end{equation}

\noindent
which we presume is reasonably constant over $\Delta t_{meas}$ .                                                                                       

    Hence in finding the average electromagnetic field of the positron 
we can see that all terms transverse to ${\hat r}$ average to zero.
Hence

\begin{equation}
    \biggl \langle {\hat r} \cdot {\bf F}_{f_0} \ \biggr \rangle \ =\
    \biggl \langle {\hat r} \cdot ({\bf B}_{g_0}\ +\ i {\bf E}_{e_0} ) \biggr \rangle \ =\ 
    {{f_0} \over {r^2} } \ \biggl \langle { {1-\beta^2} \over {\kappa^2} } \biggr \rangle \ =\ { {f_0} \over {r^2} } 
\end{equation}

\noindent
the average microscopic electric field near the bare positron is given by

\begin{equation}
    \biggl \langle  {\bf E}_{e_0} \biggr \rangle \ =\ { {e_0} \over {r^2} } \  {\hat r} \ \ \ .
\end{equation}

\noindent 
The macroscopic electric field of the positron is

\begin{equation}
    {\bf E}^{mac} \ =\ { {e_0} \over {\varepsilon r^2} } \  {\hat r}\ =\ { {e} \over {r^2} } {\hat r}
\end{equation}

\noindent
as expected.  Thus the measured charge of the positron is

\begin{equation}
        e\ =\ { {e_0} \over \varepsilon }
\end{equation}

\noindent
when the dielectric constant of the ether is taken into account.

    Hence the measured charge of the electron is not constant.  It should vary 
with the dielectric constant of the ether as the universe
expands.  A portion of the cosmological red shift could be due to the change of this dielectric constant with time.  
The expected change with time is in the observed direction. 

\subsection{Effective Magnetic Charge of the Electron}
\label{ssec-effectivemagcharge}

    If we {\it ignore the positron's line field}, and ignore, for the moment, magnetic screening due to flux 
quantization, we can find the effective magnetic charge on the positron (working with positive charge) 
using\cite{Jacksonmaxmod} the modified Maxwell equation

\begin{equation}
    \nabla \cdot {\bf B}^{mac}\ =\ 4 \pi \rho_{m-free}\ =\ 4 \pi g_0 \delta^3 ({\bf x}-{\bf y})\ =\ \mu \nabla \cdot {\bf H}\ \ \ .
\end{equation}

\noindent
The magnetic field at macroscopic distances from the positron will be

\begin{equation}
    {\bf H} \ =\ { {g_0} \over {\mu r^2} } \  {\hat r}\ =\ { {g} \over {r^2} } {\hat r} \ \ \ .
\label{eq-Hpole}
\end{equation}

\noindent
But the magnetic induction field will be

\begin{equation}
    {\bf B}^{mac} \ =\ { {g_0} \over { r^2} } \  {\hat r}\ 
\label{eq-Bmacpole}
\end{equation}

\noindent
determined by the strength of the effective free pole (magnetic charge).  

    Both of these fields will be blocked by screening currents due to flux quantization.
We do not yet have a model for these screening currents, but we can at least take these screening currents into account
through the use of a {\it screening function} $s(r)$ which cuts off the magnetic field of the positron
as it would be, on average, cut off by screening currents.

\begin{equation}
    {\bf B}^{mac} \ =\ { {g_0} \over { s r^2} } \  {\hat r}\
\label{eq-Bmacscreenedpole}
\end{equation}

The actual shape of the magnetic induction field of a positron will vary from this smooth model according to the actual positions
of the closed circuit current loops relative to the particular positron in question.  Equation (\ref{eq-Bmacscreenedpole}) 
just represents the average shape of the magnetic induction field coming from the positron itself.  Reflected fields are not included.

\subsection{Spin of the Electron}
\label{ssec-spinelectron}

    It is well known\cite{JacksonGoldhaber} that the electromagnetic angular momentum of a point monopole $g$ in the presence of a 
point charge $e$ is

\begin{equation}
{\bf L}_{eg} \ =\ \int d^3x\  {\bf x} \times \Bigl ( { 1 \over {4\pi c} } {\bf E} \times {\bf B} \Bigl )\ =\ 
{ {eg} \over c}\ {\hat z}
\label{eq-Leg}
\end{equation}

\noindent
where the angular momentum vector points in the direction from the point charge to the point pole (if both are positive).

    Since our model of the positron is almost a dyon, we can calculate the spin  of the positron if we assume that the line
is filamentary, and that it points in the ${\hat z}$-direction, where it can make no contribution to the angular momentum.
We obtain

\begin{equation}
    { \mid {\bf S}pin \mid }_{positron}\ =\ { {e_0 g_0} \over c}\ =\ { {\varepsilon e \mu g} \over c}\ =\ { {eg} \over c}\ =
\ { {\hbar} \over 2} \ \ \ .
\label{eq-spinpositron}
\end{equation}

\noindent
This calculation assumes that the line radius is small enough that the point charge and point pole
approximations can be used.  We also assume that the screening currents are sufficiently far away that they do not affect
the calculation.  The calculation is independent of the separation between the pole and the charge, but there must be some 
separation to determine the {\it direction} of the spin.

\begin{figure}
    \vbox to 1.8in{
\includegraphics{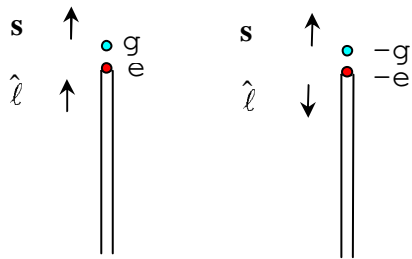}  }
    \caption{\small (Color online.)  The direction of the assumed spin of the positron (left) points from the point charge 
    to the point pole at the end of a line.  The spacing could be very small.  The assumed spin of the electron is shown on the right.}
\label{fig-electronspin}
\end{figure}

    Since there is no natural separation between the charge and the pole in our assumption (\ref{eq-origincharge}), 
we must look to experiment
to gain insight.  For a positive charge we assume that the spin is in the same direction as the magnetic moment of
the {\it line} (${\hat \ell}$).   (We do not yet have a model of the magnetic moment of the positron.)
Figure 14 shows schematically the small separation between pole and charge required to have the spin point in the
assumed direction for a positron.  Figure~\ref{fig-electronspin} also shows the situation for an electron.  For both particles the 
assumed spin direction is {\it opposite} to the line.

    It is interesting to note that the microscopic calculation of the spin (\ref{eq-Leg}-\ref{eq-spinpositron}) gives 
the same result as the macroscopic calculation

\begin{equation}
{\bf L}_{eg}^{mac} \ =\ \int d^3x\  {\bf x} \times \Bigl ( { 1 \over {4\pi c} } {\bf E}^{mac} \times {\bf H} \Bigl )\ =\
{ {eg} \over c}\ {\hat z}\ =\ {\hbar \over 2} {\hat z} \ \ \ .
\label{eq-Legmac}
\end{equation}

\noindent
The microscopic calculation is probably the most relevant since the dominant contributions to either integral 
(\ref{eq-spinpositron}) or (\ref{eq-Legmac})
come from regions very near the pole and charge.  The separation between the pole and the charge is the only length involved 
and defines the importance of contributions to the integrals.  In the final integration over the $z$-direction, half of the
integral is contributed by the region between the pole and charge.  Since this separation is apparently {\it very} small
(see equation ~\ref{eq-Rasmall}), the approximation that the screening fields do not affect the integral is probably valid.

\subsection{Mass of the Electron}
\label{ssec-masselectron}

    From Sections \ref{ssec-motionbpp} and \ref{ssec-electroncharge} 
we see that the electromagnetic field of the bare positron (again considering positive charge) 
does not carry inertia, i. e. mass.
In the average rest frame of the bare positron, however, the superconducting ether will certainly respond to the average 
electromagnetic field of the bare positron.   This response implies a potential energy, which we interpret as mass.
The wave function of the positron is the response of the ether to the electromagnetic field 
of the bare postron\cite{externalforce}

\begin{equation}
{{\bm \psi}}\ =\ \sum_j\ \Delta {\bf y}_j ({\bf x})
\label{eq-psix}
\end{equation}

\noindent
where the sum is over all line pairs which respond to the bare positron.
So the mass of the positron is just the potential energy of its wave function in its rest frame.
This response is a discontinuous function of position (as in Section~\ref{ssec-wfelectron}).  It is
only defined at equilibrium positions (in absence of distortion) of line pairs.  The wave function is a vector field, 
but it always points toward (or away from) the average position of the bare positron, 
so we can take it as a scalar without loss of information.

    So the mass of the positron is given by

\begin{equation}
     m_e c^2\ =\ \sum_j \biggl [ \int_{{\bf x}_i}^{{\bf x}_f} \tau_j d\ell_j\ -\ 
                                 \int_{{\bf x}_i}^{{\bf x}_f} \tau_0 dz_j \biggr ]
\label{eq-mesum}
\end{equation}

\noindent
where the integrals have the same end points in space, but differing paths.  The points in space ${\bf x}_i$ and ${\bf x}_f$ 
are chosen to be on line pair $j$ 
but beyond the response of line pair $j$ to the bare positron.  The first integral is over the actual path of
the distorted line pair $j$ using the actual tension $\tau_j$ which could be different from $\tau_0$ in general,
because this line pair is part of a matter wave, a quantized excitation of the ether.  
(But see below (\ref{eq-tensiontau0}) and (\ref{eq-metau0}) where we will find, in fact, that $\tau_j = \tau_0$).  The second integral
is over the straight line path (in the ${\hat z}$-direction) which the line pair would have in absence of the positron.  
Since in this case the line pair is assumed not to be in an excited state, the tension is $\tau_0$ (\ref{eq-tensiontau0}).

    The mass of the positron (or electron) in this model is clearly calculable.  At small distances there is no convergence
problem since we sum over a finite number of line pairs.  At large distances we can approximate the sum 
(\ref{eq-mesum}) by an integral.
All integrals clearly converge since we will see that the electromagnetic force causing the response of a line pair falls off 
like ${1 / {r^5} }$ where $r$ is the magnitude of the separation from the bare positron's average position.

    The zitterbewegung motion of the bare positron is far too fast to affect the ether's response.  With typical times of
$10^{-22} s$, very large energies would be required for such a response, because all responses are quantized.  The ether 
responds to the {\it average} electromagnetic field of the bare positron, as it undergoes its zitterbewegung motion in its wave
function.  The quantized changes occur over long times and can therefore have low energy.

    We calculate the response of a line pair to the average electromagnetic field of the bare 
positron in its average rest frame, where the calculation is static.  
If ${\bf p}$ is the electric dipole moment per unit length of the line pair in
question, the electric force per unit length on the line pair is

\begin{equation}
    {\bf f}_e\ =\ ({\bf p}\cdot\nabla) {\bf E}^{mac} \ \ \ .
\end{equation}

\begin{figure*}
    \vbox to 1.5in{
\includegraphics{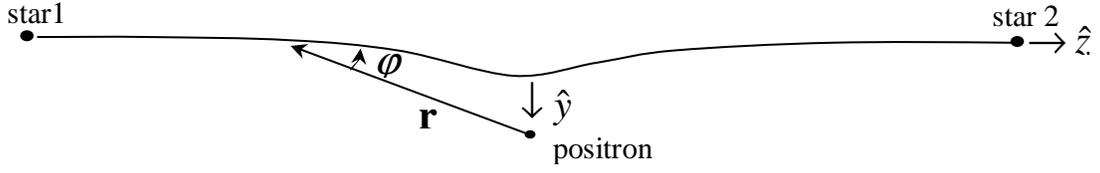}  }
    \caption{\small A line pair comes from star 1, through the wave function of a positron on its way toward star 2, where 
             it ends.  Star 2 could be very far away (over the horizon).}
\label{fig-lpstars12}
\end{figure*}

\noindent
But since the dipole moment per unit length is induced by the electric field of the positron itself we have 

\begin{equation}
    {\bf p}\ =\ \alpha (\phi) {\bf E}^{mac} 
\label{eq-pealEmac}
\end{equation}

\noindent
where $ \alpha (\phi)$ is the polarizability per unit length of the line pair, which in general depends on $\phi$, the 
angle between the electric field and the line pair\cite{Beckeralpha}.  (See Figure~\ref{fig-lpstars12}.)

\noindent
Hence the electric force per unit length on the line pair

\begin{equation}
    {\bf f}_e\ =\ \alpha (\phi) ({\bf E}^{mac}\cdot\nabla) {\bf E}^{mac}\ =\ 
   { {\alpha (\phi)} \over 2} \ \nabla (E^{mac})^2
\end{equation}

\noindent
is proportional to the gradient of the square of the electric field---independent of the sign of the 
charge of the electron or positron.
The component of the opposite force along the line pair is the gravitational force on the electron or positron.  
It is independent of the sign of the charge because it is {\it induced}.
Hence the electric force which the positron exerts on a line pair is given by

\begin{equation}
    {\bf f}_e\ =\ 
    -2 {\hat r} \ \alpha (\phi) \ { {e_0^2} \over {\varepsilon^2 r^5} }\ 
    -\ \alpha (\phi)\  { {e_0^2} \over {r^4} } \ { {\nabla \varepsilon} \over {\varepsilon^3} }
\end{equation}

\noindent
and is always {\it attractive}, as required for the force of gravity in our region of the universe.  (We assume that we can neglect
the second term in this case because $1/r \gg \mid \nabla \varepsilon \mid / \varepsilon$, but see Section \ref{ssec-1OR3})

    The effective magnetic charge of the positron will also cause a force on our line pair given by\cite{JacksonmB}

\begin{equation}
\bigl [ {\bf f}_m\bigr ]_i\ =\ \bigl [ ( {\bf m} \times \nabla)\ \times\ {\bf B}^{mac} \bigr ]_i\ =\ \sum_k m_k \nabla_i B_k^{mac}
\end{equation}

\noindent
where ${\bf m}$ is the magnetic moment per unit length of the line pair.  The induced magnetic moment is
{\it opposite} to the applied magnetic induction field

\begin{equation}
{\bf m}\ =\ - \alpha (\phi)\ {\bf B}^{mac}
\label{eq-memalBmac}
\end{equation}

\noindent
where the polarizability is the same as in the electric case.  Hence the magnetic force per unit length on a line pair is

\begin{equation}
    {\bf f}_m\ =\ - { {\alpha (\phi)} \over 2} \ \nabla (B^{mac})^2
\end{equation}

\noindent
again independent of the sign of the charge.  Using (\ref{eq-Bmacscreenedpole}) the magnetic force per unit length is

\begin{equation}
    {\bf f}_m\ =\ +2 {\hat r} \ \alpha (\phi) \ { {g_0^2} \over {s^2 r^5} }\
    +\ \alpha (\phi)\  { {g_0^2} \over {r^4} } \ { {\nabla s} \over {s^3} }
\label{eq-poleforce}
\end{equation}

\noindent
{\it repulsive} and independent of the dielectric constant of the ether.  The screening function $s(r)$ (e.g. $e^{\Lambda r}$)
must be positive and strongly increasing as a function of $r$.  It can also vary slowly with the position of the positron 
(e.g. $\Lambda$ above could vary) due to extra line pairs from a passing star.
 Because of this screening,
we assume the electric force is dominant, except {\it very} close to the positron, and in the case that the dielectric constant of
the ether is {\it very} large.

    Thus the total force per unit length on a line pair is given by

\begin{equation}
    {\bf f}\ =\ {\bf f}_e\ +\ {\bf f}_m
\end{equation}

\begin{widetext}
\begin{equation}
    {\bf f}\ =\ 
    -2 {\hat r} \ \alpha (\phi) \ { {e_0^2} \over {\varepsilon^2 r^5} }\
    -\ \alpha (\phi)\  { {e_0^2} \over {r^4} } \ { {\nabla \varepsilon} \over {\varepsilon^3} }
  \ +2 {\hat r} \ \alpha (\phi) \ { {g_0^2} \over {s^2 r^5} }\
    +\ \alpha (\phi)\  { {g_0^2} \over {r^4} } \ { {\nabla s} \over {s^3} } \ \ \ .
\label{eq-ftotal}
\end{equation} 
\end{widetext}

\begin{figure}
    \vbox to 1.5in{
\includegraphics{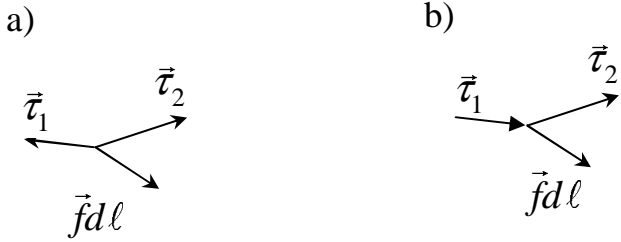}  }
    \caption{\small a) The tension in the line pair must balance the force exerted on the line pair by the positron.
                    b) This is Figure~\ref{fig-tensionbalance}(a) considering the tension always to point from left to right.}
\label{fig-tensionbalance}
\end{figure}

\noindent
In Figure~\ref{fig-lpstars12} we consider a line pair stretching out from star 1 and passing through our positron's wave function on the
way toward star 2 (which could be {\it very} far away, e.g. over the horizon).  In Figure~\ref{fig-tensionbalance}
we see a closeup of how the force per unit length ${\bf f}$ changes the tension, which is considered to point 
along the line pair in the direction from star 1 to star 2
(except in Figure~\ref{fig-tensionbalance}(a)).  

\noindent
Since the forces must balance (in the static configuration)

\begin{equation}
    {{\bm \tau}}_1\ =\ {{\bm \tau}}_2\ +\ {\bf f} d\ell
\end{equation}

\noindent
so

\begin{equation}
  { { d{{\bm \tau}} } \over {d\ell} }\ =\ -{\bf f}
\end{equation}

\noindent
and

\begin{equation}
  {{\mbox{\boldmath $\tau$}}}({\bf x})\ =\ -\int_{{\bf x}_i}^{{\bf x} } {\bf f} d\ell\ +\ {{\bm \tau}}_0
\label{eq-taux}
\end{equation}

\noindent 
where {\bf x} is any point on the distorted line pair and  ${{\bm \tau}}_0=\tau_0 {\hat z}$ 
is the initial tension vector at point ${\bf x}_i$ .
Hence

\begin{equation}
  \biggl \vert \
  {{\bm \tau}}({\bf x})\ +\ \int_{{\bf x}_i}^{{\bf x} } {\bf f} d\ell\ \biggr \vert\ =\ \tau_0
\label{eq-tau0commode}
\end{equation}

\noindent
the tension ${{\bm \tau}}$
varies such that the effective tension, {\it including} the effects of the force due to the positron,
is constant.  The transverse components change the direction of the tension, while the longitudinal components can change the
magnitude.

    It is convenient and reasonable to {\it define} the tension to 
include the longitudinal force of the positron.
This force is effectively part of the tension of the responding line pair.
Then the magnitude of the tension is always $\tau_0$ in accord with (\ref{eq-tensiontau0}),
and so that the matter wave is a gravitational/common-mode wave.  Hence we replace
$\tau_j$ by $\tau_0$ in (\ref{eq-mesum}) and obtain for the mass of the positron

\begin{equation}
    m_e c^2\ =\  \sum_j\  \Delta L_j\  \tau_0
\label{eq-metau0}
\end{equation}

\noindent
where $\Delta L_j$ is the length difference of line pair $j$ with the positron and without it.

    The response of a line pair extends far beyond the point of application of the force ${\bf f}_j$, just as the
distortion of a bow string extends far beyond the point at which it is plucked.  Just how far the response extends 
should depend on the surrounding matter density because other charged particles will exert forces on this same line pair.
These forces, which are exerted on the line pair under consideration by other charged
particles, can also be considered using (\ref{eq-taux}).  Thus the calculation is a many body problem, by its nature, 
because everything is connected.  Appropriate assumptions must be made to find a numerical solution.  
This work has not yet been done.

    Dependence on the surrounding matter density has been observed.  The mass of the electron changes from it's ``vacuum"
value when the electron is in a solid\cite{Kittelmass}.  This change of mass in a solid is thought to be a collective effect.  We
are suggesting that mass may be a collective effect even in ``vacuum".

\begin{figure*}
    \vbox to 1.5in{
\includegraphics{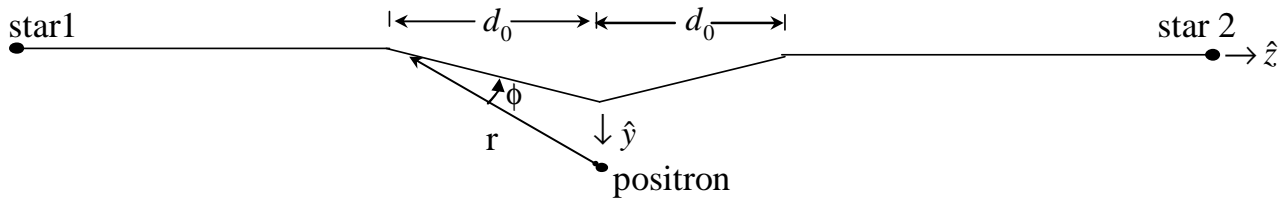}  }
    \caption{\small Toy model calculation for the mass of the electron, models the situation in Figure~\ref{fig-lpstars12}.}
\label{fig-lpstars12toy}
\end{figure*}

    In order to gain insight into this mass calculation we construct a toy model in which we assume (Figure~\ref{fig-lpstars12toy}) 
that the line pair in
question is fixed at a distance $d_0$ from the positron.  We neglect the curvature of the line pair in evaluating the total force
and consider the total force to act at the point of closest approach.  We assume the electric force dominates so that 
(taking only the first term) the components of the force per unit length which the positron exerts on the line pair are given by
  
\begin{equation}
    f_y\ =\  2 \alpha (\phi) \ { {e_0^2} \over {\varepsilon^2 r^5} } sin\phi\
\end{equation}
                                                                                                                                   
\begin{equation}
    f_z\ =\  2 \alpha (\phi) \ { {e_0^2} \over {\varepsilon^2 r^5} } cos\phi\
\end{equation}

\noindent
where we simply take $ \alpha (\phi) = R_a^2/sin\phi$ \cite{Beckeralpha}.
The total force perpendicular to the line is

\begin{equation}
    F_y\ =\ \int dz f_y\ =\ { {8 e_0^2 R_a^2} \over {3 \varepsilon^2} } {1 \over {y^4} }
\end{equation}

\noindent
where $y$ is the distance of closest approach of the line to the positron.  The total longitudinal force (along the line)
is zero but for future reference we calculate the force that the positron exerts on the line pair in the directions of star 1 and
star 2 separately:

\begin{equation}
    F_{z1}\ =\  +{ {2 e_0^2 R_a^2} \over {3 \varepsilon^2} } {1 \over {y^4} }
\label{eq-Fz1}
\end{equation}

\begin{equation}
    F_{z2}\ =\  -{ {2 e_0^2 R_a^2} \over {3 \varepsilon^2} } {1 \over {y^4} }
\end{equation}         

    If $\Delta y$ is the displacement of the line pair at closest approach, from Figure~\ref{fig-lpstars12toy}

\begin{equation}
    \biggl ({ {L_j} \over 2} \biggr )^2\ =\ (\Delta y)^2\ +\ d_0^2 
\end{equation}

\begin{equation}
    \biggl ( {{L_j} \over 2}\ -\ d_0 \biggr ) \biggl ( {{L_j} \over 2}\ +\ d_0 \biggr )\ =\ (\Delta y)^2\ 
\end{equation}

\noindent
and since $L_j \simeq 2 d_0$ we have

\begin{equation}
    \Delta L_j\ \simeq\ {{(\Delta y)^2} \over d_0} \ \ \ .
\end{equation}

\noindent
Since the line pair tension must balance $F_y$

\begin{equation}
    F_y\ =\ 2 \tau_0 \ { {\Delta y} \over (L_j/2) }\ \simeq \ 2 \tau_0 \ { {\Delta y} \over {d_0} }
\end{equation}

\noindent
we find

\begin{equation}
    \Delta y\ \simeq\ { {d_0 F_y} \over {2 \tau_0} }
\end{equation}
                                                  
\begin{equation}
    \Delta L_j\ \simeq\ { { d_0 (F_y)_j^2} \over { (2 \tau_0)^2 } } \ \ \ .
\end{equation}

    So the mass of the positron or electron in this toy model is given by

\begin{equation}
    m_e c^2\ \simeq\ \sum_j\ \Delta L_j\ \tau_0\ =\ { d_0 \over {4 \tau_0} } 
    \sum_j \biggl ( {{8 e_0^2 R_a^2} \over {3 \varepsilon^2} } \biggr )^2 { 1 \over {y_j^8} }
\end{equation}

\noindent
so, replacing the sum by an intrgral

\begin{equation}
    \sum_j\ \rightarrow\ \int dA\  d\Omega_{line-pair}\  { {dn} \over {dA d\Omega_{line-pair}} }
\label{eq-sumtoint}
\end{equation}

\noindent
we have (since $e=e_0/\varepsilon$)

\begin{equation}
    m_e c^2\ \simeq\ { {d_0} \over {\tau_0} } { {64} \over {36} } e^4 R_a^4 { {2\pi} \over {2\pi} } { {dn} \over {dA} }
            \int_{y_{min}}^{\infty} 2\pi y dy { 1 \over {y}^8}
\end{equation}

\begin{equation}
    m_e c^2\ \simeq\ { {d_0} \over {\tau_0} }\ { {16\pi} \over {27} }\  e^4 R_a^4 \ { {dn} \over {dA} }\  { 1 \over {y_{min}^6}}
    \ \ \ .
\label{eq-me}
\end{equation}

\noindent
At this stage the only quantities for which we know a numerical value in this equation are $e$ and $m_e c^2$, but the formula
illustrates what quantities are important in determining the mass of the electron in this model.  We can cross check this formula 
with others which we will obtain.  We expect order of magnitude accuracy if we were to evaluate $y_{min}$ and $d_0$ 
with care---using appropriate averaging, probably with a Monte Carlo.  The cutoff $y_{min}$ should be evaluated so the integral
in (\ref{eq-sumtoint}) approximates the required sum.  It depends, at least in principle, on ${ {dn} \over {dA} }$ and on the function being
integrated (here ${1 \over y^7}$).

    Near the Earth we will see (Section~\ref{ssec-parameters}) that the dielectric constant of the ether is estimated to be very close to $1$.
In this case we have $e_0 = e$ and $g_0 = g$, the strength of the Dirac monopole.  Hence as $r \rightarrow 0$ the force 
${\bf f}$ on a line pair is dominated by the pole force, and hence is {\it repulsive}.  So we expect the effective value of
$y_{min}$ to be the size of a typical deflection with the full pole strength.  Outside $y_{min}$ (or $r_{min}$, reflecting radial
symmetry of the configuration with many line pairs) line pairs will be concentrated as the pole is rapidly screened and
${\bf f}$ becomes attractive due to the electric force.  Since $y_{min}$ should depend only on the size of a typical deflection at
full pole strength, we will simply take $y_{min}$ to be a constant (at least in vacuum near the Earth, relevant to the calculation
of $G$ in this toy model).

    We discuss the response of our macroscopic electron in its mass shell (wave function) to an external electromagnetic field and
the possible effects of its string in Appendix~\ref{sec-response}.

\section{Gravity}
\label{sec-gravity}

\subsection{Calculation of G}
\label{ssec-calcG}
    We have seen in Section~\ref{ssec-masselectron} that the mass of the electron is proportional to 
${ {dn} \over {dA} }$ the line pair density
of the ether.  We assumed that the line pair density is isotropic reflecting the large scale structure of the universe, for which
(from (\ref{eq-dndAdOmega}) )

\begin{equation}
    \biggl ( { {dn} \over {dA d\Omega} } \biggr )_B\ =\ { 1 \over {2\pi} }  \biggl ( { {dL} \over {dV} } \biggr )_B
\end{equation}

\noindent
where the subscript references the {\it background} ether---line pairs for which star 1 and star 2 are {\it very} far away
(perhaps over the horizon).

    In contrast to the background ether, the line pair density from a star is certainly {\it not} isotropic.  Since we can
approximately (better than $1\%$ accuracy) associate a line pair coming from the star for each neutron mass ($m_n$)

\begin{equation}
    \biggl ( { {dn} \over {dA d\Omega} } \biggr )_{star}\ =\ \biggl ( { {dn} \over {dA} }  \biggr )_{star}\ 
   \delta({{\bm \Omega}} - {\hat R})
\end{equation}

\noindent
where the delta function in solid angle ensures that all line pairs point in the 
radial direction from the center of the star 
(a simplification) and where

\begin{equation}
    \biggl ( { {dn} \over {dA} }  \biggr )_{star}\ =\ { M \over {4\pi m_n R^2} }
\label{eq-dndAstar}
\end{equation}

\noindent
where $M$ is the mass of the star and $R$ is the distance from the center of the star.  In Table~\ref{table-lpdatsurface} 
we list some line-pair
densities at the surfaces of known astronomical objects, where Sagittarius $A^{\ast}$ is a star at the center of our
galaxy which is thought to be a supermassive black hole\cite{MeliasagA}.  Also listed is the star at the center of the M87 
galaxy which appears to be an even more massive black hole\cite{MeliaM87}.  For the two black holes we take

\begin{equation}
    R_s\ =\ { {2 M G} \over {c^2} }
\end{equation}

\noindent
the radius at the surface to be the Schwarzschild radius.  We also quote $\ell_{typ}$ values at the surface of the object 
using (\ref{eq-ltyp}).

\begin{table*}
\caption{Line-Pair Densities at the Surfaces of some Astronomical Objects}
\begin{tabbing}
\=object\ \ \ \ \ \ \ \ \ \ \ \ \ \ 
     \=$M(kg)$\ \ \ \ \ \ \ \ \ \ \ \ \ \ \ 
     \=$R_{surface}(m)$\ \ \ \ \ \ \ \ 
     \=${ {dn} \over {dA} } \biggl ( { {line\ pairs} \over {m^2} } \biggr )$\ \ \ \ \ 
     \=\ \ \ \ \ $\ell_{typ} (m)$\\

\>Earth                \>$5.97 \times 10^{24}$     \>$6.38 \times 10^6$     \>$6.98 \times 10^{36}$     \>$3.79 \times 10^{-19}$  \\
\>sun                  \>$1.99 \times 10^{30}$     \>$6.96 \times 10^8$     \>$1.96 \times 10^{38}$     \>$7.14 \times 10^{-20}$  \\
\>Sagittarius $A^\ast$ \>$5.2 \times 10^{36}$     \>$7.71 \times 10^9$     \>$4.17 \times 10^{42}$     \>$4.90 \times 10^{-22}$  \\
\>M87                  \>$6 \times 10^{39}$     \>$8.89 \times 10^{12}$     \>$3.62 \times 10^{39}$     \>$1.66 \times 10^{-20}$  \\
\end{tabbing}
\label{table-lpdatsurface}
\end{table*}

\begin{table*}
\caption{Line-Pair Densities at the Surface of the Earth due to the Sun, 
the Central Bulge and the Background Ether}
\begin{tabbing}
\=object\ \ \ \ \ \ \ \ \ \ \ \ \ \
     \=$M(kg)$\ \ \ \ \ \ \ \ \ \ \ \ \ \ \
     \=$R_{Earth}(m)$\ \ \ \ \ \ \ \ \ \ 
     \=${ {dn} \over {dA} } \biggl ( { {line\ pairs} \over {m^2} } \biggr )$\ \ \ \ \
     \=\ \ \ \ \ $\ell_{typ} (m)$\\

\>sun                  \>$1.99 \times 10^{30}$  \>$1.50 \times 10^{11}$     \>$4.20 \times 10^{33}$     \>$1.54 \times 10^{-17}$ \\
\>central bulge        \>$2 \times 10^{41}$     \>$2.62 \times 10^{20}$     \>$1.38 \times 10^{26}$     \>$8.5 \times 10^{-14}$ \\
\>background ether - minimum  \>\>                                          \>$7 \times 10^{38}$     \>$4 \times 10^{-20}$ \\
\>background ether - measured \>\>                                          \>$4 \times 10^{39}$     \>$1.5 \times 10^{-20}$ \\
\end{tabbing}
\label{table-lpdatEarth}
\end{table*}

    We note that the line-pair density at the surface of the Earth, from the Earth alone, is 11 orders of magnitude greater
than our previously assumed minimum line-pair density of the background ether (\ref{eq-dndAnum})!  
Assuming masses do not vary significantly
with altitude above the surface of the Earth\cite{mabove}, we infer that
the line-pair density of the background ether must be 13 orders of magnitude or more
higher than our minimum estimate (\ref{eq-dndAnum}).  This fact has significant implications for the discussion of inflation.

    In Table~\ref{table-lpdatEarth} we list the line-pair densities at the Earth due to our sun and the central bulge of our 
galaxy\cite{centralbulge}.  The mass contained in this central bulge is only perhaps a factor of 2 less than the visible mass of
our galaxy inside the orbit of the sun.  We see that the contribution of the central bulge is much smaller than the contribution of
the sun, which is much smaller than the contribution of the Earth itself.  We also list our new estimate for the minimum line-pair 
density of the background ether ($ \sim 100 \times ( { {dn} \over {dA} } )_{Earth} )$ and our rough measurement (\ref{eq-fulldensity}).

    We have painted a very static picture in discussing equation (\ref{eq-dndAstar}).  
This static picture should be viewed as a statement of
the average behavior line pairs emerging from a star.  All changes from this picture are quantized, but changes will certainly
occur.  Line pairs are attracted by charges of both signs.  So if a star is moving through a field of line pairs, the line pairs
will tend to stick to the charges, and stretch as the star continues to move. 
When the line pairs have stretched sufficiently that their tension force is strong
enough to free them (perhaps due to a changing angle), they will contract (being perfectly elastic) and move to the next row of
charges.  The line pairs will tend to form straight lines, however, because in that way they minimize their energy.  
Most line pairs are part of the background ether---so end, and hence are bound strongly, in stars which are {\it very} far away.
Assuming the
universe is still expanding, star 1 and star 2, at the ends of a typical line pair, are moving apart as the line-pair tension
$\tau_0$ continues to decrease very slowly with time throughout the universe.  
Two line pairs can bind to form a 4-line system, but the
binding is weak compared to the binding for a pair.

    In Figures \ref{fig-lpstars12} and \ref{fig-lpstars12toy} we identify the force of gravity 
from star 1 via our line pair $j$ on our standard positron to be (from (\ref{eq-Fz1}))

\begin{equation}
    {\bf F}_{g1j}\ =\ - F_{z1} {\hat z}\ =\ 
- \biggl ( { {2 e_0^2 R_a^2} \over {3 \varepsilon^2} } \biggr ) { 1 \over {y_j^4} } {\hat z} \ \ \ .
\label{eq-Fg1j}
\end{equation}

\noindent
In a similar way we identify

\begin{equation}
    {\bf F}_{g2j}\ =\ - F_{z2} {\hat z} \ 
\label{eq-Fg2j}
\end{equation}

\noindent
the force from star 2 via the same line pair on our positron.  The situation is intrinsically 3-body and equations 
(\ref{eq-Fg1j}) and
(\ref{eq-Fg2j}) represent the only self consistent way to isolate ${\bf F}_{g1j}$ and ${\bf F}_{g2j}$ respectively,
as is required to describe the force of gravity.  (In principle we should include $-{1 \over 2}F_y$ in the force from both
stars, but this force adds to zero when we consider all line pairs near the positron.)  
Both stars exert attractive forces on the positron via this line
pair $j$.  After all, line pair $j$ is {\it part of both stars}.  The force of gravity propagates instantaneously because both stars
are already at the position of the positron. The quantum nature of these forces is illustrated by the fact that many 
line pairs from star 1 will come through the positron's
wave function if it is sufficiently near star 1.  However, the nearest line pair coming from star 2 could be light years away if
star 2 is over the horizon.  Thus star 2 will not exert significant force on the positron, similar to many other stars in random
directions.

    Thus the total force of gravity due to all line pairs from star 1 on the positron is

\begin{equation}
    {\bf F}_{ge}\ =\ -\sum_j  \biggl ( { {2 e_0^2 R_a^2} \over {3 \varepsilon^2} } \biggr ) { 1 \over {y_j^4} }  {\hat z}
\end{equation}

\begin{equation}
    {\bf F}_{ge}\ =\  - \biggl ( { {2 e_0^2 R_a^2} \over {3 \varepsilon^2} } \biggr ) {\hat z}
    \int dA d\Omega { M \over {4\pi m_n R^2} }  \delta({{\bm \Omega}} - {\hat R}) { 1 \over {y^4} }
\end{equation}

\begin{equation}
    {\bf F}_{ge}\ =\ - \biggl ( { {e^2 R_a^2} \over {6 m_n y_{min}^2} } \biggr ) { M \over {R^2} } {\hat z}
\label{eq-Fge}
\end{equation}

\noindent
We observe that this force is not obviously proportional to the mass of the positron (\ref{eq-me}).  
We examine the possibility that this force obeys Newton's Law of Universal Gravitation, setting

\begin{equation}
    {\bf F}_{ge}\ =\ - \biggl ( { {e^2 R_a^2} \over {6 m_n y_{min}^2} } \biggr ) { M \over {R^2} } {\hat z}\ =
    \ -{ {G M m_e} \over {R^2} } {\hat z}
\end{equation}

\noindent
and we find that

\begin{equation}
    G m_e\ =\ { {e^2 R_a^2} \over {6 m_n y_{min}^2} } 
\end{equation}

\noindent
so

\begin{equation}
    G \ =\  { {e^2 R_a^2} \over {6 m_e m_n y_{min}^2} } \ \ \ .
\label{eq-G}
\end{equation}

\noindent
(Remember that we are only using a toy model here.)
The value of $y_{min}$, like $\varepsilon$, must be constant within our solar sysem.  The value of $y_{min}$
has an additional restriction.  It must be the same whether the positron is attracted by the sun, or the Earth, or a baseball.

    We note that the force of gravity on a single electron or positron has not been well measured\cite{Fairbank}.  It is a very 
difficult measurement so we need a consistency check for this determination of $G$.  We consider a single electron which 
combines with a single proton to form a hydrogen atom---the smallest bit of neutral matter, for which Newton's Law of Universal 
Gravitation is well established in our solar system.  We are trying to stay out of the nucleus so we simply assume that

\begin{equation}
    {\bf F}_{gp}\ =\ f \ {\bf F}_{ge}
\label{eq-Fgp}
\end{equation}

\noindent
the gravitational force on the proton from star 1 is a factor of $f$ times the gravitational force on the electron.  So negelcting
binding effects, the force on the hydrogen atom is just $1+f$ times the force on the electron.

\begin{equation}
    {\bf F}_{gH}\ =\ (1+f) \ {\bf F}_{ge}
\end{equation}

\noindent
Normal matter, neglecting binding effects, is just made up of proton-electron combinations, perhaps in the form of neutrons, 
so the total gravitational force on a mass $M^{\prime}$ ({\it assumed} proportional to mass) is just

\begin{equation}
{\bf F}_{gM^\prime}\ =\ \biggl ( { {M^{\prime} } \over {m_n} } \biggr ) {\bf F}_{gH}\ =
\ \biggl ( { {M^{\prime} } \over {m_n} } \biggr )  (1+f) \ {\bf F}_{ge}
\end{equation}

\noindent
so

\begin{equation} 
   - \biggl ( { {M^{\prime} } \over {m_n} } \biggr )  (1+f)\ \biggl ( { {e^2 R_a^2} \over {6 m_n y_{min}^2} } \biggr ) 
     { M \over {R^2} } {\hat z}\ =\ - { {G M M^{\prime} } \over {R^2} } {\hat z} 
\label{eq-proportionaltoM}
\end{equation}

\noindent
and so we also obtain

\begin{equation}
     G\ =\ (1+f) \ \biggl ( { {e^2 R_a^2} \over {6 m_n^2 y_{min}^2} } \biggr ) \ \ \ .
\end{equation}

\noindent
If we compare our two values of $G$ we find

\begin{equation}
    (1+f)\ =\ { {m_n} \over {m_e} } \ \ \ .
\end{equation}

\noindent
Hence our two values of $G$ agree using a value of $f$ which indicates that
the gravitational force exerted by star 1 is proportional to the mass of the object on
which it acts.  This proportionality is shown more generally by (\ref{eq-proportionaltoM}).

    We have not discussed the gravitational force which our positron exerts on star 1.  It is zero unless the positron's line goes
through the matter of star 1.  Newton's third law is apparently violated by the quantization of the gravitational force, but it 
should be valid on average.  (The electromagnetic reaction forces to both gravitational forces, as we have defined them, do
satisfy Newton's third law, as we have shown.)

    Our calculation of $G$ only comes from a toy model, but it  shows some important features. 
Using the cgs values

\begin{equation}
    G=6.67\times10^{-8} { {dyne-cm^2} \over {g^2} }
\end{equation}

\begin{equation}
e^2=(4.803\times 10^{-10}esu)^2=2.31\times10^{-19} dyne-cm^2
\end{equation}

\noindent
we obtain

\begin{equation}
 \biggl ( { {R_a} \over {y_{min}} } \biggr )^2 \ =\ { {6 G m_e m_n} \over {e^2} }\ =\ 2.64 \times 10^{-39} 
\end{equation}

\begin{equation}
 { {R_a} \over {y_{min}} }\ =\ 5.14 \times 10^{-20} \ \ \ .
\label{eq-Rasmall}
\end{equation}

\noindent
So apparently $R_a$, the outer radius of a line pair, is very small.  If $y_{min}$ is of order $\ell_{typ}$, $R_a \sim 10^{-39} m$!

     Since the gravitational field emerging from a star in this model is proportional to the number of line pairs emerging (hence
 the number of nucleons in the star), the gravitational force in this model  is {\it not} rigorously proportional to active mass
 (the source of the field).  We have simply {\it assumed} that the gravitational force is proportional to passive mass because we
 cannot model the nucleus.  For a rigorous treatment to an accuracy better than 1\% and a discussion of experimental tests of 
our model of gravity, see Appendix~\ref{sec-rigorousgravity}.

\subsection{\ \ \ A $1/{R^3}$ Force of Gravity}
\label{ssec-1OR3}

    In the presence of star 1, the second term in (\ref{eq-ftotal}) will lead to a gravitational force on our standard positron and we need to
either consider it, or show that it is small.  From (\ref{eq-epsminus1}) we find

\begin{equation}
    \varepsilon\ =\ 1\ +\ { {3f} \over {1-f} }\ =\ { {1+2f} \over {1-f} } 
\end{equation}

\noindent
where $f$ is the fractional concentration of superconducting fields, which implies that

\begin{equation}
   { {d\varepsilon } \over df}\  =\ { 3 \over {(1-f)^2} } \ \ \ .
\end{equation}

\noindent
Near a star, $f$ will have two contrubution---one from the background ether and the other from the star.  For the star we have

\begin{equation}
    f_{star}\ =\ { {dn} \over {dA} }\ \pi R_a^2\ =\ { {M R_a^2} \over { 4 m_n R^2 } }
\end{equation}

\noindent
where $R$ is the distance from the center of the star, and $f_{star}$ cannot be greater than $1$.  Since the superconducting fields
can overlap, the fractional concentration {\it free} of superconductor is just the product of the respective concentrations 
without superconductor

\begin{equation}
   (1-f)\ =\ (1-f_B)\ (1-f_{star})
\end{equation}

\noindent
implying that 

\begin{equation}
   \nabla f\ =\ (1-f_B)\ \nabla f_{star}\ =\ (1-f_B)\ { {M R_a^2} \over {2 m_n R^3} }\ (- {\hat R} ) \ \ \ .
\end{equation}

\noindent
Hence, adding up the contributions of all relevant line pairs, the total force of gravity on the positron due to the gradient of 
$\varepsilon$ is

\begin{equation}
    {\bf F}_{ge}^\varepsilon\ =\ 6\pi\ { {e^2} \over \varepsilon} \ \biggl ( { {dn} \over {dA} }  \biggr )_B\ 
{ {(1-f_B)} \over { (1-f)^2 } }\ { M \over m_n }\  { {R_a^4} \over {y_{min}\  R^3} }\ (- {\hat R} ) 
\label{eq-FgeOR3}
\end{equation}

\noindent
also attractive.  In calculating (\ref{eq-Fge}) we only included line pairs coming from star 1, to calculate the total gravitational force
from star 1.  The total force from the background ether was zero because it is isotropic.  But here star 1 causes the dielectric
constant of the ether to be {\it anisotropic}.  So all line pairs, even from the background ether, will exert a force on the
positron in the direction of $\nabla \varepsilon$.  The ratio of the magnitudes of the two forces of gravity is

\begin{equation}
{ {F_{ge}^\varepsilon} \over {F_{ge} } }\ =\ { {36\pi} \over {(1+2f)}}\ { (1-f_B) \over (1-f) }
\ \biggl ( { {R_a^2} \over {y_{min} R} }  \biggr )\ \biggl ( { {y_{min}} \over {\ell_{typ,B} } }  \biggr )^2 \ \ \ .
\label{eq-FgeepsOFge}
\end{equation}

\noindent
For most circumstances

\begin{equation}
    { {R_a^2} \over {y_{min}\  R} }\ \ll\ 1
\end{equation}

\noindent
and we can neglect (\ref{eq-FgeOR3}) with respect to (\ref{eq-Fge}).  However, in extreme circumstances 
where the fractional concentration of
superconductor becomes very close to 1, the $ { 1 \over {(1-f)} }$ term implies that $F_{ge}^\varepsilon$ could dominate.

\subsection{Weakening and Reversal of the Force of Gravity}
\label{ssec-weakening}

    We rewrite (\ref{eq-G}) to show the dependence of $G$ on $\varepsilon$, the dielectric constant of the ether:

\begin{equation}
    G \ =\  { {e_0^2 R_a^2} \over {6 m_e m_n\ \varepsilon^2 \  y_{min}^2} }
\end{equation}

\noindent
Actually $m_e$ (\ref{eq-me}) and by assumption (\ref{eq-Fgp}) $m_n$ both should vary with $\varepsilon$.  
So we consider the quantity

\begin{equation}
    G m_e m_n \ =\  { {e_0^2 R_a^2} \over {6\ \varepsilon^2 \  y_{min}^2} }
\end{equation}

\noindent
proportional to the gravitational force on representative masses, 
to most effectively represent the weakening of the gravitational force
as $\varepsilon$ changes.  If $y_{min}$ is constant, $G m_e m_n$ will weaken significantly 
in regions where $\varepsilon$ is important, because $\varepsilon$ becomes large near $f \rightarrow 1$.
Thus the gravitational force should weaken compared to expectations when the density of line pairs becomes high, 
for instance near a black
hole---where the gravitational force is expected to be strongest.  Since $G m_e m_n$
varies as $1/\varepsilon^2$, the effects of black
holes could be considerably weaker than expected.  This could contribute to the fact that LIGO\cite{LIGO} has not yet seen
gravitational waves.  On the other hand, the masses of black holes could be considerably larger than measured while
assuming constant $G$.

    As shown by (\ref{eq-FgeepsOFge}) in the limit $f \rightarrow 1$, the $1/{R^3}$ gravitational force 
(\ref{eq-FgeOR3}) could dominate (\ref{eq-Fge}).  
But in this limit both electric terms in (\ref{eq-ftotal}) should be dominated by the repulsive magnetic terms, which are screened but
independent of $\varepsilon$.  We will suggest in Section~\ref{sec-rotations} 
that the reversal of the  gravitational force has indeed been seen,
in a region which should have a {\it very} high density of line pairs.  If this reversal has, indeed, been seen, a reversal should
certainly have occurred in the high densities of the early universe.

\section{Newton's Second Law}
\label{sec-Newton}

    Since we have so far been unable to derive Newton's second law directly from the properties of the ether, we use 
Einstein's Principle of Equivalence of Gravitation and Inertia\cite{equivalence} 
(extended slightly) and consider a positron with acceleration ${\bf a}$.  This acceleration 
is equivalent in all respects to a gravitational field which points {\it opposite} to ${\bf a}$.
(Think of a scale in an elevator.)

\begin{equation}
    - {\bf a}\ is\ equivalent\ to\ {\bf g}\ =\ - { {GM} \over {R^2} } {\hat R}
\end{equation}

\noindent
The latter form applies if the gravitational field is due to star 1.  What we interpret as the gravitational force on the 
positron from star 1, 
we have seen is actually a force of the line pairs from star 1 on the positron. 

\begin{equation}
    {\bf F}_{line\ pairs\ on\ positron}\ =\ + m_e\ {\bf g}
\end{equation}

\noindent
Hence in the equivalent situation of acceleration ${\bf a}$, line pairs exert a force on the positron given by

\begin{equation}
    {\bf F}_{line\ pairs\ on\ positron}\ =\ - m_e\ {\bf a}
\end{equation}

\noindent
The acceleration must be due to some applied force, ${\bf F}_{applied}$.  The forces on the positron must balance:

\begin{equation}
    {\bf F}_{applied}\ +\  {\bf F}_{line\ pairs\ on\ positron}\ =\ 0
\end{equation}

\begin{equation}
    {\bf F}_{applied}\ =\ m_e\ {\bf a}
\end{equation}

\noindent
Hence the ether exerts a force equal to $-m{\bf a}$ on an accelerating mass, leading to Newton's second law of motion.

     For a deeper understanding of Newton's second law, including how it can be made consistent with the discussion of 
Section~\ref{ssec-calcG}, see Appendix~\ref{sec-understandingNewton}.

     As a result of the above discussion we see that the force exerted by the positron on the line pairs of the ether is given by

\begin{equation}
    {\bf F}_{positron\ on\ line\ pairs}\ =\ + m_e\ {\bf a} \ \ \ .
\label{eq-Fonpairs}
\end{equation}

\noindent
This force is exerted on the line pairs that go through the wave function of the positron.  The line pairs that go through the
positron's wave function respond by stretching and deforming (not by accelerating like a massive particle).

\section{Special Relativity}

    By our use of Maxwell's equations we have assumed special relativity from the beginning of this paper.  We point out how
special relativity fits naturally into our construct.

    Special relativity applies to the situation of a particle in the background ether, which is homogeneous and isotropic because
it is far from any masses.  If a photon passes this particle, its velocity is $c$, because of the characteristics of the line pair
(without rest mass) on which it is propagating---independent of the motion of the source.  This observation is 
sufficient\cite{Jacksonspecial} to build the entire special theory of relativity.

    Our concept of mass (\ref{eq-metau0}) fits very naturally into the overall construct, as potential energy of the ether.

    For a discussion of the full meaning of Lorentz invariance, see Appendix~\ref{sec-understandingNewton}.

\section{General Relativity}
\label{sec-generalrel}

    It should be possible to relate the quantized line-pair density of the ether, approximated by ${ {dn} \over {dA d\Omega} }$,
to the metric tensor of 
general relativity in static situations.  This work has not been done but has promise of uniting general relativity with
quantum theory.  The fact that $G$ is not constant must be taken into account.

    We have not yet done a serious job of studying the dynamics of the ether, e.g. using (\ref{eq-Fonpairs}).  
Such a study will be needed for a
complete model.  We note that there is an essential difference between line pairs in the ether and the structure of space-time in
general relativity.  Line pairs are quantized objects which can {\it move}, and are not fixed in space-time.

    We also note here that the concept of gravitational waves used in this paper 
is quite different from the gravitational waves of the
Standard Model\cite{Shutzgravwave}.  Our concept of gravitational waves is appropriate to our quantized model.

\section{Wave-Particle Duality}
\label{sec-wpduality}

    In this model, wave-particle duality is not a problem---it is natural.  The particle is the bare electron.  The wave is the
response of the ether  to the bare particle\cite{MCM}.

\section{Quantum Mechanics}
\label{sec-quantmech}

    Since we started with Dirac's quantization condition (\ref{eq-singv}) it is not surprising that we are consistent 
with quantum mechanics with one exception.  We construct a continuous wave function which models the 
displacement of each line pair in (\ref{eq-psix}) 
in the average rest frame ($\ast$) of the electron:

\begin{equation}
    \psi_{qm}^\ast ({\bf x})\ =\ \biggl ( \sum_j\ (-{\hat r}) \cdot \Delta {\bf y}_j ({\bf x}) \biggr )_{interpolated}
\end{equation}

\noindent
This function is a signed scalar.  
It interpolates between the displacements of the various line pairs to form a continuous function,
as in standard quantum mechanics. But contrary to the expectations of standard quantum mechanics, this wave function does not
spread out with increasing time.  The electron at rest is at equilibrium with the ether---but subject to zitterbewegung which will
manifest itself in slow motion after sufficiently long times.  The whole wave function moves, but does not spread.  
This could be a measureable difference between our model and the standard model of quantum mechanics.

    Thus we see that our electron's wave function is intrinsically localized in position space,
but has a finite spread, and therefore must have some spread
in momentum space.  Hence true plane-wave states do not exist.
Therefore we need to modify our discussion of the quantization of the photon (equation (\ref{eq-pepg}) and 
Figures \ref{fig-wfelectron}-\ref{fig-photonfixedz}) since we used a
plane wave electron for simplicity.  But there is really no essential difference from our previous discussion.  The non-plane-wave
electron state is a superposition of plane wave electrons with various values of $p_e^\mu$ and we have

\begin{equation}
    \oint p_e^\mu dx_\mu\ =\ 0
\end{equation}

\noindent
for each $p_e^\mu$ in the superposition.  This leads to the quantization of the photon (\ref{eq-pepg}) as before.

    We can also obtain the quantization of matter waves by assuming that all matter waves are localized 
(to any required accuracy consistent with the uncertainty principle) in both position and
momentum spaces.  If we simply look on a larger scale at Figure~\ref{fig-photonfixedt} 
and replace the photon by a localized matter wave, then just as for the photon we will have

\begin{equation}
    \oint p_e^\mu dx_\mu\ =\ n h \ \ \ .
\end{equation}

\noindent
where we do not know in advance the precise value of $p_e^\mu$, but presume it will correspond to a value of ${\bf p_e}$ that is
probable according to the momentum space wave function of our particle.  In Figure~\ref{fig-photonfixedt} 
(concentrating on the matter wave instead 
of the photon) we pick a path of integration along the $z$-axis, defined to be parallel to ${\bf p_e}$, 
and integrate over one complete wavelength.  From the fact
that $\ell_{typ}$ (Table~\ref{table-lpdatsurface}) is very small, we can do this to any required accuracy, 
and then turn to integrating over a $y$-leg,
which is perpendicular to the momentum.  We then use a $z$-leg which is far away, well outside the particle's wave function, 
where the contribution to the total integral is negligible.   We then complete the integral with a $y$-leg as before.
Hence we obtain as in Section~\ref{ssec-quantphoton}

\begin{equation}
    p_e\ =\ { h \over \lambda } \ \ \ \ \ \ \ \ \ \   E_e\ =\ h\ \nu \ \ \ .
\end{equation}

    Then using the standard formalism\cite{Messiah} we find the momentum space wave function of our electron

\begin{equation}
    \Phi^\ast ({\bf p})\ =\ { 1 \over {h^{3/2}} }\ \int d^3 x\ \psi_{qm}^\ast ({\bf x})\ exp(-i { {\bf p \cdot x} \over \hbar } )
\end{equation}

\noindent
in its rest frame.  From our previous discussion (Section~\ref{ssec-masselectron}) 
both $\psi_{qm}^\ast$ and $ \Phi^\ast ({\bf p})$ are spherically symmetric, if we sum over
spin states.  Hence the general wave function of the free electron can be written in the standard form, where

\begin{equation}
    \psi_{qm}({\bf x})\ =\ { 1 \over {h^{3/2}} }\ \int d^3 p\ \Phi ({\bf p})\ exp( i { {\bf p \cdot x} \over \hbar } )
\end{equation}

\noindent
where $\Phi ({\bf p})$ is no longer restricted to the rest frame.
The full wave function, including time dependence is given by

\begin{equation}
    \psi_{qm}(\tau,{\bf x})\ =\ { 1 \over {h^{3/2}} }\ \int d^3 p\ \Phi ({\bf p})\ 
    exp(-i { { p_e^\mu x_\mu} \over \hbar } )
\end{equation}

\noindent
where $p_e^\mu$ is the total 4-momentum of the electron.  In position space we have

\begin{equation}
     p_e^\mu\ \rightarrow\ i \hbar\ { {\partial\ } \over {\partial x_\mu} }
\end{equation}

\noindent
just as in the normal formalism.  This leads to Schr{\" o}dinger's equation 
and the standard quantum mechanics of angular momentum.  But until we have a model 
of the magnetic moment of the electron, we cannot fully discuss spin.  (We could, of course, 
just take the value of the magnetic moment of the electron from experiment or the Dirac equation, 
as standard quantum mechanics does.)   For instance for the hydrogen atom, 
an electron in the potential energy of a heavy proton, we obtain 

\begin{equation}
    \biggl [- {\hbar^2 \over {2m_e}} \nabla^2 \ +V(r) \biggr ] \   \psi_{qm} \ 
   =\  i\hbar\ { {\partial \psi_{qm}} \over {\partial t}}\ \ \  
\label{Sch}
\end{equation}

\noindent
the Schr{\" o}dinger equation just like in standard quantum mechanics.  The energy spectra for the two models
are identical.  In both models we have  to add the effects of spin to get the measured spectra. 

\section{Inflation}
\label{sec-inflation}

    If we have succeeded in calculating $G$, the entire concept of the Planck mass is specious because $G$ is not fundamental.
Hence general relativity and the hot big bang model, in particular, need modification.  But we can at least heuristically consider
the topic of inflation in our model by using Weinberg's result\cite{Weinbergevol} for evolution of the universe

\begin{equation}
    3\ { {\ddot R} \over { R } } \ =\ -\ { {4 \pi\ G} \over {c^4} }\  (\rho\ +\ 3 p)
\label{eq-evolutionR}
\end{equation}

\noindent
where $\rho$ is the energy density of the universe, $p$ is the pressure, and $R$ is the cosmic scale factor (sometimes called the
radius of the universe).  We suggest that the twoline potential energy (\ref{eq-twolinepot}) was the energy source for inflation.

    In the early universe, after the line pairs started to condense, we presume that they were radiation dominated, with many
excitations traveling on them, so that both $\rho$ and $p$ were positive.  The line pair density should have been extremely high
so that the fractional concentration of superconductor approached one ($f \rightarrow 1$)
and the dielectric constant of the ether was very large
($\varepsilon \rightarrow \infty$).  Thus the effective charge of the electron approached zero and from (\ref{eq-ftotal}) the gravitational
force was initially {\it repulsive}. 
Therefore, heuristically, we can think of the value of $G$ in (\ref{eq-evolutionR}) as negative and exponential expansion
resulted.

    When the universe expanded to the point where the fractional concentration of superconductor became less than 1,
$\varepsilon$ decreased, the force of gravity became attractive and the exponential expansion stopped, as long as

\begin{equation}
    \rho\ +\ 3 p\ > \ 0 \ \ \ .
\end{equation}

\noindent
This should have been true through the eras of radiation dominance and matter dominance---until fairly recent times.

    Using (\ref{eq-fulldensity}) and (\ref{eq-dndAnum}), we estimate the expansion factor during inflation to be $10^{14}$, 
with a large uncertainty.

\section{Dark Energy}
\label{sec-darkenergy}

\subsection{Explanation of Dark Energy}
\label{ssec-darkenergy}

    Dark energy consists of line pairs which stretch across the universe.  The dark energy density is the energy
density of the ether

\begin{equation}
    \rho_{\Lambda}\ =\ \tau_0\ \biggl ( {{dL} \over {dV}} \biggr )_B\ \ \ \ .
\label{eq-rhoLambda}
\end{equation}

\noindent
Note from (\ref{eq-tensiontau0}) that the energy/length due to the quantization currents has been included in $\tau_0$.

    Dark energy is completely dark.  The only way it can be detected directly is by 
its gravitational effects, such as changes in a mass moving through it.
The dark energy density decreases with time as the universe expands and the  
tension of the all the line pairs in the  universe decreases slowly.

\subsection{Pressure of Dark Energy}
\label{ssec-tensiondarkenergy}

\subsubsection{Longitudinal Tension}

    The pressure due to the longitudinal tension is given by

\begin{equation}
 -p_L\ = \ { {\tau_0 } \over {\pi R_a^2} }\  =\ { 1 \over {\pi R_a^2} }{ {dU} \over {d\ell} }\ =\ \rho_{\Lambda}\ \ \ .
\end{equation}

\noindent
So for the longitudinal tension alone we would have the ratio of pressure to energy density given by

\begin{equation}
    w_{longitudinal}\ =\ { p_L \over \rho_{\Lambda}}\ =\ -1\ \ \ \ .
\end{equation}

\noindent
If there were no transverse tension, we would take the average pressure over three directions and find

\begin{equation}
    w_{no\ transverse\ tension}\ =\ -{ 1 \over 3}\ \ \ \ .
\end{equation}

\subsubsection{Transverse Tension}

    But there is tension in one of the transverse directions.  We can easily calculate the transverse tension which the 
quantization currents need to balance in order to keep $\tau_0$ constant,  for a length $\ell$ of a line pair.  
If $\rho^\prime$ locates the center of line 2 from line 1 (Section~\ref{sec-two}), this tension is

\begin{equation}
\tau_T\ =\ { {\partial ((\tau_0-\tau_{qc}) \ell) } \over {\partial \rho^\prime} }\ 
=\ { {g_0^2} \over {a^2} } {\ell \over a } u_q K_0(u_q)\ \ \
.
\end{equation}

\noindent
The quantization currents could provide this transverse tension simply by flowing in opposite directions on the outer surfaces of
the line pair---or they could communicate with other line pairs as shown in Figure~\ref{fig-currentsforapart}.  
In the latter case the transverse tension would be transmitted to other line pairs and
 
\begin{equation}
    w_{total\ ether}\ =\ { p \over \rho_{\Lambda}}\ <\ -{ 1 \over 3}\ \ \ \ .
\end{equation}

\noindent
implying that $ \rho\ +\ 3 p\ < \ 0 $ for the ether and from (\ref{eq-evolutionR})
the ether can contribute to the current acceleration of the universe.  
Each line pair in this case is under tension in two directions.

\subsection{Estimate of the Line-Pair Density of the Background Ether}
\label{ssec-estimatelpd}

     We use (\ref{eq-me}), (\ref{eq-G}), and (\ref{eq-rhoLambda}) to estimate the line-pair density of the background ether 
in terms of the dark energy density, the mass of the electron, and Newton's constant $G$

\begin{equation}
  \biggl ( {{dn} \over {dA}} \biggr )_B^3\ =\ {{\rho_{\Lambda}} \over {d_0}} \   
  { 2700 \over {16\pi} } \ { {m_e c^2} \over {(6 G m_e m_n)^2 }} \ \ \ .
\end{equation}

\noindent
We have taken $\rho_{\Lambda}$ from the WMAP collaboration\cite{WMAP}

\begin{equation}
\rho_{\Lambda}\ =\ n_b \times (1 GeV/baryon) \times { {\Omega_\Lambda} \over {\Omega_{bar}}}\ =\ 4.15 { {GeV} \over {m^3} } \ \ \ .
\end{equation}

\noindent
In order to obtain an estimate for the properties of the background ether, we guess that $y_{min} \simeq 10\ \ell_{typ,B}$ so that

\begin{equation}
{ 1 \over {y_{min}^2} }\ \simeq\ { 1 \over {100} } \biggl ( { {dn} \over {dA} } \biggr )_B \ \ \ .
\label{eq-yminguess}
\end{equation}

\noindent
We take $d_0$ to be the mean free path in air\cite{ResHal}

\begin{equation}
    d_0\ \simeq\ 10^{-7} m \ \ \ .
\end{equation}

\noindent
We obtain

\begin{equation}
    \biggl ( {{dn} \over {dA}} \biggr )_B \ \simeq \ 4.3 \times 10^{39} {{line\ pairs} \over {m^2} } \ \ \ 
\label{eq-fulldensity}
\end{equation}

\noindent
consistent with our minumum value from Table~\ref{table-lpdatEarth}.
We note that the line pair density of the background ether is of order 20 times the line-pair density of the sun at its surface.
From (\ref{eq-dLdV}) we obtain $\ell_{ave} \simeq 10^{40} m$.

\subsection{Estimated Parameters of the Background Ether}
\label{ssec-parameters}

    Using (\ref{eq-rhoLambda}) and (\ref{eq-fulldensity}) we find the line-pair tension of the universe at the current time:

\begin{equation}
    \tau_0\ =\ { {\rho_\Lambda} \over { (dn/dA)_B } } \ =\ 1.5 \times 10^{-49} N \ \ \ .
\end{equation}

\noindent
Then again using

\begin{equation}
    y_{min}\ \simeq \ 10\ \ell_{typ,B} \ =\ 1.5 \times 10^{-19} m
\label{eq-ymintyp}
\end{equation}

\noindent
we find from (\ref{eq-Rasmall})

\begin{equation}
    R_a \simeq\ 8 \times 10^{-39} m \ \ \ .
\label{eq-Ra}
\end{equation}

\noindent
We note that our guess for $y_{min}$ (\ref{eq-yminguess}) and (\ref{eq-ymintyp}) 
looks reasonable since $y_{min}$ must be $<10^{-18} m$, or it would 
affect the measured pointlike structure of the electron.

    To estimate the fractional superconducting concentration of the background ether, we consider the fractional concentration
in one volume

\begin{equation}
    f_B\ =\ \biggl ( { {dL} \over {dV} } \biggr )_B \ \pi \ R_a^2 \ =\ 8.3 \times 10^{-37}
\label{eq-fB}
\end{equation}

\noindent
and find it to be negligible.  So the dielectric constant of the ether is 1 near the Earth and in all circumstances except the most
extreme.

\section{Dark Matter}

    Since this model predicts the weakening of the gravitational force near strong gravitational sources because of the increasing
dielectric constant of the ether, in principle it could account for dark matter.  But the very low fractional concentration of
superconductor in the background ether from (\ref{eq-fB}) (which should be similar in contribution to stars in our galaxy) 
appears to rule this out. 

\section{Rotations and AGN Jets}
\label{sec-rotations}

    We need to discuss what happens to the line pairs, which begin in a star, when it rotates.  If a line pair points along the
axis of rotation, nothing happens.  If it is off axis pointing parallel to the axis, it will twist about the axis of rotation.  But
this is no problem since line pairs can move through each other.  If a line pair is off axis and pointing perpendicular to the
axis, it will stretch and contract as the star rotates.  This is no problem since the first 
line pair is matched with another line pair
pointing in the opposite direction which stretches when the first contracts and vice versa.  Hence no net work is done by the star
in rotating

    However, as a star rotates, each bit of mass $m$ does exert a force on the line pairs in its wave function given by 
(\ref{eq-Fonpairs})

\begin{equation}
    {\bf F}_{m\ on\ line\ pairs}\ =\ -\ m\ \omega^2 {\bm \rho} \ \ \ 
\end{equation}

\noindent
(where ${\bm \rho}$ is the cylindrical radius).
Thus in extreme cases, line pairs will be concentrated on the axis of a spinning mass.  Hence we achieve a natural qualitative
understanding of AGN (Active Galactic Nuclei) Jets in this model.  The M87 galaxy (in Table~\ref{table-lpdatsurface}) 
has an AGN/black-hole at it's center
of a mass $6\times 10^{39} kg$, which sends out about $4 \times 10^{66}$ line pairs.  Many of these line pairs become concentrated 
on the axis of rotation, apparently sufficiently to reverse the force of gravity, producing jets.  The velocities in these jets are 
high\cite{Meliarot}, because the particles involved have very low mass, according to (\ref{eq-me}) because $e = e_0/\varepsilon$ 
becomes small.  (Actually if the force of gravity reverses, the dominant contributions to the mass (\ref{eq-metau0}) 
from (\ref{eq-ftotal}) are repulsive, but we can still expect the mass to be small.)

\section{The Pauli Exclusion Principle}
\label{sec-Pauli}

    We have tied a string onto every electron in the universe, therfore electrons are {\it distinguishable}.  This removes the
theoretical underpinnings of the exclusion principle, which apparently Pauli never believed\cite{YangPauli} anyway.  But something
must replace these underpinnings.  

    We suggest that the pole force of the electron does part of the job.  The pole has the strength of a Dirac monopole (if 
$\varepsilon=1$ near the Earth).  Screening currents tend to amplify the strength of the pole if you are in {\it contact} with the
electron, i.e. inside the screening currents.  So there should be {\it very} strong forces between electrons (i.e, any objects)
in contact.  We suggest the pole field of the electron is responsible for the impenetrability of
matter.  If you push down on a table, we suggest that the force which pushes back on you is the contact force of the pole fields of
the electrons in the table.  (Hence our lines sense pressure only at the ends.)

    We postulate that the rest of the job is done by the line forces between electrons.  The line force has infinite range, 
as does the spin glass spin-spin correlation length\cite{spinglass}.  We also suggest that the line force 
is responsible for the {\it exchange} force of ferromagnetism.  

    Elliot Lieb has said\cite{Lieb} that the reason the Pauli Exclusion Principle is not a force is that you cannot write it in
Hamiltonian form.  So calling it a force is perhaps appropriate for this classical model.

\section{Superconductivity}
\label{sec-superconductivity}

    The line force between two electrons provides a strong attractive force between separated electrons of opposite spin.  
Apparently any attractive
force between electrons can lead to Cooper pairs\cite{Cooper}, so we expect the line forces between two electrons to contribute to
the formation of Cooper pairs, and hence that form of superconductivity.  We note that line pairing should show an isotope effect
since nuclei are involved.  The isotope effect played a strong role in establishing the original Cooper pair mechanism.

    The quantum of flux in a Cooper pair system is $\Phi_0/2$, because the charge is due to two electrons.  We have to explain why
flux quanta of  $\Phi_0/2$ can exist in Type II superconductors, when only flux quanta of $\Phi_0$ are allowed by the ether.
We note that $\Phi_0/2$ is a point of unstable equilibrium for line pairs in the ether.  
If the curl of the phase field is filamentary, the ether currents don't know which way to
change. i.e. whether to bring the flux to $0$ or $\Phi_0$  We postulate that this position of unstable equilibrium 
with the ether currents is enough to let the Cooper pair currents keep the flux steady at $\Phi_0/2$.

    It is interesting to note that if line forces between electrons contribute strongly to the formation of Cooper pairs, then
line pairs are pretty much the same thing as Cooper pairs in a superconductor---and hence appropriate to the generation of 
mass\cite{AitchisonHey} according to some authors.

\section{The Quantum Hall Effects}
\label{sec-quantumHall}

    The integer quantum Hall effect can be taken\cite{KvK} as direct evidence for the quantization of magnetic flux in units of
$\Phi_0$.  In the fractional quantum Hall effect, it is suggested\cite{quantumhall} 
that composite fermions are formed when one electron
binds an even number of flux quanta ($\Phi_0$).  The fractional quantum Hall effect is then explained when these composite fermions 
undergo the integer quantum Hall effect, as if they were electrons.  The currents needed to contain these flux quanta are not 
obviously available, so we
suggest that the quantum Hall effects present direct evidence for our unpaired lines of flux (Section~\ref{ssec-static}).  
We suggest that the flux is contained by the superconducting fields which we postulate in Section~\ref{ssec-superconfield}.  
The force that binds the electron to the lines of flux is then given by (\ref{eq-ftotal}).  
We have seen that this same force on line pairs is the force of gravity.

\section{Experimental Comparison with the Standard Model}
\label{sec-relevantexperiments}

     We obtain Schr{\" o}dinger's equation for the hydrogen atom, and by implication for heavier atoms.  We 
obtain Newton's Second Law of motion for the electron, and by assumption for more massive objects.   Our
model of gravity agrees with Newtonian gravity within our solar system.  Therefore differences 
between the predictions of this model and the Standard Model should be hard to find.  We focus here on 
experiments which could show a difference. 

\subsection{Existing Experiments}
\label{ssec-existing}

    We show in Table~\ref{table-relevantexperiments} relevant experimental results and indicate 
if the Standard Model and this model are consistent with these experiments.

\begin{enumerate}

\item search for flux quantization outside Cooper pair superconductors:

    The most relevant experiment to this model is a search for flux quantization performed by 
Tonomura {\it et. al.}\cite{Tonomurafluxquant}.  They conclude that magnetic flux is not quantized in a torroidal ferromagnet.
If this result is correct, it rules out this model, which is based on flux quantization.  This experiment is one in a series of
beautiful electron holography experiments that has studied the Aharanov-Bohm effect in fine detail.  

    However, we suggest that this group of experiments cannot rule out this model.  The reason is that this model changes the
nature of the experimenter's probes.  We attach a line of flux onto each of their electrons.  Examination of their
methods\cite{Tonomuraholography} shows that they did not pay attention to the spins of their probe electrons.  
Consequently the lines of flux attached to their probe electrons may have spoiled their results.  Their geometry in dealing
with normal samples was likely different from that which they used for measurements with superconductors.  

    The above argument applies to any search for flux quantization which uses electrons.
    
    We consider the argument for flux quantization in Section~\ref{ssec-FQ} to be very strong.  We also note that flux quantization has
apparently been observed outside of superconductors in the quantum Hall effects.

\item  g-2 of the electron:

    We list in Table~\ref{table-relevantexperiments} the measurement of $g-2$ of the electron, 
for which the Standard Model has achieved very high precision agreement.  So far, there is no prediction for this model, 
because we do not have a model of the magnetic moment of the electron.

\item dark energy:

    Measurements of dark energy are pretty clearly not consistent with the Standard Model, but are consistent with this model of
the ether (Section~\ref{sec-darkenergy}).

\item  inflation:

    Inflation in the Standard Model requires a scalar field, which has never been observed, plus reheating\cite{Kolb} after the
scalar field does its job.  This model, at least heuristically (Section~\ref{sec-inflation}), proceeds from the string era naturally, 
with no new requirements, since the strings are still here.

\item  AGN jets:

    This model, at least qualitatively (Section~\ref{sec-rotations}), explains why the rotation axis of a black hole is special, 
why AGN jets form, and why they have very high velocities.  Explanations using the Standard Model are doubtful.

\item  spin-spin correlation length

    The spin-spin correlation length in spin glasses is infinite\cite{spinglass}.  Every spin couples equally with every other
spin regardless of how far they are apart.  This is widely considered to be ridiculous, but it is not taken as a conflict with the
Standard Model, because spin glasses are very complicated systems.  In our model an infinite correlation length is natural.
We suggest that these complicated systems which disagree with the Standard Model show a pattern.

\item  fractional quantum Hall effect:

    The fractional quantum Hall effect is another case where the Standard Model is in apparent disagreement with experiment, but
the disagreement is not taken very seriously because solids are complicated systems.  Our model has a natural interpretation of
this effect (Section~\ref{sec-quantumHall}).

\end{enumerate}

\begin{table*}
\caption{Relevant Experiments:  Are the Models Consistent with Experiment?}
\begin{tabbing}
\=experiment\ \ \ \ \ \ \ \ \ \ \ \ \ \ \ \ \ \ \ \ \ \ \ \
     \=Standard Model\ \ \ \ \ \ \ \ \ \ \ \ \ \ \ \ \ \ \ \ \ \ \ \ 
     \=this model\ \ \ \ \ \ \ \ \ \ \ \ \ \ \ \ \ \ \ \ \ \ \ \ \ \ \\
\\
\>flux quantization search\cite{Tonomurafluxquant}      \>Yes     \>No but see Section~\ref{ssec-existing} \\
\>g-2 of the electron      \>Yes      \>no prediction \\
\>dark energy     \>No      \>Yes     \\
\>inflation       \>requires scalar field, reheating      \>Yes     \\
\>AGN jets        \>doubtful                               \>Yes     \\
\>spin-spin correlation length         \>No,  but complicated      \>Yes     \\
\>fractional QH effect       \>No,  but complicated      \>Yes     \\
\end{tabbing}
\label{table-relevantexperiments}
\end{table*}

\subsection{Suggested Experiments}

     In addition we suggest the following experiments to measure a difference between this
model and the Standard Model:

\begin{enumerate}

\item Find the line:

    This model predicts that charged particles come with an attached line.  In accelerators presumably the pair to the beam
particle (with which the beam particle forms a line pair), because of energy considerations, 
is among the image charge in the beam pipe.  The stability of the beam
should be strongly affected by the image charge in the beam pipe.  Apparently this is true.  If a short section of ceramic beam
pipe were to be used, the beam would probably be lost.\cite{Sands}   It may be possible to directly observe the line in
accelerators.

\item Measure G  accurately:

     Look for a material dependence (of order a part in $10^3$) of the measured values of $G$.  See Appendix~\ref{sec-rigorousgravity}.

\item Study the Pomeron:

    If this model is correct charged particles are not isolated and unitarity will be violated.  In light of this possibility, the
rising total cross section is interesting.  This rise may be associated with the Pomeron.

\item Study the spreading of the electron wave function:

     This model predicts that the wave function of an isolated electron will not spread.

\item Look for the pole of the electron:  

     From (\ref{eq-ymintyp}) we see that measurements down to a scale of $10^{-20} m$ may prove sufficient.
Experiments able to make measurements on this scale may be able to measure the screening function $s$ in (\ref{eq-ftotal}).

\end{enumerate}

\section{Remaining Problems of This Model}

    We collect here the remaining problems that we anticipate to be the most serious for this model:

\begin{enumerate}

\item The disagreement with the flux quantization experiments of Tonomura {\it et al.} is discussesd in Section~\ref{ssec-existing}.

\item The violation of the theoretical underpinnings of the Pauli Exclusion Principle is discussed in Section~\ref{sec-Pauli}.

\item With the size of the outer radius of a line pair estimated by equation (\ref{eq-Ra}) the tension in an unpaired line 
(\ref{eq-T}) is ridiculously high.  This has observable consequences.  Other numbers appear to be fairly reasonable.  
If this model is correct, a way must be found to fix this problem.  This problem seems similar in severity to a well known
problem in the Standard Model---the energy in the electric field of a point charge is infinite.

\end{enumerate}

\section{Conclusions}

    The bare electron is modeled as the end of a superconducting line of magnetic flux.  
Electric charge must be added by hand to this bare electron, 
but with the correct charge the electron also has the correct spin, $\hbar/2$.  So the bare electron is almost a dyon, but with a 
spin-dependent magnetic charge.
If the dielectric constant of the ether is 1 near the Earth, the effective magnetic charge of the electron 
has the strength of a Dirac monopole, but is strongly screened by the ether---superconducting line pairs from all the other
electrons in the universe.
The bare electron interacts with the ether and undergoes zitterbewegung, very rapid random walk motion, traveling near the
speed of light in its average rest frame.
The wave function of the electron is the response of the ether to the average 
electromagnetic field of the bare electron, explaining wave-particle duality in a natural way.
Quantum mechanics results from this wave nature.  The mass of
the electron is the potential energy of its wave function (in its rest frame).
The charge and mass of the electron are predicted to vary with the dielectric constant of the ether.  This may affect the
interpretation of the cosmological red shift.

    The photon is modeled as a helical wave, one wavelength long, traveling on a line pair.  The spin of the helical photon is
calculated to be $\hbar$.

    The ether is a mesh of superconducting line pairs, of a new type which do not conduct electrons or Cooper pairs.  
These line pairs are under tension, so could represent dark energy.  In general,
quantization conditions arise because the ether is a multiply connected superconductor.  

    Using a toy version of our model with measured values of Newton's constant $G$, the mass of the electron and the dark 
energy density of the universe, we estimate the line pair density near the Earth to be $4 \times 10^{39} line\ pairs/m^2$.
This is a factor of $10^{14}$ larger than would be expected without inflation.

    The gravitational force exerted by a star on a positron is due to the electromagnetic field of the positron interacting with
the superconducting line pairs coming from the star.  This force is independent of the sign of the positron's charge because it is
{\it induced}.  The electric force is attractive but the magnetic force is repulsive.  Hence the gravitational force is predicted
to weaken, then reverse, as the dielectric constant of the ether becomes large.

    The ether exerts a force equal to $-m{\bf a}$ on an accelerating mass---which leads to Newton's second law of motion.

    We have proven that both the principles of gauge invariance and microscopic causality are violated by the superconducting
fields of this model.  Because this model is acausal, it is also nonlocal and even self interactions are finite.

    We suggest that the Pauli Exclusion Principle is actually due to a 
very strong spin-dependent magnetic force: the pole force and line
force between electrons.  We suggest that the pole force, a contact force, is responsible for the impenetrability of matter.

\begin{acknowledgements}

    I thank my colleagues at Stony Brook for helpful discussions.  This includes Chris Quigg who introduced me to the Dirac string
long ago.  Any mistakes, of course, are my own.  This work was supported by the National Science Foundation. 
\end{acknowledgements}

\appendix
\section{Electromagnetic Field of the Moving Line}
\label{sec-EMmovingline}

\subsection{The Instantaneous Radius of Curvature}
\label{ssec-instradius}

     It is useful to define the angle $\theta$ so that $tan(\theta)$ gives the instantaneous slope of the string at any point
for solution (\ref{eq-yzline}) moving in the $y-z$ plane.

\begin{equation}
     \hat{\ell}\ =\ \hat{y} sin(\theta)+\hat{z} cos(\theta)\ =\ \hat{y} \ell_y +\hat{z} \ell_z
\end{equation}

\noindent
Here $\hat{\ell}$ is the direction of the magnetic field (\ref{eq-Bfieldline}) at ${\bf x}$, the point in question.  Also $\hat{\ell}$ is the
direction of the centerline at the relevant point (${\bf y}={\bf x}-{\bm \rho}$).  (For simplicity, we assume 
here that this point is unique as will be true in the case of sufficiently large radius of curvature.)  

     We follow the discussion of Section~\ref{ssec-supercurrent} in seeking the derivatives of $\hat \ell$ with respect to 
position and time at the field point
${\bf x}$.  This unit vector gives the direction of the magnetic field at ${\bf x}$ and is assumed to be parallel transported from
point ${\bf y}$ on the centerline to ${\bf x}$ for this purpose.  Note that the derivatives of ${\hat \ell}$ will be {\it
different} at point ${\bf x}$ than on the centerline (point {\bf y}) because the radius of curvature will in general be different
for a curving line.

     If $\hat{\ell}$ changes, the
direction of its change must be perpendicular to itself:

\begin{equation}
 d \hat{\ell}\ =\ d\theta [\hat{y} cos(\theta)-\hat{z} sin(\theta)]\ =\ 
d\theta[{\hat y} \ell_z-{\hat z} \ell_y ]\ =\ 
d\theta\ \hat{\theta}
\end{equation}

\noindent
We define the instantaneous radius of curvature $R$ by the equation

\begin{equation}
 d \theta\ =\ -\  d\ell/R
\end{equation}

\noindent
where we define the radius to be {\it negative} if the change in $\theta$ is positive, assuming positive $d\ell$.
We make this choice of sign because a line with decreasing $\theta$ in $y-z$ motion matches the configuration of helical 
motion---and it is convenient to consider this radius of curvature positive.

    Since $\ell$ changes only in the $\hat{\ell}$ direction, we have

\begin{equation}
d{\hat{\ell}}\ =\ {(-\hat{\theta}) \over R} {\hat{\ell}} {\bf{\cdot}} {\bf{dx}}\ \ \ .
\end{equation}

\noindent
Hence we can evaluate the partial derivatives of the components of $\hat{\ell}$, for 
solution (\ref{eq-yzline}) motion in the $y-z$ plane:

\begin{equation}
{{\partial \ell_z} \over {\partial z}}\ =\ { {+\ell_y \ell_z} \over R }
\end{equation}

\begin{equation}
{{\partial \ell_z} \over {\partial y}}\ =\ { {+{\ell_y}^2} \over R }
\end{equation}

\begin{equation}
{{\partial \ell_y} \over {\partial z}}\ =\ { {-{\ell_z}^2} \over R }
\end{equation}

\begin{equation}
{{\partial \ell_y} \over {\partial y}}\ =\ { {-\ell_y \ell_z} \over R }
\end{equation}

    Since ${{\hat \ell}^\ast}$ does not change in the rest frame of the string, we use Section~\ref{ssec-supercurrent} to find
the partial derivatives of the components of ${\hat \ell}$ with respect to time.

\begin{equation}
{{\partial \ell_z} \over {\partial \tau}}\ =\ {-{\ell_y^2 \beta} \over R }
\end{equation}

\begin{equation}
{{\partial \ell_y} \over {\partial \tau}}\ =\ { {+\ell_y \ell_z \beta} \over R }
\end{equation}

     At a given point in time the instantaneous radius of curvature changes with the 
change in the position of
the field point ${\bf x}$ according to the equation

\begin{equation}
dR\ =\ +{\hat \theta} \cdot d{\bf x}
\end{equation}

\noindent
because at a given time the center of curvature is fixed,
and therefore for solution (\ref{eq-yzline}) the partial derivatives with respect to $y$ and $z$ are
given by

\begin{equation}
{{\partial R } \over {\partial y}}\ =\ +\ell_z
\end{equation}

\begin{equation}
{{\partial R } \over {\partial z}}\ =\ -\ell_y\ \ \ 
\end{equation}

\noindent
and the partial derivative with respect to $\tau$ is

\begin{equation}
{{\partial R } \over {\partial \tau}}\ =\ -\ \ell_z \beta\ \ \ .
\end{equation}

    From (\ref{eq-lhat}) we can find general expressions for the direction of the change with decreasing $\theta$ (hence
$-{\hat \theta}$, defining positive $R$) and the expression for $R_0$, the radius of curvature of the centerline.
Taking the differential of (\ref{eq-lhat}) with respect to $z$ we find

\begin{equation}
d{\hat \ell}\ =\ \biggl ( { {{\hat y} + {\hat z} F^\prime } \over {\sqrt{1 + (F^\prime)^2} } }  \biggr ) \  
\biggl ( { {F^{\prime\prime}} \over {\sqrt{1 + (F^\prime)^2 } } } \biggr )\ 
 { {dz} \over {\sqrt{1 + (F^\prime)^2 } } } 
\end{equation}

\begin{equation}
-{\hat \theta}\ =\ {{-({{\hat y} + {\hat z} F^\prime })} \over {\sqrt{1 + (F^\prime)^2} } }
\ =\ -({\hat y} \ell_z-{\hat z} \ell_y) 
\end{equation}

\begin{equation}
   { 1 \over {R_0} }\ =\ { {-{F^{\prime\prime}}} \over {\sqrt{1 + (F^\prime)^2 } } } 
\end{equation}

\noindent
and so noting

\begin{equation}
 { {dz} \over {\sqrt{1 + (F^\prime)^2 } } }\ =\ {\hat \ell} \cdot {d{\bf x}}
\end{equation}

\noindent
we find

\begin{equation}
d{\hat \ell}\ =\ { {(-{\hat \theta})} \over {R_0} } {\hat \ell} \cdot {d{\bf x}}
\end{equation}

\noindent
for a displacement in the $z$-direction along the centerline.  We note that the $y$-component of $(-{\hat \theta})$
is always negative, but that the radius of curvature in $y-z$ motion can change dynamically between positive and 
negative values.

\subsection{Relation between the Radius of Curvature and the Shape of the Trajectory}
\label{ssec-radiusandshape}

     If we consider the position of the centerline as a function of $z$ (as an independent
variable---{\it not} a field coordinate) we can relate the radius of curvature on the centerline
to the shape (second derivative) of the trajectory:

\begin{equation}
     tan\theta\ =\ {dy \over dz}
\end{equation}  

\begin{equation}
    { {d(tan\theta)} \over dz }\ =\ { 1 \over {cos^2\theta} } { {d\theta} \over {dz} }\ =\ { {d^2y} \over {dz^2}}
\end{equation}

\begin{equation}
   { 1 \over {R_0} }\ =\ -\ { d\theta \over d\ell }\ =\ -\ \ell_Z^3 { {d^2y} \over {dz^2}}
\end{equation}

\noindent
We use this equation to show how the radius of curvature is transformed in going from the lab to
the line's rest frame:

\begin{equation}
     R\ =\ \gamma R^\ast
\end{equation}

\subsection{General Expression for the Curl of ${\hat \ell}$ }
\label{ssec-generalcurl}

\begin{figure}
    \vbox to 2.5in{
\includegraphics{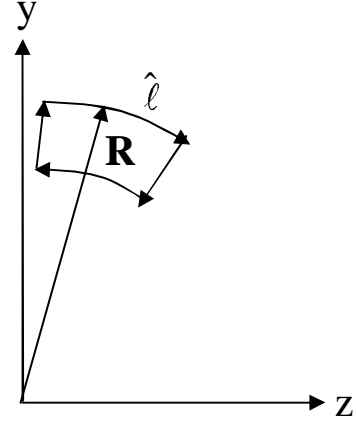}  }
    \caption{\small Path integral for the calculation of $\nabla \times {\hat \ell}$.}
\label{fig-curl}
\end{figure}

    We use a polar coordinate system with origin at the center of curvature to find ${\bf \nabla} \times {\hat \ell}$.
From the definition of the curl

\begin{equation}
    {\bf \nabla} \times {\hat \ell}\ =\ { { {\hat n} \oint_C {\hat \ell} \cdot d{\bf x} } \over {A_C} }
\end{equation}

\noindent
where ${\hat n}$ is the normal (given by the right hand rule) to the circulation of ${\hat \ell}$ around the
closed path $C$, divided by the area enclosed.  (See Figure~\ref{fig-curl}.)  We find

\begin{equation}
    {\bf \nabla} \times {\hat \ell}\ =\ { { {\hat n}\ \big [ (R+dR)d\phi - R d\phi \big ] } \over {dR\ (R+dR/2)\ d\phi} }
    \ =\ { { {\hat n}\ dR\ d\phi } \over {dR\ R\ d\phi} }
\end{equation}

\begin{equation}
    {\bf \nabla} \times {\hat \ell}\ =\ { {\hat n} \over R}
\label{eq-curl}
\end{equation}

\noindent
where ${\hat n}={\hat x}$ in the case of Figure~\ref{fig-curl}.

\subsection{Common Mode Geometry}
\label{ssec-commonmodegeom}

    In common mode geometry Appendix~\ref{ssec-instradius} refers to ${\hat \ell}_1$ which points generally in the $+{\hat z}$-direction.
We need to discuss how ${\hat \ell}_2$ differs, since ${\hat \ell}_2\ =\ -{\hat \ell}_1$.  From (\ref{eq-lhat}) we find

\begin{equation}
    {\hat \ell}_2\ =\ { {-{\hat z}\ +\ {\hat y} F^\prime } \over {\sqrt{1 + (F^\prime)^2} } }
\end{equation}

\begin{equation}
d{\hat \ell}_2\ =\ -\ \biggl ( { {{\hat y} + {\hat z} F^\prime } \over {\sqrt{1 + (F^\prime)^2} } }  \biggr ) \
\biggl ( { {F^{\prime\prime}} \over {\sqrt{1 + (F^\prime)^2 } } } \biggr )\
 { {dz} \over {\sqrt{1 + (F^\prime)^2 } } }
\end{equation}

\begin{equation}
d{\hat \ell}_2\ =\ +\ \biggl ( { {{\hat y} + {\hat z} F^\prime } \over {\sqrt{1 + (F^\prime)^2} } }  \biggr ) \
\biggl ( { {F^{\prime\prime}} \over {\sqrt{1 + (F^\prime)^2 } } } \biggr )\
 { {\hat \ell}_2 \cdot d{\bf x} }
\end{equation}

\begin{equation}
=\ -\ ({\hat y}\ell_{1z}-{\hat z} \ell_{1y} )\  
\biggl ( { -{F^{\prime\prime}} \over {\sqrt{1 + (F^\prime)^2 } } } \biggr )\
 { {\hat \ell}_2 \cdot d{\bf x} }
\end{equation}

\begin{equation}
=\ +\ ({\hat y}\ell_{2z}-{\hat z} \ell_{2y} )\
    { 1 \over {R_0} }
 { {\hat \ell}_2 \cdot d{\bf x} }
\end{equation}

\begin{equation}
d{\hat \ell}_2\ =\ 
    (+{\hat \theta}_2)\  { 1 \over {R_0} }\  { {\hat \ell}_2 \cdot d{\bf x} }
\end{equation}

\noindent
and all the spatial first derivatives are reversed (when referenced to a fixed coordinate system, as must be true)
compared to the case of ${\hat \ell}_1$ and expressed in terms of ${\hat \ell}_1$ or ${\hat \ell}_2$ as appropriate.
But the radius of curvature on the centerline $R_0$ (and by extension $R$) is the same for both lines.

\subsection{Helical Mode Geometry}
\label{ssec-helicalmodegeom}

    For helical geometry from (\ref{eq-lhat1}) and (\ref{eq-lhat2}) we find, in the rest frame of {\it both} lines

\begin{equation}
    d{{\hat \ell}^\ast}_1\ =\  { {kd_0\ d{ {\hat x}^\ast } } \over {\sqrt{1 + (kd_0)^2} } } 
    \ =\ (-{ {\hat y}^\ast} ) \ { {k^2 d_0\  dz} \over {\sqrt{1 + (kd_0)^2} } }
\end{equation}

\begin{equation}
    d{{\hat \ell}^\ast}_1\ =\  (-{ {\hat y}^\ast} )\ k^2 d_0\ { {{\hat \ell}^\ast}_1 } \cdot d{\bf x}\ \ \ .
\end{equation}

\noindent
So for line 1

\begin{equation}
    { 1 \over {{R_0}^\ast} }\ =\ k^2 d_0
\end{equation}

\noindent
the radius of curvature is a constant on the centerline.
In the rotating rest frame we use the time of an observer at the center of rotation 
who is at rest in the lab.  The derivatives of ${{\hat \ell}^\ast}_1$ can be calculated from

\begin{equation}
    d{{\hat \ell}^\ast}_1\ =\  (-{ {\hat y}^\ast} )\ { 1 \over {R^\ast_1} }\ { {{\hat \ell}^\ast}_1 } \cdot d{\bf x}
\end{equation}

\noindent
but if we only need the curl, we can use (\ref{eq-curl}).

    For line 2

\begin{equation}
    d{{\hat \ell}^\ast}_2\ =\  { {kd_0\ d{ {\hat x}^\ast } } \over {\sqrt{1 + (kd_0)^2} } }
    \ =\ (-{ {\hat y}^\ast} ) \ { {k^2 d_0\  dz} \over {\sqrt{1 + (kd_0)^2} } }
\end{equation}

\begin{equation}
    d{{\hat \ell}^\ast}_2\ =\  (+{ {\hat y}^\ast} )\ k^2 d_0\ { {{\hat \ell}^\ast}_2 } \cdot d{\bf x}
\end{equation}

\noindent
and we see that the radius of curvature on the centerline is also a constant for line 2, with the same
constant value of ${R_0}^\ast$ as for line 1.  The derivatives of ${{\hat \ell}^\ast}_2$ can be calculated from

\begin{equation}
    d{{\hat \ell}^\ast}_2\ =\  (+{ {\hat y}^\ast} )\ { 1 \over {R^\ast_2} }\ { {{\hat \ell}^\ast}_2 } \cdot d{\bf x} \ \ \ .
\end{equation}

\subsection{Calculations of the Electromagnetic Field}
\label{ssec-calcEMfield}

    In order to calculate the electromagnetic field from (\ref{eq-supercurrentyz}), we need
to make use of the fact that $\Delta y$ and $\Delta z$ are related by the fact that
${\bm \rho}$ is a cylindrical radius.  Selecting events which are
simultaneous in the lab to define ${\hat \ell}$ and ${\bm \rho}$, 
and requiring ${\hat \ell}^\ast \cdot {\bm \rho}^\ast=0$ in the rest frame
of the line, we find

\begin{equation}
\gamma^2 \ell_y \Delta y\ +\ \ell_z \Delta z\ =\ 0\ \ \ .
\end{equation}

\noindent
Using this relationship and (\ref{eq-supercurrentyz}) in (\ref{eq-generalsol}) we find for the electromagnetic field:

\begin{displaymath}
{\bf E}(\tau,{\bf x})= {{-\gamma \beta 2g_0} \over {a^2}}{\hat x} \Biggl\{
\ell_z K_0(\rho^\ast/a) \bigl\{ 1 - {a^2 \over R^2}  (1 - 3\ell_y^2\beta^2 \bigr\}
\end{displaymath}

\begin{equation}
-\ {{K_1(\rho^\ast/a)} \over \rho^\ast} {a \over R} \gamma^2 \Delta y (1-2{\ell_y}^2\beta^2) \Biggr\}
\end{equation}

\begin{displaymath}
{\bf B}(\tau,{\bf x})={{\gamma 2g_0} \over {a^2}} \Biggl\{ 
K_0(\rho^\ast/a)\hat \ell \bigl\{ 1 - {a^2 \over R^2}(1-\ell_y^2\beta^2) \bigr\}
\end{displaymath}

\begin{displaymath}
+{K_1(\rho^\ast/a) \over {\rho^\ast}} {a \over R} 
\bigl\{ \hat{y}\Delta z[(\ell_z^2-\ell_y^2)\beta^2+1] - \hat{z}\Delta y\gamma^2[1 - 2\ell_y^2\beta^2] \bigr\}
\end{displaymath}

\begin{equation}
-K_0(\rho^\ast/a) {a^2 \over R^2} 2\ell_y\beta^2[\hat{y}\ell_z^2-\hat{z}\ell_y\ell_z] \Biggr\}
\end{equation}

\noindent
So the terms in $a/R$ and $(a/R)^2$ are not consistent with (\ref{eq-Efieldline}) and (\ref{eq-Bfieldline}).

    In helical geometry if we attempt to use (\ref{eq-generalsol}) to try to calculate the electromagnetic field
in the rest frame of line 1, we obtain

\begin{equation}
    {\bf E}^\ast({\bf x}^\ast)\ =\ 0
\end{equation}

\begin{equation}
    {\bf B}^\ast({\bf x}^\ast)\ =
\ { {2g_0} \over {a^2} }\ K_0({\rho^\ast}_1/a)\ {{\hat \ell}^\ast}_1 \biggl ( 1\ -\ { {a^2} \over {{R^\ast}^2} } \biggr )
\end{equation}

\noindent
which is consistent with (\ref{eq-eline*}) but differs from (\ref{eq-bline*}) by a term which is second order in $(a/{R^\ast})$ 
and consequently very small.

\subsection{Energy of Bending}
\label{ssec-energyofbending}

We consider the dependence of the energy of a curved but static line on the radius of curvature.  
We use a curvilinear coordinate system which bends with the line and has $\theta$, $\rho$ and $\phi$
as independent variables with the elements of length $Rd\theta$, $d\rho$, and $\rho d \phi$.  The 
cylindrical radius of curvature $R$ at the point ${\bf x}$ is given by

\begin{equation}
R\ =\ R_0\ +\ \rho\ cos(\phi)
\end{equation} 

\noindent
where $R_0$ is the radius of curvature of the centerline.  We calculate the magnetic energy in a 
length $\Delta\ell$ of the line where

\begin{equation}
\Delta\ell\ =\ R_0\ \Delta\theta\ \ \ .
\end{equation}

\noindent
We find

\begin{displaymath}
\Delta U\ =\ {1 \over 8\pi}\ \int\ R\ d\theta\ d\rho\ \rho\ d\phi\ B^2
\end{displaymath}

\begin{equation}
=\ {{g_0^2\Delta\ell} \over a^4} \int_0^\infty\ d\rho\ \rho {K_0(\rho/a)}^2\ +\ 
{{g_0^2 \Delta\theta} \over {2\pi a^4}} \bigl\lbrack sin(\phi) {{\Biggr \vert}_0^{2\pi}} 
\int_0^\infty d\rho \rho^2 {K_0(\rho/a)}^2\ \ \ .
\end{equation}

\noindent
So we obtain 

\begin{equation}
{{dU} \over {d\ell}}\ =\ {g_0^2 \over {2a^2}}
\end{equation}

\noindent
as in (\ref{eq-T0}).  The energy of the line is independent of its curvature.  One side of the centerline
stretches, but the other side contracts, with no net change in energy.

\section{General Motion in the Common Mode}
\label{sec-generalmotionincommonmode}

    For common mode motion with the line separation perpendicular to the motion,
we have shown that the interaction terms in (\ref{eq-intcommode}) sum to zero and therefore solutions
of type (\ref{eq-yzline}) can propagate at the speed of light on the two lines of an isolated pair.

    Now we would like to consider a line pair moving in the $y-z$ plane with the line
separation also in the $y-z$ plane.  In common mode the separation of the centerlines is
fixed at $2d_c$ perpendicular to both centerlines.  This geometry has the complication that
the corresponding points on the two lines have {\it different} values of $z=\sigma$ though
each is a solution of type (\ref{eq-yzline}).  So 

\begin{equation}
    y_1\ =\ y\ +\ d\ \ \ \ \ \ \ \ \ \ \ \ \ \ y_2\ =\ y\ -\ d
\end{equation}

\begin{equation}
    d\ =\ (0,\ d_c{\hat d})
\end{equation}

\noindent
where $d_c$ is a constant and ${\hat d} \cdot {\hat \ell}=0$.

    Using arguments similar to Appendix~\ref{ssec-instradius} and remembering that here $z$ is an independent
variable---not a field coordinate, we find

\begin{equation}
    { {\partial {\hat d} } \over { \partial z} }\ =\ +\ { {\hat \ell} \over {R_0} }\ { {d\ell} \over {dz} }
\end{equation}

\begin{equation}
    { {\partial {\hat d} } \over { \partial \tau} }\ =\ -\ { {\hat \ell} \over {R_0} }\ { {d\ell} \over {dz} }
\end{equation}

\noindent
where $R_0$ is the radius of curvature of the {\it common mode centerline} of both lines.  So

\begin{equation}
    {\dot d}^\mu\ =\ (0,\ -{\hat \ell})^\mu\ { d_c  \over {R_0} }\ { {d\ell} \over {dz} }
\end{equation}

\begin{equation}
    {d^\prime}^\mu\ =\ (0,\ +{\hat \ell})^\mu\ { d_c  \over {R_0} }\ { {d\ell} \over {dz} }\ \ \ .
\end{equation}

\noindent
We find that (\ref{eq-Lm1}) and (\ref{eq-Lm2}) become

\begin{widetext}
\begin{equation}
 {\cal L}_{m1}\ =\ {T\over c}\ \biggl \vert {{d\ell} \over {dz}} \biggr \vert\ \biggl \vert 1+{d_c\over{R_0}} \biggr \vert \
\sqrt{1-{\beta_\perp}^2}
= {T\over c} \biggl \vert 1 + { {d_c} \over {R_0}} \biggr \vert \ 
\Bigl\{ (y^\prime \cdot {\dot y})^2 \ -\ (y^\prime)^2\ (\dot y)^2 \Bigl\}^{1/2}\
\end{equation}

\begin{equation}
 {\cal L}_{m2}\ =\ {T\over c}\ \biggl \vert {{d\ell} \over {dz}} \biggr \vert\ \biggl \vert 1-{d_c\over{R_0}} \biggr \vert \
\sqrt{1-{\beta_\perp}^2}
= {T\over c} \biggl \vert 1 - { {d_c} \over {R_0}} \biggr \vert \
\Bigl\{ (y^\prime \cdot {\dot y})^2 \ -\ (y^\prime)^2\ (\dot y)^2 \Bigl\}^{1/2}\
\end{equation}
\end{widetext}

\noindent
so the total Lagrangian becomes

\begin{equation}
 {\cal L}_m ={\cal L}_{m1}+{\cal L}_{m2}
= {2T\over c}\Bigl\{ (y^\prime \cdot {\dot y})^2 \ -\ (y^\prime)^2\ (\dot y)^2 \Bigl\}^{1/2}
\end{equation}

\noindent
identical to ${\cal L}_m$ in (\ref{eq-Lm}).

    As in Section~\ref{sssec-commonmodesol} the common mode interaction terms add to zero 
when both lines are referenced to the common mode
centerline.  The two lines do not bend so that they have precisely the same shape.  They have the same center of curvature when
referenced to the common mode centerline, so that their individual centerlines have {\it different} radii of curvature but the
radii of curvature of the fields are identical at any point

    Hence we get the same equations of common mode motion with the separation perpendicular to the plane of the motion
as we get if the separation is in the plane of the motion.  Therefore we presume that the orientation of ${\hat d}$ makes no 
difference to the common mode motion, and common mode solutions
(gravitational waves) of type (\ref{eq-circularline}) can also propagate at the speed of light on an isolated line pair.

\section{Motion of the Bare Electron}
\label{sec-motionbareelectron}

\subsection{Orientation of the Bare Positron-Pole}
\label{ssec-orientation}

\begin{figure}
    \vbox to 3.0in{
\includegraphics{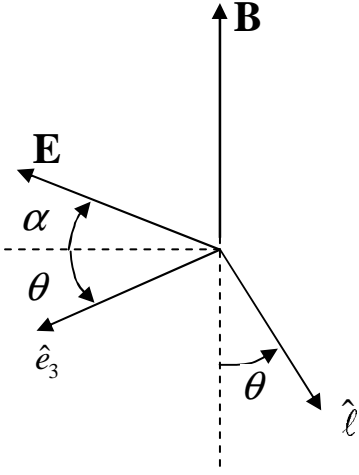}  }
    \caption{\small Diagram of the orientation of the end of a line relative to the electric and magnetic fields.}
\label{fig-orientation}
\end{figure}

    Following the reasoning of Section~\ref{ssec-motionbpp} 
we have shown how the equations of motion of the the bare positron-pole 
(\ref{eq-bpptmotion}-\ref{eq-bpp3motion}) are solved in the case that ${\bf E}\cdot{\bf B}=0$ (where ${\bf E}$ and ${\bf B}$ 
refer to external fields).
If ${\hat \ell}$ is at an angle $\pi-\theta$ from the magnetic induction field ${\bf B}$ and ${\bf E}\cdot{\bf B}=EBsin\alpha$
as shown in Figure~\ref{fig-orientation}, then from (\ref{eq-bpplmotion}) and (\ref{eq-bpp3motion}) we find

\begin{equation}
-{ {B_3} \over {E_\ell} }\ =\ \beta_\perp \ =\ { { B_\ell E_3 + B_0 \sqrt{B_0^2 + E_3^2 - B_\ell^2} } \over {B_0^2 + E_3^2} }
\end{equation}

\begin{widetext}
\begin{equation}
 { { sin(\theta) B} \over {- sin(\theta+\alpha) E} }\ =\ 
{ {BE[-cos(\theta) cos(\theta+\alpha)] + B_0 \sqrt{B_0^2 + E^2 cos^2(\theta+\alpha) - B^2 cos^2(\theta) } } 
\over {B_0^2 + E^2cos^2(\theta+\alpha) } }
\label{eq-transcendental}
\end{equation}
\end{widetext}

\noindent
which is the transcendental equation which must be solved for $\theta$ given $E$, $B$, and $\alpha$.  If $\alpha$ is small, 
we presume that $\theta$ will be small and $\beta_\perp$ is fixed, since all the cosine terms in 
(\ref{eq-transcendental}) are equal to $1$.  Then we have

\begin{equation}
    \theta\ B\ =\ -\ (\theta + \alpha) E \beta_\perp
\end{equation}

\begin{equation}
    \theta\ =\ - { {\alpha \beta_\perp E} \over {B + \beta_\perp E} }
\end{equation}

\noindent
showing that the orientation of $\hat \ell$ is easily determined if $\alpha$ is small.

\subsection{Motion of the Positron}
\label{ssec-motionelectron}

    The electric charge at the end of each line is easily incorporated into the Principle of Least Action using
the electric charge current in (\ref{eq-actionnoh}).  
The only unnatural requirement for this model is putting the charge and pole at (nearly)
the same position at the end of each line, so that the end of a line becomes a almost a dyon, with a spin dependent magnetic
charge.  With both electric and magnetic charge, equation (\ref{eq-motionepparts}), 
the boundary condition to be satisfied at the end of each line, becomes:

\begin{widetext}
\begin{equation}
{{\partial {\cal L}_{m}} \over {\partial y^\prime_\mu} }
+{ {g_0} \over c}\  {\dot y}_\alpha \bigl\langle {\tilde F}^{\alpha \mu} \bigr\rangle_{self}
+{ {g_0} \over c}\  {\dot y}_\alpha \bigl\langle {\tilde F}^{\alpha \mu} \bigr\rangle_{external}
+{ {e_0} \over c}\  {\dot y}_\alpha \bigl\langle { F}^{\alpha \mu} \bigr\rangle_{self}
+{ {e_0} \over c}\  {\dot y}_\alpha \bigl\langle { F}^{\alpha \mu} \bigr\rangle_{external}
=0
\end{equation}

\noindent
We evaluate the boundary condition just outside the end of each line so that the electric charge is included.  The pole self field
is continuous so we can evaluate it just inside the line as we did in Section~\ref{ssec-selfintbareelectronpole}.  
The self field of the electric charge will be
finite if we average over the cross section of the end of the line.  Then, since we switch to the point charge approximation to
evaluate the motion of the bare positron, we can argue that the electric self energy is zero from symmetry.  It is easy to see that
the cross terms in the self energy are zero.  Evaluating them in the rest frame of the positron where ${\dot y}_\alpha^\ast=(1,0)$,
 we see that the pole only interacts with a magnetic field, whereas the charge only interacts with an electric field.  Hence the
self-energy terms are unchanged from the case of the positron-pole and the equation of motion becomes

\begin{equation}
\beta_\ell\ {u^\mu_\perp}\ +\ \sqrt{1-{\beta_\perp}^2}\ (0,{\hat \ell})^\mu \
=\ -{ 1 \over {B_0}}\ [ ( {\bm \beta} \cdot {\bf B},
\ {\bf B} - {\bm \beta} \times {\bf E} )\ +\ f
( {\bm \beta} \cdot {\bf E},
\ {\bf  E} + {\bm \beta} \times {\bf B} )
]
\label{eq-eqofmotionwcharge}
\end{equation}

\noindent
where

\begin{equation}
    f\ \equiv\ { {e_0} \over {g_0} } \ =\ 2\ { {e_0^2} \over {\hbar c} }
\end{equation} 

\noindent 
is a small number if $\varepsilon \simeq 1$ near the Earth---implying that the motion of the bare positron is dominated by the
pole, which is much stronger than the charge.  We can show this directly by combining terms in 
(\ref{eq-eqofmotionwcharge}) to find

\begin{equation}
\beta_\ell\ {u^\mu_\perp}\ +\ \sqrt{1-{\beta_\perp}^2}\ (0,{\hat \ell})^\mu \
=\ -{ 1 \over {B_0}}\ ( {\bm \beta} \cdot \{ {\bf B} + f{\bf E} \},
\ \{ {\bf B} + f{\bf E} \} - {\bm \beta} \times \{ {\bf E} - f{\bf B} \} )
\end{equation}
\end{widetext}

\noindent
that the motion reduces to the case of the bare positron-pole.  We simply make the substitutions

\begin{equation}
    {\bf B}\ \rightarrow\ {\bf B}^\prime\ =  {\bf B} + f  {\bf E}
\end{equation}

\begin{equation}
    \ {\bf E}\ \rightarrow\ {\bf E}^\prime\ =  {\bf E} - f  {\bf B}
\end{equation}

\noindent
and the equation of motion of the bare positron reduces to the equation of motion of the bare positron-pole, in terms of
these transformed fields. 

\section{Response of the Electron}
\label{sec-response}

\subsection{Response to the Electromagnetic Field}
\label{ssec-responseEM}

     This model of the positron (using positive charge for convenience) has an electric charge of $e_0$, a screened magnetic pole
field of effective magnetic charge $g_0$, and a line field that ensures that the total magnetic flux leaving the positron is
zero.  Hence our positron (assuming $\varepsilon = 1$ near the Earth) has the expected electric charge but no true magnetic charge.
It has a magnetic dipole moment, which we are currently unable to model.  Therefore our model of the positron should behave like a
physical positron if we assume that the positron's measured magnetic dipole moment is consistent with our model.

\subsection{Can You Pull on the Positron's String?}
\label{ssec-pullstring}

     The short answer to this question is no.  The string tension is quantized and therefore can not transmit an increase in force
 to the positron.  The string is perfectly elastic.  If you pull on it in the direction of the string, you simply increase the
 string's length and do not affect the positron's motion.  However, if you pull on the string perpendicularly to the string, you
 can affect the orientation of the string, and thereby change the positron's motion.  So this fact brings out the importance of
 spin and developing a model of the magnetic moment that will be needed to make this model complete.

\section{Rigorous Treatment of Gravity}
\label{sec-rigorousgravity}

     Following the line of reasoning which led to (\ref{eq-dndAstar}) and hence to 
(\ref{eq-G}), the line pair density coming from a star of arbitrary material is given by

\begin{equation}
    \biggl ( { {dn} \over {dA} }  \biggr )_{star}\ =\ { M \over {4\pi m_u R^2} } {\biggl < { A \over A_t} \biggr >}_i
\end{equation}

\noindent
where $m_u$ is one atomic mass, $A$ is the mass number of the material (number of nucleons) and $A_t$ is the relative atomic
mass of the material.  The brackets indicate the mass weighted mean and the subscript $i$ indicates that the ratio of the line 
pair density from a star to its mass depends on the material of the active mass (source of the field).  Then the value of 
$G_i$ becomes

\begin{equation}
    G_i \ =\  { {e^2 R_a^2} \over {6 m_e m_u y_{min}^2} }\ {\biggl < { A \over A_t} \biggr >}_i \ \ \ .
\end{equation}

\noindent
The size of the material dependence factor is typically $10^{-3}$.  This factor approaches $1\%$ for hydrogen, but 
it is difficult to measure $G$ with a gas.  There are wild fluctuations\cite{Quinn} in existing measurements of $G$ and this model
may be able to help\cite{McCarthy} bring understanding to these measurements.

     This model can not assess whether or not the gravitational force is proportional to passive mass (on which the field acts)
because we do not model the nucleus. 

\section{Understanding Newton's Second Law}
\label{sec-understandingNewton}

     The discussion of Section~\ref{ssec-calcG} indicates that when a line pair passes a charged particle, it will exert an attractive force on
 this particle, and this force will switch directions as the line pair passes.  Hence the force of a line pair on a particle
{\it will} depend on the velocity of the line pair.  If line pairs in our surroundings dominantly come from stars we can see,
the ``fixed" stars, we expect that the force of the ether on a charged particle {\it will} depend on the velocity of the particle,
instead of only on the acceleration of the particle.

     However, we have seen that the average line pair (Section~\ref{ssec-estimatelpd}) is about a factor of $10^{14}$ larger than the size of our
 visible universe, probably because of inflation.  This indicates that the total volume of our universe is about a factor of
$(10^{14})^3=10^{42}$ greater than the volume of our visible universe!  Matter which went over our horizon early in the
inflationary process will now be moving at velocities far beyond the velocity of light relative to us, but may have stars which are
the source of line pairs near us.

     Thus we postulate that the line pairs near us are moving isotropically with uniformly distributed velocities at least up to
 the velocity of light.  This causes Lorentz invariance.  Apparently these line pairs are not accelerating with respect to our
 inertial system\cite{bbframe}.


\begin{thebibliography}{99}

\bibitem{WMAP} C. Bennet {\it et al.}, ApJS {\bf 148} (2003), p. 24.
\bibitem{standard}  M. K. Gaillard, P. D. Grannis, and F. J. Sciulli, Reviews of
        Modern Physics {\bf71}, S96 (March 1999).
\bibitem{d0comp}B. Abbott {\it et al.}, \D0 Collaboration, Phys. Rev. Letters
 {\bf{82}}, 2457 (1999).
\bibitem{colorflux} See, for example, Donald H. Perkins,
        {\it Introduction to High Energy Physics}, Addison-Wesley (1987), p. 21 and
        B. Hatfield, {\it Quantum Field Theory of Point Particles and Strings},                                                                                                                                   
        Addison-Wesley (1992), p. 479.
\bibitem{Copenhagen} N. Bohr, Phys. Rev. {\bf48}, 696 (1935).  N. Bohr,
         ``Discussion with Einstein on Epistomological Problems in Atomic Physics",
         in {\it Albert Einstein: Philosopher-Scientist}, P. A. Schilp, Ed.,
         Library of Living Philosophers, Evanston, Ill. (1949), p. 232.
\bibitem{Einstein} A. Einstein, B. Podolsky, and N. Rosen, Phys. Rev. {\bf 47}, 777,
         (1935).  Albert Einstein,``Reply to Criticism", in 
         {\it Albert Einstein: Philosopher-Scientist}, P. A. Schilp, Ed.,
         Library of Living Philosophers, Evanston, Ill. (1949), p. 672.
\bibitem{Mermin} See, for example, N. D. Mermin, Reviews of Modern Physics {\bf 65},
         803 (1993).         
\bibitem{Dirac2} P. A. M. Dirac, Phys. Rev. {\bf 74}, 817 (1948).
\bibitem{Jackson}  J. D. Jackson, {\it Classical Electrodynamics}, John Wiley
         \& Sons, Inc. (1975).  (We follow the conventions of the second edition rather
         than the third because the third edition does not use a consistent set of units.)
\bibitem{gaugeTonomura} A. Tonomura, {\it The Quantum World Unveiled by Electron Waves}, 
        World Scientific (1998), p. 133.
\bibitem{bbframe} D. J. Fixsen {\it et al.}, Astrophys. J. {\bf 420}, 445 (1994).
\bibitem{Douglas} Michael R. Douglas, JHEP 05(2003)46, p. 55.
             Roberto Valandro, arXiv:0801.05842v2, Ph. D. thesis, p. 10 
             This is a readable introduction to the Landscape.
\bibitem{Mach}  Ernst Mach, {\it The Science of Mechanics}, written in May, 1883, 
        translated by Thomas J. McCormack, Open Court (1960), p. 284.
\bibitem{Dirac1} P. A. M. Dirac, Proc. R. Soc. London {\bf A133}, 60 (1931).
\bibitem{single} The fact that a spinor changes into its negative under a $2 \pi$
        rotation is irrelevant here.  We are not rotating the wave function.
\bibitem{Londonflux} F. London, {\it Superfluids}, Vol. I, John Wiley \& Sons, Inc. (1950), p. 152.
\bibitem{Saffouri} Some parts of our point of view have been expressed by
        M. H. Saffouri, Nuovo Cimento {\bf 96A}, 1 (1986). 
\bibitem{Faraday} We generally follow the ``line" terminology of Faraday and Maxwell 
        (but not ``line of force").  See J. C. Maxwell,
        {\it A Treatise on Electricity and Magnetism}, Preface to the First Edition, 
        p. ix, Feb. 1, 1873, Dover (1954).
\bibitem{Jackson2} J. D. Jackson, {\it Classical Electrodynamics}, John Wiley
         \& Sons, Inc. (1975), pp. 258-9.
\bibitem{Lamb} H. Lamb, {\it Hydrodynamics}, Dover (1945).
\bibitem{Helmholtz} H. Helmholtz (translated by P. G. Tait), Phil. Mag. and J. of
        Science {\bf 33}, 485 (1867).
\bibitem{Kelvin1} W. Thomson, Phil. Mag. {\bf 10}, 155 (1880).
\bibitem{Kelvin2} W. Thomson, Phil. Mag. and J. of Science {\bf 34}, 15 (1867).
\bibitem{Abrikosov} A. A. Abrikosov, Rev. Mod. Phys. {\bf 76}, 975 (2004).
\bibitem{TypeII} D. Saint-James, G. Sarma, and E. J. Thomas, {\it Type II Superconductivity},
                  Pergammon (1969), p. 48.
\bibitem{Devices} T. Van Duzer and C. W. Turner, {\it Principles of Superconductive Devices and Circuits},
                 Elsevier (1981), p. 310.
\bibitem{Arfken} see, for instance, G. Arfken, {\it Mathematical Methods for Physicists}, 
         Academic Press (1970), p. 510.
\bibitem{London2} F. London, {\it Superfluids}, Vol. I, John Wiley \& Sons, Inc.(1950), p. 29.
\bibitem{diamag}  J. H. Van Vleck, {\it The Theory of Electric and Magnetic Susceptibilities},
                  Oxford University Press (1932), p. 90.  
                  Ya. G. Dorfman, {\it Diamagnetism and the Chemical Bond}, Elsevier (1965), p. 1.
                  A. Tonomura, {\it The Quantum World Unveiled by Electron Waves}, World Scientific (1998), p. 123.
\bibitem{Kittel} C. Kittel, {\it Introduction to Solid State Physics}, $5^{th}$ edition, pp. 373-4 (1976).
\bibitem{curlrho0} The fact that $\Phi(\rho)= \oint {\bf A} \cdot {\bf dx} \rightarrow 0$ as $\rho \rightarrow 0$
                   implies that $\rho=0$ is {\it not} a special point in taking the curl of ${\bf A}$
                   even though the curl of ${\bf A}_{long\ distance}$ is proportional to a delta function.
                   Equation (\ref{eq-bline}) correctly represents the singularity at $\rho = 0$.
\bibitem{Yang}  The phase field is a particular gauge field.  See C. N. Yang, Phys. Rev. Letters {\bf{33}}, 445 (1974) and
                T. T. Wu and C. N. Yang, Phys. Rev. {\bf{D12}}, 3843 (1975).
\bibitem{phaseindep} We presume that the phase fields of two lines add linearly.  Hence the only regions which are important in
                    determining whether or not the wave functions of the lines are single valued, are the regions with 
                    non-zero curl.                     
\bibitem{London} F. London, {\it Superfluids}, Vol. I, John Wiley \& Sons, Inc. (1950), p. 64.  
        We actually quote the relativistic generalization of the standard London 
        equations.   London gives this relativistic generalization here.
\bibitem{Jackson3} J. D. Jackson, {\it Classical Electrodynamics}, John Wiley
         \& Sons, Inc. (1975), p. 118.
\bibitem{Qworld} A. Tonomura, {\it The Quantum World Unveiled by Electron Waves}, World Scientific (1998), p. 154.
\bibitem{Weinberg}  S. Weinberg, {\it Gravitation and Cosmology}, John Wiley \& Sons, Inc. (1972), p. 469.
\bibitem{stringbreak} A.C. Philips, {\it The Physics of Stars}, John Wiley \& Sons, Inc. (1999), p. 62.
\bibitem{Kolb} E. W. Kolb, {\it Dynamics of the Inflationary Era}, FERMILAB-CONF/303-A (1999).
\bibitem{Einstein2} Albert Einstein,``Autobiographical Notes", in
         {\it Albert Einstein: Philosopher-Scientist}, P. A. Schilp, Ed.,
         Library of Living Philosophers, Evanston, Ill. (1949), p. 25.
\bibitem{Maxwell} J. C. Maxwell, {\it A Treatise on Electricity and Magnetism}, volume II 
        (written in 1873), Dover (1954) p.822.
\bibitem{Linde} A. D. Linde, {\it Inflation and Quantum Cosmology}, Academic Press (1990), p. 15.
\bibitem{HindKib} M. B. Hindmarsh and T. W. B. Kibble, Rep. Prog. Phys. {\bf 58}, 487 (1995).
\bibitem{Peebles} P. J. E. Peebles, {\it Principles of Physical Cosmology}, Princeton University Press
        (1993), p. 103.
\bibitem{Hofstadter}  R. Hofstadter, H. R. Fechter, and J. A. McIntyre, Phys. Rev. {\bf 92}, 978 (1953).
\bibitem{highfield} S. Chakrabarty, D. Bandyopadhay, S. Pal, Phys. Rev. Lett. {\bf 78}, 2898 (1997).
\bibitem{Vilenkin} A. Vilenkin, Physics Reports {\bf 121}, 283 (1985).
\bibitem{Jacksonmac} J. D. Jackson, {\it Classical Electrodynamics}, John Wiley
         \& Sons, Inc. (1975), p. 228.
\bibitem{Jacksondeldotd} J. D. Jackson, {\it Classical Electrodynamics}, John Wiley
         \& Sons, Inc. (1975), p. 226.
\bibitem{Tonomuraphi0} A Tonomura {\it et al.}, Phys. Rev. Lett. {\bf 51}, 331 (1983).  
A. Tonomura, {\it The Quantum World Unveiled by Electron Waves}, World Scientific (1998), p. 118.
\bibitem{Beckersphere} R. Becker, {\it Electromagnetic Fields and Interactions}, Blaisdell (1964), p. 94.
\bibitem{Goddard} P. Goddard, J. Goldstone, C. Rebbi and C.B. Thorn, Nuclear Physics {\bf B56} (1973) pp. 111-114.
\bibitem{Barutarb} A. O. Barut, {\it Electrodynamics and Classical Theory of Fields and Particles}, 
        Dover (1964,1980), p. 60.
\bibitem{BJD} J. D. Bjorken and S. D. Drell, {\it Relativistic Quantum Mechanics}, McGraw-Hill (1964), p. 52.
\bibitem{BarutBornzin} A. O. Barut and G. L. Bornzin, Nuclear Physics {\bf B81}, 480 (1974).  There is a sign mistake in
        equation (12) of this reference.
\bibitem{Jackson4} J. D. Jackson, {\it Classical Electrodynamics}, John Wiley
         \& Sons, Inc. (1975), p. 222.
\bibitem{polefield}  The nonrelativistic approximation certainly fails at the ends of the line.
         So (4.24) should be thought of as valid away from the ends of the line, if the transverse motion
         is nonrelativistic.
\bibitem{MTW}  C. W. Misner, K. S. Thorne, and J. A. Wheeler, {\it Gravitation}, Freeman (1973), p. 165.
\bibitem{BjDrellaction} J. D. Bjorken and S. D. Drell, {\it Relativistic Quantum Fields}, McGraw-Hill (1965), p. 70.
\bibitem{Qworld2} A. Tonomura, {\it The Quantum World Unveiled by Electron Waves}, World Scientific (1998), p. 143.
\bibitem{MTWform} C. W. Misner, K. S. Thorne, and J. A. Wheeler, {\it Gravitation}, Freeman (1973), p. 53.
\bibitem{Peeblesworldtime} P. J. E. Peebles, {\it Principles of Physical Cosmology}, Princeton University Press
        (1993), p. 59.  M. B. Hindmarsh and T. W. B. Kibble, Rep. Prog. Phys. {\bf 58}, 486 (1995).
\bibitem{JacksonLW} J. D. Jackson, {\it Classical Electrodynamics}, John Wiley
         \& Sons, Inc. (1975), p. 654.
\bibitem{Schwingerdyon} J. Schwinger, Science {\bf 165}, 757 (1969).
\bibitem{Jacksonmaxmod} J. D. Jackson, {\it Classical Electrodynamics}, John Wiley
         \& Sons, Inc. (1975), p. 252.
\bibitem{JacksonGoldhaber} J. D. Jackson, {\it Classical Electrodynamics}, John Wiley
         \& Sons, Inc. (1975), p. 256.  A. S. Goldhaber, Phys. Rev. {\bf 140B}, 1407 (1965). 
\bibitem{externalforce} We presume, but do not prove, that the response can be derived from (\ref{eq-actionnoh}) 
         using the superconducting current which is
         induced on the outer surface of each line pair (and extending a few penetration depths into the pair).
\bibitem{Beckeralpha} R. Becker, {\it Electromagnetic Fields and Interactions}, Blaisdell (1964), p. 111.
         If we approximate a line pair by a series of superconducting spheres we can see that the polarizability
         per unit length of a line pair
         should be approximately given by $\alpha(\phi) \sim R_a^2/sin\phi$ where $R_a=k_a a$ is the cylindrical 
         outer radius of a line pair.  See Figure~\ref{fig-lpstars12}.
\bibitem{JacksonmB} J. D. Jackson, {\it Classical Electrodynamics}, John Wiley
         \& Sons, Inc. (1975), p. 185.
\bibitem{Kittelmass} C. Kittel, {\it Introduction to Solid State Physics}, $5^{th}$ edition, p. 218 (1976).
\bibitem{MeliasagA} Fulvio Melia, {\it The Black Hole at the Center of our Galaxy}, Princeton University Press (2003), p. 49.
\bibitem{MeliaM87} Fulvio Melia, {\it The Black Hole at the Center of our Galaxy}, Princeton University Press (2003), p. 169.
\bibitem{mabove} This variation cannot be measured via the gravitational force because of the weak principle of equivalence.
         Use of another type of force is required.
\bibitem{centralbulge} M. Zeilik, S. Gregory, and E. Smith, {\it Introductory Astronomy and Astrophysics}, Saunders (1992), p. 279.
         D. Elmegreen, {\it Galaxies and Galactic Structure}, Prentice Hall (1998), p. 150.
\bibitem{Fairbank} F. C. Witteborn and W. M. Fairbank, Phys. Rev. Lett. {\bf 19}, 1049 (1967).
\bibitem{LIGO} B. Barish and R. Weiss, {\it LIGO and the Detection of Gravitational Waves}, Physics Today {\bf 52}, 42 (1999).
\bibitem{equivalence}  S. Weinberg, {\it Gravitation and Cosmology}, John Wiley \& Sons, Inc. (1972), p. 68.
\bibitem{Jacksonspecial} J. D. Jackson, {\it Classical Electrodynamics}, John Wiley \& Sons, Inc. (1975), p. 503.
\bibitem{Shutzgravwave} B. F. Shutz, {\it First Course in General Relativity}, Cambridge University Press (1998), p. 214.
\bibitem{MCM} This idea was first suggested to me by M. C. McCarthy.
\bibitem{Messiah} A. Messiah, {\it Quantum Mechanics}, John Wiley \& Sons, Inc. (1958), p. 118.
\bibitem{ResHal} R. Resnick and D. Halliday, {\it Physics for Students of Science and Engineering}, John Wiley \& Sons, Inc.
(1961), p. 515.
\bibitem{Weinbergevol}  S. Weinberg, {\it Gravitation and Cosmology}, John Wiley \& Sons, Inc. (1972), p. 472.
\bibitem{Meliarot} Fulvio Melia, {\it The Black Hole at the Center of our Galaxy}, Princeton University Press (2003), p. 171.
\bibitem{YangPauli} C. N. Yang, colloquium, May 4, 1999, State University of New York at Stony Brook.
\bibitem{spinglass} J. Mydosh, {\it Spin Glasses: An Experimental Introduction}, Taylor and Francis (1993), p. 144.
\bibitem{Lieb} Elliot Lieb, colloquium, March 2, 1999, State University of New York at Stony Brook.  
\bibitem{Cooper}  Leon Cooper, private communication.
\bibitem{AitchisonHey} I. Aitchison and A. Hey, {\it Gauge Theories in Particle Physics}, Adam Hilger Ltd. (1982), p. 205.
\bibitem{KvK}  Klaus von Klitzing, Rev. Mod. Phys. {\bf 58}, 522 (1985).
\bibitem{quantumhall} J. K. Jain and R. K. Kamila, {\it Composite Fermions: Particles of the Lowest Landau Level}, p. 12
                      in {\it Composite Fermions}, ed. by O Heinonen, World Scientific (1998).
\bibitem{Tonomurafluxquant} A. Tonomura {\it et al.}, Phys. Rev. Lett. {\bf 51}, 331 (1983).
\bibitem{Tonomuraholography} A. Tonomura, {\it Electron Holography}, Springer (1998).
\bibitem{Sands} Matthew Sands, private communication.
\bibitem{Quinn} T.J. Quinn, Nature(London) {\bf 408} 919 (2000).
\bibitem{McCarthy} R. McCarthy, in preparation for Physical Review.


\end{thebibliography}
\end{document}